\documentclass[a4paper,onesided,12pt]{report} 
\usepackage{EMU_ThesisPHD2}
\usepackage{amsmath}
\usepackage{amssymb}
\usepackage{graphicx}
\usepackage{esint}
\usepackage{pslatex}
\usepackage{graphics}
\usepackage{cite}

\degree{ Ph.D. in Physics }
\subyear{2016}
\Dept{Physics}
\InstituteDirector{Prof. Dr. Cem Tanova}
\DeptChair{Prof. Dr. Mustafa Halilsoy}
\cosuperi{Assoc. Prof. Dr. Habib Mazharimousavi}
\supervisor{Prof. Dr. Mustafa Halilsoy}
\examineri{Prof. Dr. Ayhan Bilsel}
\examinerii{Prof. Dr. Durmu\c{s} A. Demir}
\examineriii{Assoc. Prof. Dr. Tahsin \c{C}agr{\i} \c{S}i\c{s}man}
\examineriv{Assoc. Prof. Dr. \.{I}zzet Sakall{\i}}
\dateofapproval{2016}
\begin{document}
\pagenumbering{roman}
\title{Studies on Thin-shells and Thin-shell Wormholes}
\author{Ali {\"O}vg{\"u}n}
\makephdtitle 
\makeapprovalpage

\begin{abstract}
\vspace{-2.0cm}

\noindent The study of traversable wormholes is very hot topic for the past
30 years. One of the best possible way to make traversable wormhole is using the thin-shells to cut and paste two spacetime which has tunnel from one region of space-time to another, through which a traveler might freely pass in wormhole throat. These geometries need an exotic matter which involves a stress-energy tensor that violates the null energy condition. However, this method can be used to minimize the amount of the exotic matter.  The goal of this thesis study is to study on thin-shell and thin-shell wormholes in general relativity in 2+1 and 3+1 dimensions. We also investigate the stability of such objects. 


\end{abstract}

\begin{ozet}

\vspace{-2.0cm}

 \noindent Solucan delikleri bilim ve bilim kurgu d\"{u}nyas\i n\i n en pop\"{u}ler
konular\i ndan biridir. 30 sene boyunca pop\"{u}leritesini daha da art\i
rd\i . Olas\i\ solucan deli\u{g}i yapabilmek i\c{c}in en kullan\i \c{s}l\i\
ve kararl\i\ y\"{o}ntemlerden biri Einsteinin yer\c{c}ekimi kuram\i\ i\c{c}%
erisinde ince-kabuklu uzay solucan deli\u{g}i yapmakt\i r. Bu kuramlarda 
\"{o}nemli olan\i\ ge\c{c}i\c{s}i yapacak olan\i n, solucan deli\u{g}inin bo%
\u{g}az\i ndan serbest\c{c}e ge\c{c}i\c{s}ine olanak vermesi ve belirli \c{s}%
artlar\i\ sa\u{g}lamas\i d\i r, ve egzotik madde miktar\i n\i\ en d\"{u}\c{s}%
\"{u}k seviyeye \c{c}ekebilmektir. Bu \c{c}al\i \c{s}mada kara deliklerin
etraf\i nda olu\c{s}abilecek ince-kabuklu zar\i , ve bunlar\i\ kullanarak
kararl\i\ yap\i da solucan deli\u{g}i olu\c{s}turmaya \c{c}al\i \c{s}t\i k.

\noindent \textbf{Anahtar Kelimeler}: Solucan Delikleri, Kara Delikler,
 \end{ozet}



\newpage \vspace*{11.5cm} \centerline{"Imagination is more important than knowledge." - Albert Einstein}

\begin{acknowledgements} 
\vspace{-1.2cm}

\noindent I would like to express my deep gratitude to Prof. Dr. Mustafa Halilsoy, my supervisor, for his patient guidance, enthusiastic encouragement and useful critiques of this research work. I also thank to my co-supervisor Assoc. Prof. Dr. Habib Mazharimousavi for willingness to spend his time discussing to me and for help on some difficult calculations.

\noindent I would also like to thank Assoc. Prof. Dr. \.{I}zzet Sakall{\i}, for his advice and assistance in keeping my progress on schedule. 
My grateful thanks are also extended to Prof. Dr. {\"O}zay G{\"u}rtu\u{g}.

\noindent I would like to thank my friends in the Department of Physics and Chemistry, the Gravity and General Relativity Group for their support and for all the fun we have had during this great time. I wish to thank also other my friends for their support and encouragement throughout my study.

\noindent Finally, I would like to thank my family. 

\textbf{This Ph.D\ thesis is based on the following  13 SCI and 1 SCI-expanded papers :}

\begin{enumerate}
\item Thin-shell wormholes from the regular Hayward black hole, M. Halilsoy,
\textbf{A. Ovgun} and S. H. Mazharimousavi, \textit{Eur. Phys. J. C 74, 2796 (2014).}

\item Tunnelling of vector particles from Lorentzian wormholes in 3+1
dimensions, I. Sakalli and \textbf{A. Ovgun}, \textit{Eur. Phys. J. Plus 130, no. 6, 110 (2015).}

\item On a Particular Thin-shell Wormhole, \textbf{A. Ovgun} and I. Sakalli,
\textit{arXiv:1507.03949 (accepted for publication in Theoretical and Mathematical Physics).}

\textbf{Other papers by the author:}

\item Existence of wormholes in the spherical stellar systems, \textbf{A. Ovgun} and
M. Halilsoy, \textit{Astrophys Space Sci  361, 214 (2016).}

\item Gravitinos Tunneling From Traversable Lorentzian Wormholes, I. Sakalli
and \textbf{A. Ovgun}, \textit{Astrophys. Space Sci. 359, 32 (2015).
}
\item Gravitational Lensing Effect on the Hawking Radiation of Dyonic Black
Holes, I. Sakalli, \textbf{A. Ovgun} and S. F. Mirekhtiary. \textit{Int. J. Geom. Meth. Mod. Phys. 11, no. 08, 1450074 (2014).}

\item Uninformed Hawking Radiation, I. Sakalli and \textbf{A. Ovgun}, \textit{Europhys. Lett. 110, no. 1, 10008 (2015).}

\item Hawking Radiation of Spin-1 Particles From Three Dimensional Rotating
Hairy Black Hole, I. Sakalli and \textbf{A. Ovgun}, \textit{J. Exp.Theor. Phys. 121, no. 3, 404 (2015).}

\item Quantum Tunneling of Massive Spin-1 Particles From Non-stationary
Metrics, I. Sakalli and \textbf{A. Ovgun.}, \textit{Gen. Rel. Grav. 48, no. 1, 1 (2016).}

\item Entangled Particles Tunneling From a Schwarzschild Black Hole immersed
in an Electromagnetic Universe with GUP, \textbf{A. Ovgun}, \textit{Int. J. Theor. Phys. 55, 6, 2919 (2016).}

\item Hawking Radiation of Mass Generating Particles From Dyonic Reissner
Nordstrom Black Hole, I. Sakalli and \textbf{A. Ovgun}, \textit{arXiv:1601.04040 (accepted for publication in Journal of Astrophysics and
  Astronomy).}

\item Tunneling of Massive Vector Particles From Rotating Charged Black
Strings, K. Jusufi and \textbf{A. Ovgun}, \textit{Astrophys Space Sci 361, 207 (2016).}

\item Massive Vector Particles Tunneling From Noncommutative Charged Black
Holes and its GUP-corrected Thermodynamics, \textbf{A. Ovgun} and K. Jusufi,
 \textit{Eur. Phys. J. Plus 131, 177 (2016).}

\item Black hole radiation of massive spin-2 particles in (3+1) dimensions, I. Sakalli, \textbf{A. Ovgun}, \textit{Eur. Phys. J. Plus 131, 184 (2016).}

\end{enumerate}

\end{acknowledgements}

\tableofcontents

\chapter{INTRODUCTION}

\pagenumbering{arabic}

\section{General Relativity}
\vspace{-0.75cm}
100 years ago, Albert Einstein presented to the world his theory of General
Relativity (GR) which is said that space and time are not absolute, but can
be distorted or warped by matter. Einstein's genius was in his willingness
to confront the contradictions between different branches of physics by
questioning assumptions so basic that nobody else saw them as assumptions.
One prediction of the GR is that in particular, a cataclysmic event such as
the collapse of a star could send shock waves through space, gravitational
waves (GWs),discovered and announced in February 2016 by LIGO (Laser
Interferometer Gravitational-Wave Observatory) \cite{ligo,ligo1}. The GWs
the other proposal that matter can warp space and time leads to many other
predictions, notably that light passing a massive body will appear bent to a
distant observer, whereas time will appear stretched. Each of these
phenomena has been observed; the bending of distant starlight by the sun was
first observed in 1919, and the synchronization of clocks in GPS satellites
with earthbound clocks has to take account of the fact that clocks on Earth
are in a strong gravitational field. Another prediction of GR is that if
enough matter is concentrated in a small volume, the space in its vicinity
will be so warped that it will curve in on itself to the extent that even
light cannot escape\cite{3}. Astronomical observations suggest that each
galaxy has a super massive Black Hole (BH) at its centre\cite{blag}.
Mind-blowing concepts wormholes (WHs) can be considered his last prediction,
which is also known as Einstein--Rosen bridge\cite{erosen}. It is a
short-cut connecting two separate points in spacetime. Moreover, GR is used
as a basis of nowadays most prominent cosmological models\cite{ao17}.
Einstein's equations start to break down in the singularities of BHs. Before
going to study deeply BHs and WHs, lets shortly review the GR.

The GR is the\ Einstein's theory of gravity and it is a set of non-linear
partial differential equations (PDEs). The Einstein field equations are \cite{3}
\begin{equation}
G_{\mu \nu }=8\pi GT_{\mu \nu }
\end{equation}
where $G$ is the Newton constant, $G_{\mu \nu }$ is a Einstein tensor and $%
T_{\mu \nu }$ is a energy momentum tensor, which includes both energy and
momentum densities as well as stress (that is, pressure and shear), which
must satisfy the relation of $\nabla ^{\mu }T_{\mu \nu }=0$. This
relation can be called as the equation of motion for the matter fields.
Furthermore, the Einstein tensor $G_{\mu \nu }$ is also divergence free $%
\nabla ^{\mu }G_{\mu \nu }=0.$
Noted that the Einstein tensor $G_{\mu \nu }$ is 
\begin{equation}
G_{\mu \nu }=R_{\mu \nu }-\frac{1}{2}g_{\mu \nu }R \label{EFE}
\end{equation}%
where $R_{\mu \nu }$ is called the Ricci tensor \cite{3}
\begin{equation}
R_{\mu \nu }=\partial _{\rho }\Gamma _{\nu \mu }^{\rho }-\partial _{\nu
}\Gamma _{\rho \mu }^{\rho }+\Gamma _{\rho \lambda }^{\rho }\Gamma _{\nu \mu
}^{\lambda }-\Gamma _{\nu \lambda }^{\rho }\Gamma _{\rho \mu }^{\lambda }.
\end{equation}%
The Ricci scalar of curvature scalar $R$ is the contraction of the Ricci
tensor.%
\begin{equation}
R=g^{\mu \nu }R_{\mu \nu }
\end{equation}%
where \ the Christoffel symbols $\Gamma _{\nu \mu }^{\rho }$ are defined by 
\begin{equation}
\Gamma _{\mu \nu }^{\rho }=\frac{1}{2}g^{\rho \sigma }\left( g_{\sigma \mu,
\nu}+g_{\sigma \nu, \mu }-g_{\mu \nu, \sigma }\right)
\end{equation}
in which $g^{\sigma \mu }$ is the inverse of the metric function and it is
the main ingredient of the Einstein field equations.

GR is based on two important postulates. The first one is the principle of
special relativity. The second one is the equivalence principle. Einstein's
`Newton's apple' which in sighted to gravitation was the equivalence
principle. As an example, consider two elevators one is at rest on the Earth 
and the other is accelerating in space. Inside the
elevator suppose that there is no windows so it is impossible to realize the
difference between gravity and acceleration. Thus, it gives the same results
as observed in uniform motion unaffected by gravity. In addition, gravity
bends light in which a photon crossing the elevator accelerating into space,
the photon appears to fall downward.
\vspace{-0.75cm}
\section{Black Holes}
\vspace{-0.75cm}
A BH is an object that is so compact that its gravitational force is strong
enough to prevent light or anything else from escaping. It has a singularity
where all the matter in it is squeezed into a region of infinitely small
volume. There is an event horizon\ which is an imaginary sphere that
measures how close to the singularity you can safely get\cite{ao12,ao11}.

By far the most important solution in this case is that discovered by Karl Schwarzschild, 
which describes spherically symmetric vacuum spacetimes.
The fact that the Schwarzschild metric is not just a good solution,
 but is the unique spherically symmetric vacuum solution, is known as Birkhoff's theorem.
It only has a mass, but no electric charge, and no spin. Karl Schwarzschild discovered this BH geometry at the close of
1915\cite{sch}, within weeks of Einstein presenting his final theory of GR.
Real BHs are likely to be more complicated than the Schwarzschild geometry:
real BHs probably spin, and the ones that astronomers see are not isolated,
but are feasting on material from their surroundings.

The no-hair theorem states that the geometry outside (but not inside!) the horizon of an isolated BH is
characterized by just three quantities: Mass, Electric charge, Spin. BHs are
thus among the simplest of all nature's creations. When a BH first forms
from the collapse of the core of a massive star, it is not at all a no-hair
BH. Rather, the newly collapsed BH wobbles about, radiating GWs. The GWs
carry away energy, settling the BH towards a state where it can no longer
radiate. This is the no-hair state.

In this thesis, we will use some of the exact solution of the Einstein field equations to construct WHs and thin-shells. 
In GR, Birkhoff's theorem states that any spherically
symmetric solution of the vacuum field equations must be static and
asymptotically flat. This means that the exterior solutions (i.e. the
spacetime outside of a spherical, non-rotating, gravitating body) must be
given by the Schwarzschild metric.
\begin{equation}
ds^{2}=-\left( 1-\frac{2M}{r}\right) dt^{2}+\frac{1}{\left( 1-\frac{2M}{r}%
\right) }dr^{2}+r^{2}(d\theta ^{2}+\sin^{2} \theta d\phi ^{2})
\end{equation}
Noted that it is singular at $r_{s}=2M.$To see that this is a true
singularity one must look at quantities that are independent of the choice
of coordinates. One such important quantity is the Kretschmann scalar,
which is given by
\begin{equation}
R^{\alpha \beta \gamma \delta }R_{\alpha \beta \gamma \delta }=\frac{48M^{2}%
}{r^{6}}
\end{equation}
Note that Schwarzschild solution has those properties: spherically
symmetric, static, coordinates adapted to the time-like Killing vector field 
$X_{a}$, time-symmetric and time translation invariant, a
hyper-surface-orthogonal time-like Killing vector field $X^{a}$,
asymptotically flat and geometric mass $\frac{GM}{c^{2}}$.
\vspace{-0.75cm}
\section{Wormholes}
\vspace{-0.75cm}
Wormholes (WHs) is a hypothetical connection between widely separated regions of space–time \cite{morris,yurtsever}.
Although Flamm's work on the WH physics dates back to 1916, in connection
with the newly found Schwarzschild solution \cite{flamm}, WH solutions were
firstly considered from physics standpoint by Einstein and Rosen (ER) in
1935, which is known today as ER bridges connecting two identical sheets 
\cite{erosen}. Then in 1955 Wheeler used "geons" (self-gravitating bundles
of electromagnetic fields) by giving the first diagram of a
doubly-connected-space \cite{wheeler}. Wheeler added the term "wormhole" to
the physics literature, however he defined it at the quantum scale. After
that, first traversable WH was proposed by Morris-Thorne in 1988 \cite%
{morris}. Then Morris, Thorne, and Yurtsever investigated the requirements
of the energy condition for WHs \cite{yurtsever}. After while, Visser
constructed a technical way to make thin-shell WHs which thoroughly surveyed
the research landscape as of 1988 \cite{visser1,visser2}. After this, there
are many papers written to support this idea \cite{witt,MV1,MV2,MV3,MV4,MV5}%
. WHs are existed in the theory of GR, which is our best description of the
Universe. But experimentally there is no \ evidence and no one has any idea
how they would be created. 
\begin{figure}[h!]

\centering
\includegraphics[width=0.80\textwidth]{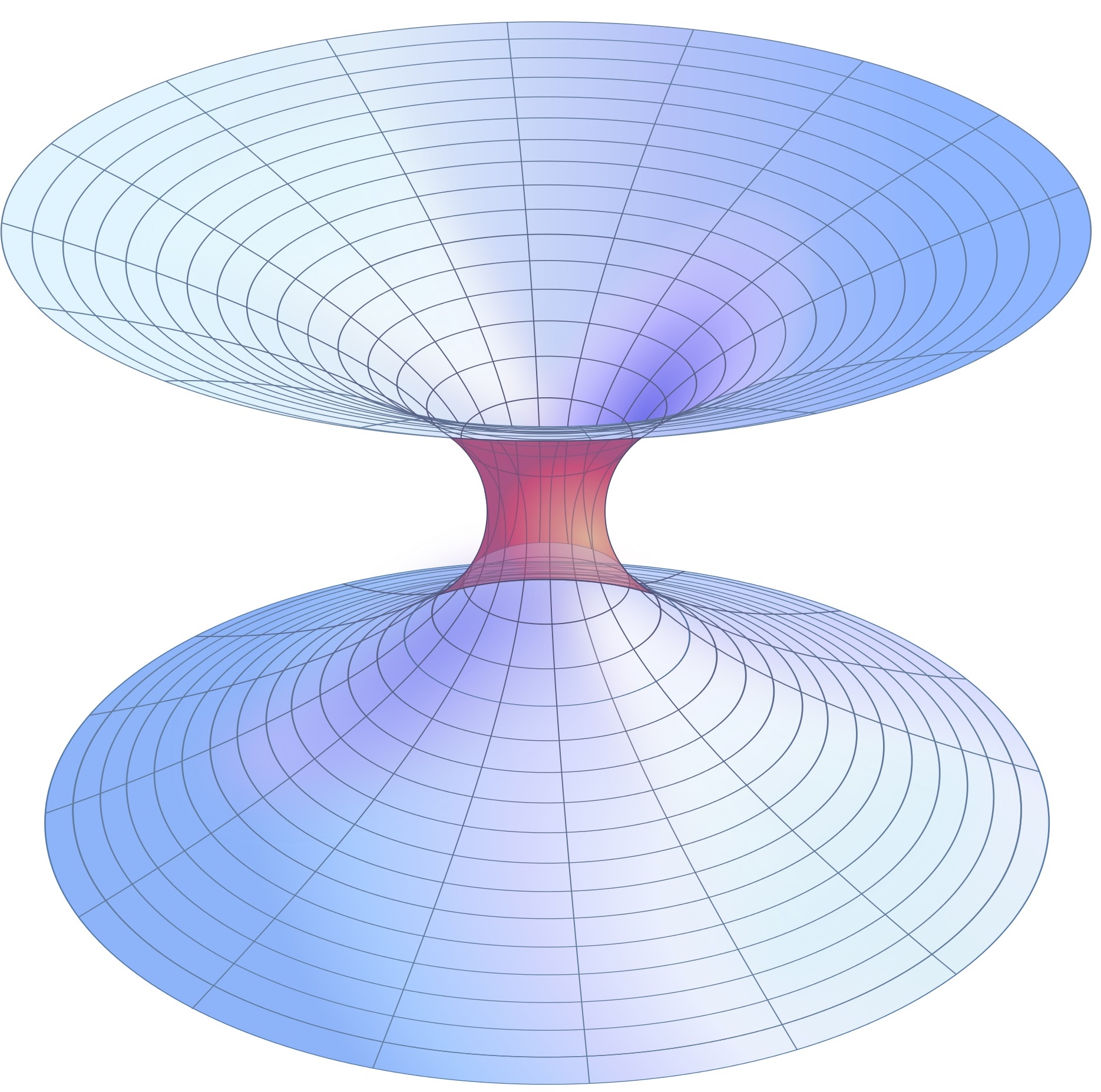} \label{fig:wormhole}
\caption{Wormhole}
\end{figure}
\vspace{-0.75cm}
\subsection{Traversable Lorentzian Wormholes}
\vspace{-0.75cm}
A WH is any compact region on the space time without any singularity,
however with a mouth to allow entrance\cite{loboreview1,loboreview2}. \
Firstly one considers an interesting exotic spacetime metric and solves the
Einstein field equation, then finds the exotic matter \ needed as a source
responsible for the respective geometry. It is needed the exotic matter
which violates the null energy condition,$\left( T_{\mu \nu }k^{\mu }k^{\nu
}\geq 0\right) $, where $k^{\mu }$ is a null vector. It also violates the
causality by allowing closed time-like curves. Furthermore, the other
interesting outcome is that time travel is possible without excess the speed
of light. Those outcomes are based on theoretical solutions and useful for
\textquotedblleft gedanken-experiments\textquotedblright. As it is well
known the Casimir effect has similar features that violates this
condition in nature\cite{cassimir}. Now, lets give an example of a
traversable WH metric which is given by
\begin{equation}
ds^{2}=-c^{2}dt^{2}+dl^{2}+(k^{2}+l^{2})(d\theta ^{2}+\sin^{2}\theta d\phi
^{2})
\end{equation}
The first defined traversable WH is Morris Thorne WH\cite{morris} 
\begin{equation}
ds^{2}=-e^{2f(r)}dt^{2}+\frac{1}{1-\frac{b(r)}{r}}dr^{2}+r^{2}(d\theta
^{2}+\sin^{2}\theta d\phi ^{2})
\end{equation}
where $f(r)$ is the red-shift function\ (the lapse function) \ that change
in frequency of electromagnetic radiation in gravitational field and $b(r)$
is the shape function. At $t=const.$ and $t=\frac{\pi }{2}$, the 2-curved
surface is embedded into 3-dimensional Euclidean space 
\begin{equation}
d\tilde{s}^{2}=\frac{1}{1-\frac{b(r)}{r}}dr^{2}+r^{2}d\phi
^{2}=dz^{2}+dr^{2}+r^{2}d\phi ^{2}
\end{equation}%
To be a solution of a WH, one needs to impose that the geometrical throat flares out
conditions, the minimality of the wormhole throat, which are given by \cite{visser1}
\begin{equation}
\frac{d^{2}r}{dz^{2}}=\frac{b-b^{\prime }r}{2b^{2}}>0  \label{flareout}
\end{equation}
with new radial coordinate $l$ (proper distance), while $r>b$%
\begin{equation}
ds^{2}=-e^{2f(r)}dt^{2}+dl^{2}+r(l)^{2}(d\theta ^{2}+\sin^{2}\theta d\phi
^{2})\end{equation}%
Although there is a coordinate singularity where the metric coefficient $%
g_{rr}$ diverges at the throat, the proper radial distance which runs from $-\infty$ to $\infty $can be redefined as
\begin{equation}
\frac{dl}{dr}=\pm \left( 1-\frac{b}{r}\right) ^{-\frac{1}{2}}.
\end{equation}
\textbf{Properties of the metric:}
\vspace{-0.75cm}

-No event horizons ($g_{tt}$ $=-e^{2f(r)}\neq 0$),

-$f(r)$ must be finite everywhere,

- Spherically symmetric and static:
Static means that the metric does not change over time and irrotate, on the other way,
 stationary spacetime means that the metric does not change in time, however it can rotate. Hence, the Kerr metric is a stationary spacetime and not not static, the Schwarzschild solution is an example of static spacetime.

-Solutions of the Einstein field equations,

-Physically reasonable stress energy tensor,

-Radial coordinate $r$ such that circumference of circle centered around
throat given by $2\pi r,$

-$r$ decreases from $+\infty $ to $b=b_{0}$ (minimum radius) at throat, then
increases from $b_{0}$ to $+\infty ,$

-At throat exists coordinate singularity where $r$ component diverges,
-Reasonable transit times,

-Proper radial distance $l(r)$ runs from $-\infty $ to $+\infty $ and vice
versa,

-Throat connecting two asymptotically flat regions of spacetime,

-Bearable tidal gravitational forces,

-Stable against perturbations,

-Physically reasonable construction materials.

The Einstein field equations are \cite{ao13}
\begin{equation}
G_{t}^{t}=-\frac{b^{\prime }}{r^{2}},  \label{gtt}
\end{equation}%
\begin{equation}
G_{r}^{r}=\frac{-b}{r^{3}}+2(1-\frac{b}{r})\frac{f^{\prime }}{r},
\end{equation}%
\begin{equation}
G_{\theta }^{\theta }=G_{\phi }^{\phi }=\left( 1-\frac{b}{r}\right) \left[
f^{\prime \prime }++{f^{\prime }}^{2}+\frac{f^{\prime }}{r}-\left( f^{\prime
}+\frac{1}{r}\right) \left\{ \frac{b^{\prime }r-b}{2r(r-b)}\right\} \right] ,
\end{equation}
The only non-zero components of $T_{\nu }^{\mu }$ are 
\begin{equation}
T_{t}^{t}=-\rho,  \label{Tr}
\end{equation}%
\begin{equation}
T_{r}^{r}=p_{r},
\end{equation}
\begin{equation}
T_{\theta }^{\theta }=T_{\phi }^{\phi }=p_{t}.  \label{Ttheta}
\end{equation}
\begin{eqnarray}
\rho &=&\frac{1}{8\pi }\,\frac{b^{\prime }}{r^{2}}\,,  \label{rhoWH} \\
p_{r} &=&\frac{1}{8\pi }\,\left[ -\frac{b}{r^{3}}+2\left( 1-\frac{b}{r}%
\right) \frac{f^{\prime }}{r}\right] \,,  \label{tauWH} \\
p_{t} &=&\frac{1}{8\pi }\left( 1-\frac{b}{r}\right) \left[ f^{\prime \prime
}+f^{\prime 2}-\frac{b^{\prime }r-b}{2r^{2}(1-b/r)}f^{\prime }-\frac{%
b^{\prime }r-b}{2r^{3}(1-b/r)}+\frac{f^{\prime }}{r}\right] \,.
\label{pressureWH}
\end{eqnarray}%
At the throat they reduce to the simplest form 
\begin{eqnarray}
\rho (r_{0}) &=&\frac{1}{8\pi }\,\frac{b^{\prime }(r_{0})}{r_{0}^{2}}\,, \\
p_{r} (r_{0}) &=&\frac{1}{8\pi r_{0}^{2}}\,, \\
p_{t} (r_{0}) &=&\frac{1}{8\pi }\,\frac{1-b^{\prime }(r_{0})}{2r_{0}^{2}}%
\;(1+r_{0}f^{\prime }(r_{0}))\,.
\end{eqnarray}
Note that the sign of $B^{\prime }(r)$ and energy density $\rho (r)$ must be
same to minimize the exotic matter, since the condition of $B^{\prime }(r)>0$
must be satisfied. Moreover, the next condition is the flare outward of the
embedding surface ( $B^{\prime }(r)<B(r)/r$ ) at/near the throat. Therefore
it shows that $p_{r}(r)+\rho (r)<0$ on this regime, in which the radial
pressure is named by $p_{r}(r)$. So, the exotic matter which supports this
WH spacetime violates the null energy condition.

The four velocity is calculated by $U^{\mu }=dx^{\mu }/{d\tau }%
=(U^{\,t},0,0,0)=(e^{-f(r)},0,0,0)$ and four acceleration is $a^{\mu
}=U^{\mu }{}_{;\nu }\,U^{\nu }$, so one obtains 
\begin{eqnarray}
a^{t} &=&0\,,  \\
a^{r} &=&\Gamma _{tt}^{r}\,\left( {\frac{dt}{{d\tau }}}\right) ^{2}={f(r)^{\prime }\,(1-b/r)\,,  \label{radial-acc}
}\end{eqnarray}%
in which the prime $"^{\prime }"$ is "$\frac{dt}{{dr}}$". Furthermore, the
geodesics equation \ is used for a radial moving particle to obtain the
equation of motion as following 
\begin{equation}
{\frac{{d^{\,2}r}}{d{\tau }^{2}}}=-\Gamma _{tt}^{r}\,\left( {\frac{dt}{{%
d\tau }}}\right) ^{2}=-a^{r}\,.  \label{radial-accel}
\end{equation}
In addition, static observers are geodesic for $f^{\prime }(r)=0$. WH
has an attractive feature\ if $a^{r}>0$ and repulsive feature\ if $a^{r}<0$.
The sign of $f^{\prime }$ is important for the behaviour of the particle
geodesics. The shape function is obtained by 
\begin{equation}
b(r)=b(r_{0})+\int_{r_{0}}^{r}\,8\pi \,\rho (r^{\prime })\,r^{\prime
2}\,dr^{\prime }=2m(r)\,.
\end{equation}%
This is used to find the effective mass of the interior of the WH that gives%
\begin{equation}
m(r)=\frac{r_{0}}{2}+\int_{r_{0}}^{r}\,4\pi \,\rho (r^{\prime })\,r^{\prime
2}\,dr^{\prime }\,,
\end{equation}
Also \ at the limit of infinity, we have 
\begin{equation}
\lim_{r\rightarrow \infty }m(r)=\frac{r_{0}}{2}+\int_{r_{0}}^{\infty }\,4\pi
\,\rho (r^{\prime })\,r^{\prime 2}\,dr^{\prime }=M\,.
\end{equation}
\vspace{-0.75cm}
\subsection{Energy Conditions}
\vspace{-0.75cm}
WHs are supported by exotic matter, and the suitable energy condition is
defined as diagonal~\cite{loboreview0}, 
\begin{equation}
T^{\mu}_{\nu}=\mathrm{diag}(\rho ,p_{1},p_{2},p_{3})\,,  \label{diagonalT}
\end{equation}%
in which $\rho $ and $p_{j}$ denotes to the mass density and the three
principal pressures, respectively. Perfect fluid of the Stress-energy tensor
is obtained if $p_{1}=p_{2}=p_{3}$. It is believed that the normal matters
obey these energy conditions, however, it violates the energy conditions and
needs certain quantum fields (Casimir effect) or dark energy.
\begin{description}
\item[\textbf{Null energy condition (NEC)}] 

The NEC is 
\begin{equation}
T_{\mu \nu }k^{\mu }k^{\nu }\geq 0\,.  \label{eq:nec}
\end{equation}

\item where $k^{\mu }$ is null vector. Using the Eq.(\ref{diagonalT}), we
obtain 
\begin{equation}
\rho +p_{i}\geq 0\,.  \label{eq:necpf}
\end{equation}

\item[\textbf{Weak energy condition (WEC)}] 

The WEC is 
\begin{equation}
T_{\mu \nu }U^{\mu }U^{\nu }\geq 0\,.  \label{eq:wec}
\end{equation}%
where the timelike vector is given by $U^{\mu }$. Eq.(\ref{eq:wec}) is for
the measured energy density by moving with four-velovity $U^{\mu }$ of any
timelike observer. It must be positive and the geometric defination refer to the Einstein field equations Eq.(\ref{EFE}) $G_{\mu \nu }U^{\mu }U^{\nu }\geq 0$. It can be written as 
\begin{equation}
\rho \geq 0\quad \mbox{and},\quad \rho +p_{i}\geq 0\,.  \label{eq:wecpf}
\end{equation}%
The WEC involves the NEC.

\item[\textbf{Strong energy condition (SEC)}] 

The SEC asserts that 
\begin{equation}
\left( T_{\mu \nu }-\frac{T}{2}\;g_{\mu \nu }\right) U^{\mu }U^{\nu }\geq
0\,,  \label{eq:sec}
\end{equation}%
in which $T$ is the trace of the stress energy tensor. Because $T_{\mu \nu }-\frac{T}{2}\;g_{\mu \nu }=\frac{R_{\mu \nu}}{8 \pi}$, according to Einstein field equations Eq. (\ref{EFE})
the SEC is a statement about the Ricci tensor.
Then by using the diagonal stress energy tensor \ given in Eq. (\ref%
{diagonalT}), the SEC reads 
\begin{equation}
\quad \rho +p_{i}\geq 0.  \label{eq:secpf}
\end{equation}%
The SEC involves the NEC, but not necessarily the WEC.

\item[\textbf{Dominant energy condition (DEC)}] 

The DEC is 
\begin{equation}
T_{\mu \nu }U^{\mu }U^{\nu }\geq 0\quad \mbox{and}\quad T_{\mu \nu }U^{\nu }:%
\mbox{is not spacelike}  \label{eq:dec}
\end{equation}%
The energy density must be positive. Moreover, the energy flux should be timelike
or null. The DEC involves the WEC, and automatically the NEC, but not
necessarily the SEC. It becomes 
\begin{equation}
\rho \geq 0.  \label{eq:decpf}
\end{equation}
\end{description}

It can be verified that WHs violate all the energy conditions. Therefore
using the Eq.s (\ref{rhoWH})-(\ref{tauWH}) with $k^{\mu }=(1,1,0,0)$ we \
obtain 
\begin{equation}
\rho-p_{r}=\frac{1}{8\pi }\,\left[ \frac{b^{\prime }r-b}{r^{3}}%
+2\left( 1-\frac{b}{r}\right) \frac{f^{\prime }}{r}\right] .
\label{NECthroat}
\end{equation}%
Thanks to the flaring out condition of the throat from Eq. (\ref{flareout})
: $(b-b^{\prime 2}>0$, one shows that $b(r_{0})=r=r_{0}$ at the throat and
because of the finiteness of $f(r)$, from Eq. (\ref{NECthroat}) we have $\rho-p_{r}<0$. Hence all the energy conditions are violated and this matter is named as the  exotic
matter.
\vspace{-0.75cm}
\subsection{Hawking Radiation of the Traversable Wormholes}
\vspace{-0.75cm}
Since Einstein, Hawking's significant addition to understanding the
universe is called the most significant\cite{Hawking}. Hawking showed that
the BHs are not black but grey which emit radiation. This was the discovery
of Hawking radiation, which allows a BH to leak energy and gradually fade
away to nothing. However, the question of what happens to the information
is remains. All the particles should fall into BH and we do not know what happened to them. The particles that come out of a BH seem to be
completely random and bear no relation to what fell in. It appears that the
information about what fell in is lost, apart from the total amount of mass
and the amount of rotation. If that information is truly lost that strikes
at the heart of our understanding of science.

To understand whether that information is in fact lost, or whether it can be
recovered, Hawking and colleagues, including Andrew Strominger, from Harvard, are currently working to understand \textquotedblleft supertranslations\textquotedblright\ to explain the
mechanism by which information is returned from a BH and encoded on the
hole's \textquotedblleft event horizon\textquotedblright \cite{hawking2}. In
the literature, there exist several derivations of the Hawking radiation%
\cite%
{Wilczek1,Damour,Mann,review,ao3,ao4,ao5,ao6,ao7,ao8,ao9,ao10,ao14,ao15,ao16}.

The transmission probability $\Gamma$ is defined by 
\begin{equation}
\Gamma =e^{-2ImS/\hslash },  \label{ab1}
\end{equation}
where $S$ is the action of the classically forbidden trajectory. Hence, we
have found the Hawking temperature from the tunneling rate of the emitted
particles.

For studying the HR of traversable WHs, we consider a general spherically
symmetric and dynamic WH with a past outer trapping horizon. The traversable
WH metric can be transformed into the generalized retarded
Eddington-Finkelstein coordinates as following\cite{kim}%
\begin{equation}
ds^{2}=-Cdu^{2}-2dudr+r^{2}\left( d\theta ^{2}+Bd\varphi ^{2}\right) ,
\label{met2}
\end{equation}
where $C=$ $1-2M/r$\ and $B=\sin ^{2}\theta $. The gravitational energy is $%
M=\frac{1}{2}r(1-\partial ^{a}r\partial _{a}r)$\ which is also known as a
Misner-Sharp energy. $\ $It reduces to $M=\frac{1}{2}r$ on a trapping horizon%
\cite{haw}. Furthermore, \ there is a past marginal surface at the $C=0$ (at
horizon: $r=r_{0}$) for the retarded coordinates \cite{kim2}.

Firstly, we give the equation of motion for the vector particles which is
known as the Proca equation in a curved space-time \cite{ao9,ao6,Kruglov1,K2}%
:\begin{equation}
\frac{1}{\sqrt{-g}}\partial_{\mu} \left( \sqrt{-g}\psi ^{\nu ; \mu }\right)+\frac{m^{2}}{\hbar ^{2}}\psi ^{\nu }=0,  \label{proca}
\end{equation}
in which the wave functions are defined as $\ \psi _{\nu }=(\psi _{0},\psi
_{1},\psi _{2},\psi _{3}).$ By the help of \ the method of WKB
approximation, the following HJ ans\"{a}tz is substituted into Eq. (\ref%
{proca})
\begin{equation}
\psi _{\nu }=\left( c_{0},c_{1},c_{2},c_{3}\right) e^{\frac{i}{\hbar }%
S(u,r,\theta ,\phi )},  \label{4psi}
\end{equation}
with the real constants $\left( c_{0},c_{1},c_{2},c_{3}\right) $.
Furthermore, we define the action $S(u,r,\theta ,\phi )$ as following 
\begin{equation}
S(u,r,\theta ,\phi )=S_{0}(u,r,\theta ,\phi )+\hbar S_{1}(u,r,\theta ,\phi
)+\hbar ^{2}S_{2}(u,r,\theta ,\phi )+....  \label{5psi}
\end{equation}
Because of the (\ref{met2}) is symmetric, the Killing vectors are $\partial
_{\theta }$\ and $\partial _{\phi }$.\ Then one can use the separation of
variables method to the action $S_{0}(u,r,\theta ,\phi )$: 
\begin{equation}
S_{0}=Eu-W(r)-j\theta -k\phi ,  \label{6psi}
\end{equation}%
It is noted that $E$ and $(j,k)$ are energy and real angular constants,
respectively. After inserting Eqs. (\ref{4psi}), (\ref{5psi}), and (\ref%
{6psi}) into Eq. (\ref{proca}), \ a matrix equation $\Delta \left(
c_{0,}c_{1},c_{2},c_{3}\right) ^{T}=0$ (to the leading order in $\hbar $) is
obtained$,$ which has the following non-zero components:
\begin{eqnarray}
\Delta _{11} &=&2B\left[ \partial _{r}W(r)\right] ^{2}r^{2},\   \notag \\
\Delta _{12} &=&\Delta _{21}=2m^{2}r^{2}B+2B\partial
_{r}W(r)Er^{2}+2Bj^{2}+2k^{2},  \notag \\
\Delta _{13} &=&-\frac{2\Delta _{31}}{r^{2}}=-2Bj\partial _{r}W(r),\   \notag
\\
\Delta _{14} &=&\frac{\Delta _{41}}{Br^{2}}=-2k\partial _{r}W(r),  \notag \\
\Delta _{22} &=&-2BCm^{2}r^{2}+2E^{2}r^{2}B-2j^{2}BC-2k^{2}C,\   \label{7n}
\\
\Delta _{23} &=&\frac{-2\Delta _{32}}{r^{2}}=2jBC\partial _{r}W(r)+2EjB, 
\notag \\
\Delta _{24} &=&\frac{\Delta _{42}}{Br^{2}}=2kC\partial _{r}W(r)+2kE,  \notag
\\
\Delta _{33} &=&m^{2}r^{2}B+2BEr^{2}\partial _{r}W(r)+r^{2}BC\left[ \partial
_{r}W(r)\right] ^{2}+k^{2},  \notag \\
\Delta _{34} &=&\frac{-\Delta _{43}}{2B}=-kj,  \notag \\
\Delta _{44} &=&-2r^{2}BC\left[ \partial _{r}W(r)\right] ^{2}-4BEr^{2}%
\partial _{r}W(r)-2B(m^{2}r^{2}+j^{2}).  \notag
\end{eqnarray}
The determinant of the $\Delta $-matrix ($\mbox{det}\Delta =0$) is used to
get%
\begin{equation}
\mbox{det}\Delta =64Bm^{2}r^{2}\left\{ \frac{1}{2}r^{2}BC\left[ \partial
_{r}W(r)\right] ^{2}+BEr^{2}\partial _{r}W(r)+\frac{B}{2}\left(
m^{2}r^{2}+j^{2}\right) +\frac{k^{2}}{2}\right\} ^{3}=0.  \label{8n}
\end{equation}
Then \ the Eq. (\ref{8n}) \ is solved for $W(r)$ 
\begin{equation}
W_{\pm }(r)=\int \left( \frac{-E}{C}\pm \sqrt{\frac{E^{2}}{C^{2}}-\frac{m^{2}%
}{C}-\frac{j^{2}}{CB^{2}r^{2}}-\frac{k^{2}}{Cr^{2}}}\right) dr.  \label{9}
\end{equation}%
The above integral near the horizon ($r\rightarrow r_{0}$) reduces to
\begin{equation}
W_{\pm }(r)\simeq \int \left( \frac{-E}{C}\pm \frac{E}{C}\right) dr.
\label{10int}
\end{equation}
As shown in the Eq. (\ref{ab1}), the probability rate of \ the
ingoing/outgoing particles only depend on the imaginary part of the action.
Eq. (\ref{10int}) has a pole at $C=0$ on the horizon. Using the contour
integration in the upper $r$ half-plane, one obtains
\begin{equation}
W_{\pm }=i\pi \left( \frac{-E}{2\kappa |_{H}}\pm \frac{E}{2\kappa |_{H}}%
\right) .  \label{11}
\end{equation}
From which
\begin{equation}
ImS=ImW_{\pm },  \label{12}
\end{equation}
that the $\kappa |_{H}=\partial _{r}C/2$ is the surface gravity. Note that
the $\kappa |_{H}$ is positive quantity because the throat is an outer
trapping horizon \cite{kim,kim2}. When we define the probability of incoming
particles $W_{+}$ to $100\%$ such as $\Gamma _{absorption}\approx
e^{-2ImW}\approx 1$. Consequently $W_{-}$ stands for the outgoing particles.
Then we calculate the tunneling rate of the vector particles as \cite%
{Mann,K2} 
\begin{equation}
\Gamma =\frac{\Gamma _{emission}}{\Gamma _{absorption}}=\Gamma
_{emission}\approx e^{-2ImW_{-}}=e^{\frac{2\pi E}{\kappa |_{H}}}.
\label{13tn}
\end{equation}
The Boltzmann factor $\Gamma \approx $ $e^{-\beta E}$ where $\beta $ is the
inverse temperature is compared with the Eq. (\ref{13tn}) to obtain the
Hawking temperature $T|_{H}$ of the traversable WH as
\begin{equation}
T|_{H}=-\frac{\kappa |_{H}}{2\pi },  \label{14}
\end{equation}%
However $T|_{H}$ is negative, as also shown by \cite{kim,kim2}. The main
reason of this negativeness is the phantom energy \cite{kim,ao6}, which is
located at the throat of WH.\ Moreover, as a result of the phantom energy,
the ordinary matter can travel backward in time because in QFT particles and 
anti-particles are defined via sign of time.

Surprisingly, we derive the the negative $T|_{H}$ that past outer trapping
horizon of the traversable WH radiate thermal phantom energy (i.e. dark
energy). Additionally, the radiation of phantom energy has an effect of
reduction of the size of the WH's throat and its entropy. Nonetheless, this
does not create a trouble. The total entropy of universe always increases,
hence it prevents the violation of the second law of thermodynamics.
Moreover, in our different work, we show that the gravitino also tunnels
through WH and we calculate \ the tunneling rate of the emitted gravitino
particles from traversable WH.

\chapter{ROTATING THIN-SHELLS IN (2+1)-D}

The procedure of dealing with a given \ surface of discontinuity is well
known since the Newtonian theory of gravity\cite{5}. Firstly, the continuity
of the gravitational potential should be checked and then from the surface
mass the discontinuity of the gravitational field might be occured. Those
boundary conditions are derived from the field equation. Notwithstanding,
for the case of GR there is different problem because of the nonlinearity of
the field equations as well as the principle of general covariance. To solve
this headache, on a hypersurface splitting spacetimes one must introduce
specific boundary conditions to the induced metric tensor and the extrinsic
curvature. This method is called the Darmois-Israel formalism or\ the
thin-shell formalism\cite{ISRAEL1}. There are many different using
areas of this method such as dynamic thin-layers, connecting branes, quantum
fields in thin-shell spacetimes, WHs, collapsing shells and radiating spheres%
\cite%
{lake,counter,eid,ES2,sotiriou,israeljunction,MV1,MV2,MV3,MV4,MV5,MV6plot,loboreview0,loboreview1,loboreview2,simeo,bhar}%
. It is well known that this method and the searching for distributional
solutions to Einstein's equations are same. This method is also used for the
stars that are expected to display interfacial layers much smaller than
their characteristic sizes with nontrivial quantities, such as surface
tensions and surface energy densities. Another example is compact stars with
interfaces separating their cores and their crusts, (i.e. strange quark
stars and neutron stars)\cite{col}. Note that the Darmois-Israel formalism
gives the nontrivial properties of transitional layers fully taking into
account GR, but just under the macroscopic point of view. Furthermore, the
thin-shell formalism would be the proper formalism for approaching any
gravitational system that presents discontinuous behaviors in their physical
parameters.

On the other hand, this method can be used to screen the BH's hairs against
the outside observer at infinity\cite{MV45,nor3}. One should show that the
thin-shell is a stable under the perturbations. Our main aim is to apply
such a formalism to the general charge carrying rotating BH solution in 2+1
dimensions. The BH has a thin-shell which has a radius greater than the
event horizon\cite{lopes,lopes2}. Once this situation is occurred, we check
the stability analysis.
\vspace{-0.75cm}
\section{Construction of the Rotating Thin-Shells}
\vspace{-0.75cm}
The metric for the general rotating BH solution in 2+1 dimensions is given as \cite{lopes2}%
\begin{equation}
ds_{B}^{2}=-U(r)dt^{2}+\frac{1}{U(r)}dr^{2}+r^{2}\left[ d\phi +h(r)dt\right]
^{2}
\end{equation}
By using the famous cut-and-paste method introduced by help of
Darmois-Israel junction conditions, the thin-shell WH is constructed.
Firstly we take two copies of the bulk 
\begin{equation}
\mathcal{M}_{\pm }=\{x^{\mu }|t\geq t\left( \tau \right) \text{and }r\geq
a\left( \tau \right) \}
\end{equation}%
with the line elements given above. Then we paste them at an identical
hypersurface 
\begin{equation}
\Sigma _{\pm }=\Sigma =\left\{ x^{\mu }|t=t\left( \tau \right) \text{ and }%
r=a\left( \tau \right) \right\} .
\end{equation}%
For convenience we move to a comoving frame to eliminate cross terms in the
induced metrics by introducing 
\begin{equation}
d\phi +h_{\pm }^{{}}\left( a\right) dt=d\psi .
\end{equation}
Then for interior and exterior of WH, it becomes 
\begin{equation}
ds_{\pm }^{2}=-U_{\pm }(r)dt^{2}+\frac{dr^{2}}{U_{\pm }\left( r\right) }%
+r^{2}\left[ d\psi +\left( h_{\pm }^{{}}\left( r\right) -h_{\pm }^{{}}\left(
a\right) \right) dt\right] ^{2}.
\end{equation}
The geodesically complete manifold is satisfied at the hypersurface $\Sigma $
which we shall call the throat. We define the throat for the line element by%
\begin{equation}
ds_{\Sigma }^{2}=-d\tau ^{2}+a^{2}d\psi ^{2}.
\end{equation}%
First of all the throat must satisfy the Israel junction conditions so 
\begin{equation}
-U(a)\dot{t}^{2}+\frac{\dot{a}^{2}}{U\left( a\right) }=-1
\end{equation}%
and it is found that%
\begin{equation}
\dot{t}=\frac{dt}{d\tau }=\frac{1}{U}\sqrt{\dot{a}^{2}+U}
\end{equation}%
and 
\begin{equation}
\ddot{t}=-\frac{\dot{U}}{U^{2}}\sqrt{\dot{a}^{2}+U}+\frac{2\dot{a}\ddot{a}+%
\dot{U}}{2U\sqrt{\dot{a}^{2}+U}}
\end{equation}%
in which a $dot$ stands for the derivative with respect to the proper time $%
\tau .$
Second step is the satisfaction of the Einstein's equations in the form of
Israel junction conditions on the hypersurface which are \ 
\begin{equation}
k_{i}^{j}-k\delta _{i}^{j}=-8\pi GS_{i}^{j},
\end{equation}%
in which $k_{i}^{j}=K_{i}^{j\left( +\right) }-K_{i}^{j\left( -\right) },$ $%
k=tr\left( k_{i}^{j}\right) $ and the extrinsic curvature with embedding
coordinate $X^{i}:$ 
\begin{equation}
K_{ij}^{\left( \pm \right) }=-n_{\gamma }^{\left( \pm \right) }\left( \frac{%
\partial ^{2}x^{\gamma }}{\partial X^{i}\partial X^{j}}+\Gamma _{\alpha
\beta }^{\gamma }\frac{\partial x^{\alpha }}{\partial X^{i}}\frac{\partial
x^{\beta }}{\partial X^{j}}\right) _{\Sigma }  \label{K}
\end{equation}
The parametric equation of the hypersurface $\Sigma $ is given by 
\begin{equation}
F\left( r,a\left( \tau \right) \right) =r-a\left( \tau \right) =0,
\end{equation}
\begin{figure}[h!]

\centering
\includegraphics[width=0.80\textwidth]{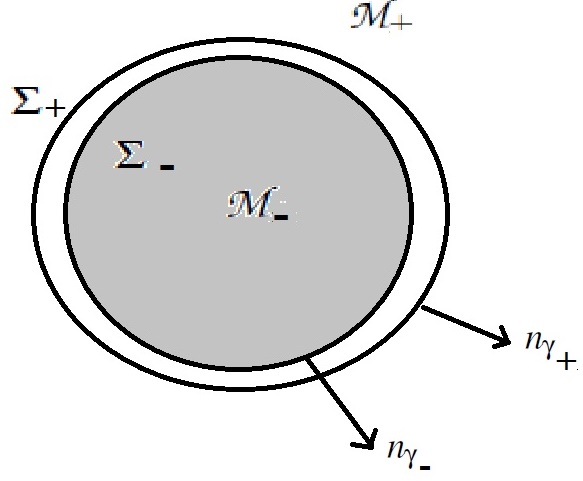} \label{fig:thinshel}
\caption{Thin-Shells}

\end{figure}
and the normal unit vectors to $\mathcal{M}_{\pm }$ defined by 
\begin{equation}
n_{\gamma }=\left( \pm \frac{1}{\sqrt{\Delta }}\frac{\partial F}{\partial
x^{\gamma }}\right) _{\Sigma },
\end{equation}%
where 
\begin{equation}
\Delta =\left\vert g^{\alpha \beta }\frac{\partial F}{\partial x^{\alpha }}%
\frac{\partial F}{\partial x^{\beta }}\right\vert
\end{equation}%
\begin{equation}
\Delta =g^{tt}\frac{\partial F}{\partial t}\frac{\partial F}{\partial t}%
+g^{rr}\frac{\partial F}{\partial r}\frac{\partial F}{\partial r}
\end{equation}%
\begin{equation}
\Delta =-\frac{1}{U}(-\frac{\dot{a}}{\dot{t}})^{2}+U
\end{equation}%
so they are found as follows 
\begin{equation}
\Delta =\frac{U^{2}}{\dot{a}^{2}+U}
\end{equation}%
\begin{equation}
\frac{1}{\sqrt{\Delta }}=\frac{\sqrt{\dot{a}^{2}+U}}{U}
\end{equation}
The normal unit vector must be satisfied $n^{\gamma }n_{\gamma }=1,$
non-zero normal unit vectors are calculated as
\begin{equation}
n_{t}=\pm \frac{1}{\sqrt{\Delta }}\left( -\frac{da}{dt}\right) _{\Sigma }=-%
\frac{1}{\sqrt{\Delta }}\frac{\dot{a}}{\dot{t}}
\end{equation}
\begin{equation}
n_{r}=\frac{1}{\sqrt{\Delta }}
\end{equation}
\begin{equation}
n_{\gamma }=\frac{1}{\sqrt{\Delta }}\left( -\frac{\dot{a}}{\dot{t}}%
,1,0\right)
\end{equation}%
it reduces to%
\begin{equation}
n_{\gamma }=\frac{\sqrt{\dot{a}^{2}+U}}{U}\left( -\frac{\dot{a}U}{\sqrt{\dot{%
a}^{2}+U}},1,0\right) =\left( -\dot{a},\frac{\sqrt{\dot{a}^{2}+U}}{U}%
,0\right) .
\end{equation}
Before calculating the components of the extrinsic curvature tensor, we
redefine the metric of the bulk in 2+1 dimensions given as
\begin{equation}
ds_{B}^{2}=-U(r)dt^{2}+\frac{dr^{2}}{U\left( r\right) }+r^{2}\left[ d\psi
+\omega (r)dt\right] ^{2},
\end{equation}
where 
\begin{equation}
\omega (r)=h\left( r\right) -h\left( a\right) .
\end{equation}
It becomes 
\begin{equation}
ds_{B}^{2}=\left[ -U(r)+r^{4}\omega (r)\right] dt^{2}+\frac{dr^{2}}{U(r)}%
+r^{2}d\psi ^{2}+2r^{2}\omega (r)dtd\psi .  \label{metr}
\end{equation}
Note that the line element on the throat is 
\begin{equation}
ds_{\Sigma }^{2}=-d\tau ^{2}+a^{2}d\psi ^{2}.
\end{equation}
and the corresponding Levi-Civita connections which is defined as 
\begin{equation}
\Gamma _{\mu \nu }^{\alpha }=\frac{1}{2}g^{\alpha \beta }\left( \frac{%
\partial g_{\beta \mu }}{\partial x^{\nu }}+\frac{\partial g_{\beta \nu }}{%
\partial x^{\mu }}-\frac{\partial g_{\mu \nu }}{\partial x^{\beta }}\right)
\end{equation}%
For the metric given in Eq.(\ref{metr}) these are calculated as 
\begin{equation}
\Gamma _{tr}^{t}=\Gamma _{rt}^{t}=\frac{U^{\prime }-r^{2}\omega \omega
^{\prime }}{2U},
\end{equation}%
\begin{equation}
\Gamma _{tt}^{r}=\frac{U}{2}\left[ u^{\prime }-2r\omega ^{2}-2r^{2}\omega
\omega ^{\prime }\right] ,
\end{equation}%
\begin{equation}
\Gamma _{tr}^{\psi }=\frac{\omega ^{2}\omega ^{\prime }r^{3}+\omega ^{\prime
}Ur-U^{\prime }\omega r+2\omega U}{4f},
\end{equation}%
\begin{equation}
\Gamma _{t\psi }^{r}=-\frac{Ur(r\omega ^{\prime }+2\omega )}{2},
\end{equation}%
\begin{equation}
\Gamma _{\psi \psi }^{r}=-Ur
\end{equation}%
\begin{equation}
\Gamma _{r\psi }^{t}=\frac{-r^{2}\omega ^{\prime }}{2U}
\end{equation}%
\begin{equation}
\Gamma _{r\psi }^{\psi }=\frac{\omega \omega ^{\prime }r^{3}+2U}{4U}
\end{equation}%
and%
\begin{equation}
\Gamma _{rr}^{r}=-\frac{U^{\prime }}{2U}.
\end{equation}
One finds the extrinsic curvature tensor components using the definition\
given in Eq.(\ref{K})
\begin{equation}
K_{\tau \tau }=-n_{t}\left( \frac{\partial ^{2}t}{\partial \tau ^{2}}+\Gamma
_{\alpha \beta }^{t}\frac{\partial x^{\alpha }}{\partial \tau }\frac{%
\partial x^{\beta }}{\partial \tau }\right) _{\Sigma }-n_{r}\left( \frac{%
\partial ^{2}r}{\partial \tau ^{2}}+\Gamma _{\alpha \beta }^{r}\frac{%
\partial x^{\alpha }}{\partial \tau }\frac{\partial x^{\beta }}{\partial
\tau }\right) _{\Sigma }
\end{equation}%
\begin{equation}
=-n_{t}\left( \ddot{t}+2\Gamma _{tr}^{t}\dot{t}\dot{a}\right) _{\Sigma
}-n_{r}\left( \ddot{a}+\Gamma _{tt}^{r}\dot{t}^{2}+\Gamma _{rr}^{r}\dot{a}%
^{2}\right) _{\Sigma }
\end{equation}%
Note that on the hyperplane i.e $r=a$ , $\omega =0$. The extrinsic curvature
for tau tau is 
\begin{equation}
K_{\tau \tau }=-n_{t}\left( \ddot{t}+\frac{U^{\prime }}{U}\dot{t}\dot{a}%
\right) _{\Sigma }-n_{r}\left( \ddot{a}+\frac{UU^{\prime }}{2}\dot{t}^{2}-%
\frac{U^{\prime }}{2U}\dot{a}^{2}\right) _{\Sigma }
\end{equation}%
After substituting all the variables it becomes 
\begin{equation}
K_{\tau \tau }=-\frac{\left( \ddot{a}+\frac{U^{\prime }}{2}\right) }{\sqrt{%
\dot{a}^{2}+U}}.
\end{equation}%
Also it is found that the psi-psi component of the extrinsic curvature is 
\begin{equation}
K_{\psi \psi }=-n_{t}\left( \frac{\partial ^{2}t}{\partial \psi ^{2}}+\Gamma
_{\alpha \beta }^{t}\frac{\partial x^{\alpha }}{\partial \psi }\frac{%
\partial x^{\beta }}{\partial \psi }\right) _{\Sigma }-n_{r}\left( \frac{%
\partial ^{2}r}{\partial \psi ^{2}}+\Gamma _{\alpha \beta }^{r}\frac{%
\partial x^{\alpha }}{\partial \psi }\frac{\partial x^{\beta }}{\partial
\psi }\right) _{\Sigma },
\end{equation}%
\begin{equation}
K_{\psi \psi }=+n_{r}\Gamma _{\psi \psi }^{r}=+n_{r}\text{ }aU,
\end{equation}%
\begin{equation}
K_{\psi \psi }=a\sqrt{\dot{a}^{2}+U}.
\end{equation}%
Lastly the tau-psi component of the extrinsic curvature is also found as 
\begin{equation}
K_{\tau \psi }=-n_{t}\left( \frac{\partial ^{2}t}{\partial \tau \partial
\psi }+\Gamma _{\alpha \beta }^{t}\frac{\partial x^{\alpha }}{\partial \tau }%
\frac{\partial x^{\beta }}{\partial \psi }\right) _{\Sigma }-n_{r}\left( 
\frac{\partial ^{2}r}{\partial \tau \partial \psi }+\Gamma _{\alpha \beta
}^{r}\frac{\partial x^{\alpha }}{\partial \tau }\frac{\partial x^{\beta }}{%
\partial \psi }\right) _{\Sigma },
\end{equation}
\begin{equation}
=-n_{t}\left( \Gamma _{r\psi }^{t}\frac{\partial r}{\partial \tau }\frac{%
\partial \psi }{\partial \psi }\right) _{\Sigma }-n_{r}\left( \Gamma _{r\psi
}^{r}\frac{\partial r}{\partial \tau }\frac{\partial \psi }{\partial \psi }%
\right) _{\Sigma },
\end{equation}%
\begin{equation}
=-n_{t}\left( \Gamma _{r\psi }^{t}\dot{a}\right) -n_{r}\left( \Gamma _{t\psi
}^{r}\dot{t}\right) =-n_{t}\left( -\frac{a^{2}\omega ^{\prime }}{2U}\dot{a}%
\right) -n_{r}\left( -\frac{a^{2}}{2}U\omega ^{\prime }\dot{t}\right) ,
\end{equation}%
\begin{equation}
K_{\tau \psi }=\frac{a^{2}\omega ^{\prime }}{2}.
\end{equation}
We can write them also in the following form 
\begin{equation}
K_{\tau }^{\tau }=g^{\tau \alpha }K_{\tau \alpha }=\frac{2\ddot{a}+U^{\prime
}}{2\sqrt{\dot{a}^{2}+U}},
\end{equation}%
\begin{equation}
K_{\psi }^{\psi }=g^{\psi \alpha }K_{\psi \alpha }=\frac{\sqrt{\dot{a}^{2}+U}%
}{a},
\end{equation}%
\begin{equation}
K_{\tau }^{\psi }=g^{\psi \alpha }K_{\tau \alpha }=\frac{\omega ^{\prime }}{2%
},
\end{equation}%
and 
\begin{equation}
K_{\psi }^{\tau }=g^{\tau \tau }K_{\tau \psi }=-\frac{a^{2}}{2}\omega
^{\prime }.
\end{equation}
For a thin-shell with different inner and outer spacetime, they become
\begin{equation}
\text{ }K_{\tau }^{\tau \pm }=\frac{2\ddot{a}+U^{\prime }}{2\sqrt{\dot{a}%
^{2}+U}},
\end{equation}%
\begin{equation}
\text{\ \ }K_{\psi }^{\psi \pm }=\frac{\sqrt{\dot{a}^{2}+U}}{a},
\end{equation}%
\begin{equation}
\text{\ }K_{\tau }^{\psi \pm }=\frac{\omega ^{\prime }}{2},
\end{equation}%
and 
\begin{equation}
K_{\psi }^{\tau \pm }=-\frac{a^{2}}{2}\omega ^{\prime }.
\end{equation}
As a result one obtains 
\begin{equation}
K_{\tau \tau }=-\frac{\ddot{a}+\frac{U^{\prime }}{2}}{\sqrt{\dot{a}^{2}+U}}
\end{equation}%
\begin{equation}
K_{\psi \psi }=a\sqrt{\dot{a}^{2}+U}
\end{equation}%
\begin{equation}
K_{\tau \psi }=\frac{a^{2}\omega ^{\prime }}{2}
\end{equation}
\vspace{-0.75cm}
\section{Israel Junction Conditions For Rotating Thin-Shells}
\vspace{-0.75cm}
In this section, we briefly review the Darmois-Israel junction conditions \cite{ISRAEL1,israeljunction}.

The action of gravity is 
\begin{equation}
S_{Gr}=S_{EH}+S_{GH}
\end{equation}%
where the first term is Einstein-Hilbert action and second term is
Gibbons-Hawking boundary action term.
\begin{equation}
S_{Gr}=\frac{1}{16\pi G}\int_{M}\sqrt{-g}Rd^{4}x+\frac{1}{8\pi G}%
\int_{\Sigma }\sqrt{-h}Kd^{3}x.
\end{equation}
The variation of this action is 
\begin{equation}
\delta S_{Gr}=\frac{1}{16\pi G}\int_{M}\sqrt{-g}G_{ab}\delta g^{ab}d^{4}x+%
\frac{1}{16\pi G}\int_{\Sigma }\sqrt{-h}n_{\alpha }J^{\alpha }d^{3}x
\end{equation}%
\begin{equation*}
+\frac{1}{8\pi G}\int_{\Sigma }\delta \sqrt{-h}Kd^{3}x+\frac{1}{8\pi G}%
\int_{\Sigma }\sqrt{-h}\delta Kd^{3}x.
\end{equation*}
\begin{equation}
\delta S_{Gr}=\frac{1}{2\kappa }\left[ \int_{M}\sqrt{-g}G_{\mu \nu }\delta
g^{\mu \nu }d^{4}x+\int_{\Sigma }\sqrt{-h}(K_{ab}-h_{ab}K)\delta h^{ab}d^{3}x%
\right]
\end{equation}%
where%
\begin{equation}
t_{ab}=\frac{2}{\sqrt{-h}}\frac{\delta S_{Mat}}{\delta h^{ab}}
\end{equation}
\begin{equation}
K_{ab}-h_{ab}K=-8\pi Gt_{ab}
\end{equation}%
\begin{equation*}
D_{a}t^{ab}=-\frac{1}{8\pi G}\left[ D_{a}K_{ab}-D_{a}\left( h_{ab}K\right) %
\right]
\end{equation*}%
\begin{equation*}
=-\frac{1}{8\pi G}R_{cd}n^{c}h^{db}=-T_{cd}n^{c}h^{db}
\end{equation*}
We have found that 
\begin{equation}
\text{ }K_{\tau }^{\tau \pm }=\frac{\ddot{a}+\frac{U_{\pm }^{\text{ }\prime }%
}{2}}{\sqrt{U_{\pm }}\sqrt{\Theta }},
\end{equation}%
\begin{equation}
K_{\psi }^{\psi \pm }=\frac{\sqrt{U_{\pm }}}{a}\sqrt{\Theta },
\end{equation}%
where $\Theta =1+\frac{\dot{a}^{2}}{U_{\pm }},$ 
\begin{equation}
K_{\tau }^{\psi \pm }=\frac{\omega ^{\prime }}{2},
\end{equation}%
and 
\begin{equation}
K_{\psi }^{\tau \pm }=-\frac{a^{2}}{2}\omega ^{\prime }.
\end{equation}
\begin{equation}
K^{\pm }=K_{i}^{i\pm }=\frac{\ddot{a}+\frac{U_{\pm }^{\text{ }\prime }}{2}}{%
\sqrt{U_{\pm }}\sqrt{\Theta }}+\frac{\sqrt{U_{\pm }}}{a}\sqrt{\Theta }
\end{equation}%
\begin{equation}
-8\pi GS_{i}^{j}=[K_{i}^{j}]-[K]
\end{equation}
in which $[A]=A^{+}-A^{-}.$
Also $\ $%
\begin{equation}
S_{i}^{j}=\left( 
\begin{array}{cc}
S_{\tau }^{\tau } & S_{\psi }^{\tau } \\ 
S_{\tau }^{\psi } & S_{\psi }^{\psi }%
\end{array}%
\right)
\end{equation}
\begin{equation}
-8\pi GS_{\tau }^{\tau }=\text{ }K_{\tau }^{\tau }-K=\text{ }K_{\tau }^{\tau
}-\text{ }K_{\tau }^{\tau }-K_{\psi }^{\psi }=-K_{\psi }^{\psi }
\end{equation}
\begin{equation}
8\pi GS_{\tau }^{\tau }=K_{\psi }^{\psi }
\end{equation}%
\begin{equation}
8\pi GS_{\tau }^{\tau }=K_{\psi }^{\psi +}-K_{\psi }^{\psi -}
\end{equation}%
\begin{equation}
S_{\tau }^{\tau }=\frac{1}{8\pi Ga}\left( \sqrt{U_{+}+\dot{a}^{2}}-\sqrt{%
U_{-}+\dot{a}^{2}}\right)
\end{equation}%
Then for other components%
\begin{equation}
-8\pi GS_{\psi }^{\psi }=K_{\psi }^{\psi }-K=K_{\psi }^{\psi }+\text{ }%
K_{\tau }^{\tau }-K_{\psi }^{\psi }=+\text{ }K_{\tau }^{\tau }
\end{equation}%
\begin{equation}
S_{\psi }^{\psi }=-\frac{1}{8\pi G}\left( K_{\tau }^{\tau +}-K_{\tau }^{\tau
-}\right)
\end{equation}%
\begin{equation}
S_{\psi }^{\psi }=\frac{1}{8\pi G}\left[ -\frac{\ddot{a}+\frac{U_{+}^{\text{ 
}\prime }}{2}}{\sqrt{\dot{a}^{2}+U_{+}}}+\frac{\ddot{a}+\frac{U_{-}^{\prime }%
}{2}}{\sqrt{\dot{a}^{2}+U_{-}}}\right]
\end{equation}%
and the last component is 
\begin{equation}
S_{\tau }^{\psi }=S_{\psi }^{\tau }=-\frac{1}{8\pi G}\left( K_{\psi }^{\tau
+}-K_{\psi }^{\tau -}\right) =-\frac{a^{2}}{8\pi G}\left( -\omega
_{+}^{\prime }+\omega _{-}^{\prime }\right)
\end{equation}
The special condition of $\omega _{+}^{\prime }=\omega _{-}^{\prime },$ $%
\omega _{+}=\omega _{-}$ so $S_{\tau }^{\psi }=S_{\psi }^{\tau }=0.$%
Therefore it implies that the upper-shell and the lower-shell are
corotating. The surface stress-energy tensor is
\begin{equation}
S_{b}^{a}=\left( 
\begin{array}{cc}
-\sigma & 0 \\ 
0 & p%
\end{array}%
\right)
\end{equation}
where%
\begin{equation}
\sigma =-\frac{1}{8\pi Ga}\left( \sqrt{U_{+}+\dot{a}^{2}}-\sqrt{U_{-}+\dot{a}%
^{2}}\right)   \label{sag}
\end{equation}%
and 
\begin{equation}
p=\frac{1}{8\pi G}\left[ -\frac{\ddot{a}+\frac{U_{+}^{\text{ }\prime }}{2}}{%
\sqrt{\dot{a}^{2}+U_{+}}}+\frac{\ddot{a}+\frac{U_{-}^{\prime }}{2}}{\sqrt{%
\dot{a}^{2}+U_{-}}}\right] 
\end{equation}
The case of the static is obtained by assuming $\dot{a}=0$ and $\ddot{a}=0$,
\begin{equation}
\sigma =-\frac{1}{8\pi Ga_{0}}\left( \sqrt{U_{+}}-\sqrt{U_{-}}\right)
\end{equation}%
and 
\begin{equation}
p=\frac{1}{8\pi G}\left[ -\frac{\frac{U_{+}^{\text{ }\prime }}{2}}{\sqrt{%
U_{+}}}+\frac{\frac{U_{-}^{\prime }}{2}}{\sqrt{U_{-}}}\right] .
\end{equation}
\vspace{-0.75cm}
\section{ Energy Conservation}
\vspace{-0.75cm}
The Darmois-Israel junction condition for connecting a hypersurface $%
\mathcal{M}_{+}$ with a hypersurface $\mathcal{M}_{-}$ can be written as 
\begin{equation}
\left[ g_{ij}\right] =0  \label{eqn-jumpg}
\end{equation}%
and 
\begin{equation}
\left[ K_{ij}\right] =0  \label{eqn-jumpK}
\end{equation}
The boundary surface $\Sigma $ is defined when both (\ref{eqn-jumpg}) and (%
\ref{eqn-jumpK}) are satisfied. If only (\ref{eqn-jumpg}) is satisfied then
we refer to $\Sigma $ as a thin-shell.  \\
Conditions (\ref{eqn-jumpg}) and (\ref{eqn-jumpK}) require a common
coordinate system on $\Sigma$ and this is easily done if one can set $%
\xi_+^a = \xi_-^a$. Failing this, establishing (\ref{eqn-jumpg}) requires a
solution to the three dimensional metric equivalence problem.
After the signs of normal vector is choosen, there is no ambiguity in (\ref%
{eqn-jumpK}) and (\ref{eqn-jumpg}) and (\ref{eqn-jumpK}) are used in
conjunction with the Einstein tensor $G_{\alpha \beta }$ to calculate the
identities 
\begin{equation}
\left[ G_{\alpha \beta }n^{\alpha }n^{\beta }\right] =0  \label{eqn-Gnn}
\end{equation}%
and 
\begin{equation}
\left[ G_{\alpha \beta }\frac{\partial x^{\alpha }}{\partial \xi ^{i}}%
n^{\beta }\right] =0.  \label{eqn-Gfn}
\end{equation}
This shows that for timelike $\Sigma $ the flux through $\Sigma $ (as
measured comoving with $\Sigma $) is continuous. \\
The Israel formulation of thin shells follows from the Lanczos equation 
\begin{equation}  \label{eqn-Lanc}
S_{ij} = \frac{\Delta}{8 \pi}(\left[K_{ij} \right] - g_{ij} \left[{K_i}^i %
\right])
\end{equation}
and we refer to $S_{ij}$ as the surface stress-energy tensor of $\Sigma$.
The ``ADM'' constraint 
\begin{equation}
\nabla_j K^j_i - \nabla_i K = G_{\alpha \beta} \frac{\partial x^\alpha}{%
\partial \xi^i} n^\beta
\end{equation}
along with Einstein's equations then gives the conservation \emph{identity} 
\begin{equation}  \label{eqn-claw}
\Delta \nabla_i S^i_j = \left[T_{\alpha \beta} \frac{\partial x^\alpha}{%
\partial \xi^i} n^\beta \right].
\end{equation}
The ``Hamiltonian'' constraint 
\begin{equation}
G_{\alpha \beta} n^\alpha n^\beta = (\Delta (^3 R) + K^2 - K_{ij} K^{ij})/2
\end{equation}
gives the evolution \emph{identity} 
\begin{equation}  \label{eqn-elaw}
-S^{i j} \overline{K}_{i j} = \left[ T_{\alpha \beta} n^\alpha n^\beta %
\right].
\end{equation}
The dynamics of the thin-shell are not understood from the identities (\ref%
{eqn-claw}) and (\ref{eqn-elaw}). The evolution of the thin-shell is
obtained by the Lanczos equation(\ref{eqn-Lanc}) \cite{eid}.
The $p$ and $\sigma $ are used to satisfy the energy condition 
\begin{equation}
\frac{d}{d\tau }\left( \sigma a\right) +p\frac{d}{d\tau }\left( a\right) =N
\end{equation}
where
\begin{equation}
N=\frac{1}{8\pi G}\frac{\dot{a}\left[ U_{-}^{\prime }\sqrt{U_{+}+\dot{a}^{2}}%
-U_{+}^{\prime }\sqrt{U_{-}+\dot{a}^{2}}\right] }{\sqrt{U_{+}+\dot{a}^{2}}%
\sqrt{U_{-}+\dot{a}^{2}}}
\end{equation}
Note that "$^{\prime }$" prime stands for the derivative respect to $a$ and
the energy on the shell is not conserved.
\vspace{-0.75cm}
\section{Stability Analyses of Thin-Shells}
\vspace{-0.75cm}
Another relation between the energy density and pressure which is much
helpful is the energy conservation relation which is given by%
\begin{equation}
\frac{\partial \sigma }{\partial \tau }+\frac{\dot{a}}{a}\left( p+\sigma
\right) =0.  \label{energy}
\end{equation}%
This relation must be satisfied by $\sigma $ and $p$ even after the
perturbation which gives an inside to the problem.

The idea is to perturb the shell while it is at the equilibrium
point $a=a_{0}$. By using the Eq.(\ref{sag}), we apply the perturbation and
find the equation of motion of the shell \ (note that $8\pi G=1$) is%
\begin{equation}
\dot{a}^{2}+V_{eff}=0
\end{equation}
 where 
\begin{equation}
V_{eff}={\frac{U_{-}^{2}+\left( -2a^{2}\sigma ^{2}-U_{+}\right)
U_{-}\,+\left( a^{2}\sigma ^{2}-U_{+}\right) ^{2}}{4\sigma ^{2}a^{2}}.}
\end{equation}
This one dimensional equation describes the nature of the equilibrium point
whether it is a stable equilibrium or an unstable one. To see that we expand 
$V_{eff}$ about $a=a_{0}$ and keep the first non-zero term which is 
\begin{equation}
V_{eff}\left( a\right) \sim \frac{1}{2}V_{eff}^{\prime \prime }\left(
a_{0}\right) \left( a-a_{0}\right) ^{2}.
\end{equation}%
One can easily show that $V_{eff}^{\prime }\left( a_{0}\right)
=V_{eff}\left( a_{0}\right) =0$ and therefore everything depends on the sign
of $V_{eff}^{\prime \prime }\left( a_{0}\right) .$ Let's introduce $%
x=a-a_{0} $ and write the equation of motion again%
\begin{equation}
\dot{x}^{2}+\frac{1}{2}V_{eff}^{\prime \prime }\left( a_{0}\right) x^{2}=0
\end{equation}%
which after a derivative with respect to time it reduces to%
\begin{equation}
\ddot{x}+\frac{1}{2}V_{eff}^{\prime \prime }\left( a_{0}\right) x=0.
\end{equation}%
This equation explicitly implies that if $\frac{1}{2}V_{eff}^{\prime \prime
}\left( a_{0}\right) >0$ the $x$ will be an oscillating function about $x=0$
with the angular frequency $\omega _{0}=\sqrt{\frac{1}{2}V_{eff}^{\prime
\prime }\left( a_{0}\right) }$ but otherwise i.e., $\frac{1}{2}%
V_{eff}^{\prime \prime }\left( a_{0}\right) <0$ the motion will be
exponentially toward the initial perturbation. Therefore our task is to find 
$V_{eff}^{\prime \prime }\left( a_{0}\right) $ and show that under what
condition it may be positive for the stability and negative for the
instability of the shell.
 Can naturally formed absorber thin-shells, in cosmology to hide the reality from our telescopes? This can be revise the ideas of no-hair black hole theorem.
Thin-shell can be used to find exhibiting a remarkable property of QCD-like charge confinement to cover the Dirac and Yang-Mills fields. The charged interior spacetime is completely con-
fined within the finite-spacial-size – analog of QCD quark confinement. Naturally this takes us away from classical physics into the realm of gravity coupled QCD. 
\vspace{-0.75cm}
\section{Example of BTZ Thin-Shells}
\vspace{-0.75cm}
Let's set the lapse function of $U$ for the inner\ shell which is de Sitter
spacetime with mass $M_{2}$ \ and outer shells which is a BTZ BH with mass $%
M_{1}$and charge $Q_{1}$ to be as follows\cite{MV45,BTZ,BTZ1}
\begin{equation}
U_{-}=-M_{2}+\frac{a^{2}}{\ell ^{2}}
\end{equation}
and 
\begin{equation}
U_{+}=-M_{1}+\frac{a^{2}}{\ell ^{2}}-Q_{1}^{2}\ln \left( a^{2}+s^{2}\right)
\end{equation}
where s and l are constants.
\ After some calculations, we obtain that the energy density and the
pressures can be recast as%
\begin{equation}
\sigma =-S_{\tau }^{\tau }=\frac{1}{8\pi a}\left( \sqrt{U_{-}\left( a\right)
+\dot{a}^{2}}-\sqrt{U_{+}\left( a\right) +\dot{a}^{2}}\right) \label{stabsig0}
\end{equation}%
and 
\begin{equation}
p=S_{\theta }^{\theta }=\frac{2\ddot{a}+U_{+}^{\prime }\left( a\right) }{%
16\pi \sqrt{U_{+}\left( a\right) +\dot{a}^{2}}}-\frac{2\ddot{a}%
+U_{-}^{\prime }\left( a\right) }{16\pi \sqrt{U_{-}\left( a\right) +\dot{a}%
^{2}}}.
\end{equation}
For a static configuration of radius a, we obtain (assuming $\dot{a}=0$ and $%
\ddot{a}=0$)
\begin{equation}
\sigma _{0}=\frac{1}{8\pi a_{0}}\left( \sqrt{U_{-}\left( a_{0}\right) }-%
\sqrt{U_{+}\left( a_{0}\right) }\right) 
\end{equation}%
and 
\begin{equation}
p_{0}=\frac{U_{+}^{\prime }\left( a_{0}\right) }{16\pi \sqrt{U_{+}\left(
a_{0}\right) }}-\frac{U_{-}^{\prime }\left( a_{0}\right) }{16\pi \sqrt{%
U_{-}\left( a_{0}\right) }}.
\end{equation}
 To obtain the stability criterion, one starts by rearranging Eq. (%
\ref{stabsig0}) in order to obtain the equation of motion
\begin{equation}
\dot{a}^{2}+V(a)=0
\end{equation}%
where the $V(a)$ is the potential as 
\begin{equation}
V\left( a\right) =\frac{U_{-}\left( a\right) +U_{+}\left( a\right) }{2}%
-\left( \frac{U_{-}\left( a\right) -U_{+}\left( a\right) }{16\pi a\sigma }%
\right) ^{2}-\left( 4\pi a\sigma \right) ^{2}.
\end{equation}
Now we impose the energy conservation condition which must be satisfied
after the perturbation and try to find out weather the motion of the shell
is oscillatory or not. This openly means a relation between $p$ and $\sigma
. $ Finally in order to have the thin-shell stable against radial
perturbation, $V_{eff}^{\prime \prime }\geq 0$ at the equilibrium point
i.e., $a=a_{0}$ where $V_{eff}=V_{eff}^{\prime }=0.$To keep our study as
general as possible we assume $p$ to be an arbitrary function of $\beta $
and $\sigma $ i.e., 
\begin{equation}
p\simeq p_{0}+\beta \sigma \text{ }
\end{equation}%
where $p_{0}=cons.$ In Fig.2.2 we plot $V^{\prime \prime }\left( a_{0}\right) 
$ for the specific value of $m=1.0$ and $Q=0.2.$ As one observes in the
region with $\omega >0$ the thin-shell is stable while otherwise it occurs
for $\omega <0.$ 
\begin{figure}[h!]

\centering
\includegraphics[width=0.80\textwidth]{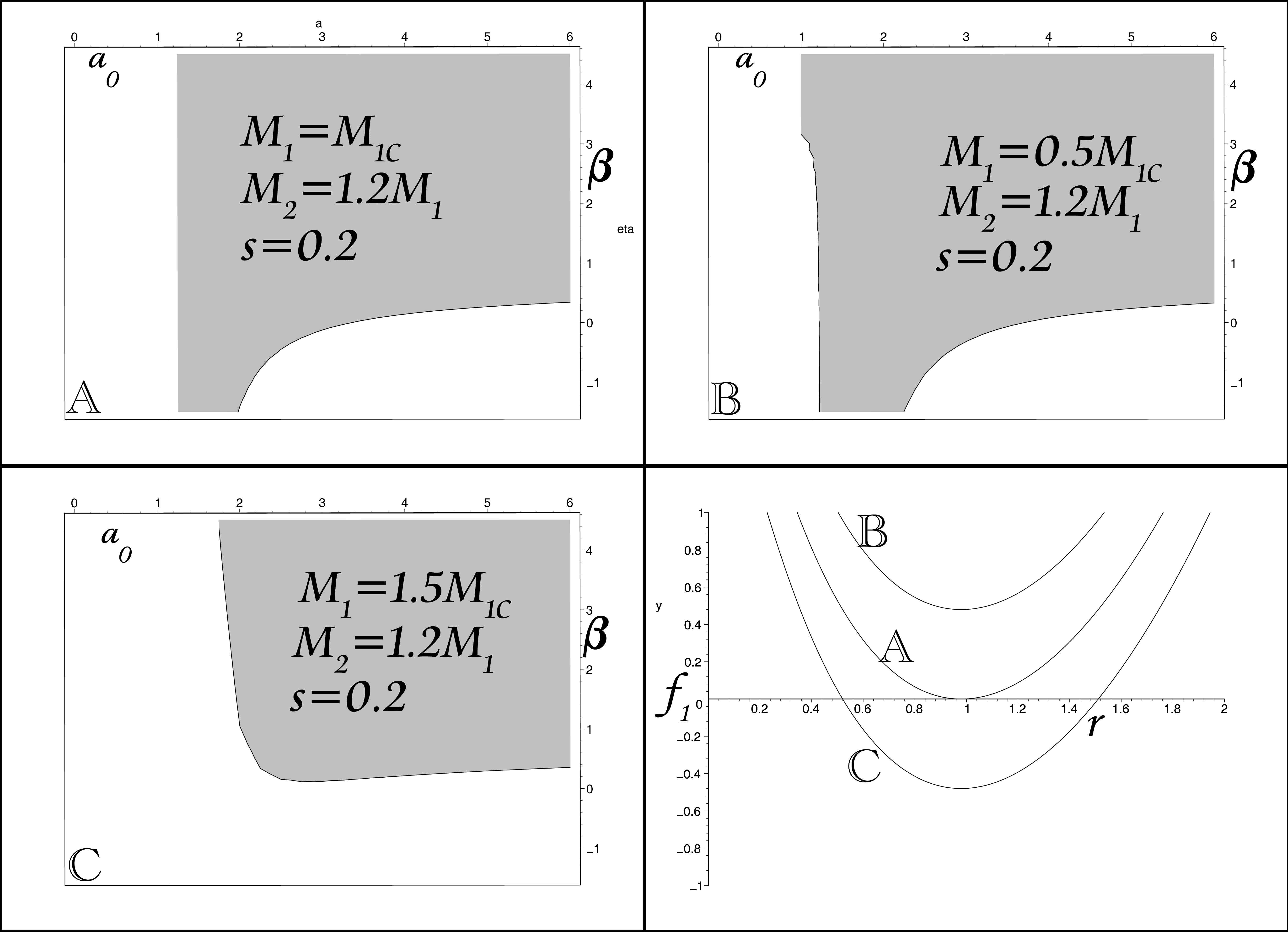} \label{fig:thinshell1BTZ}
\caption{Stability regions of the BTZ thin-shell}
\end{figure}
\vspace{-0.75cm}
\section{Generalization of Rotating Thin-Shells}
\vspace{-0.75cm}
The generalized spacetime metric is given as follows%
\begin{equation}
ds_{b}^{2}=-f(r)^{2}dt^{2}+g\left( r\right) ^{2}dr^{2}+r^{2}\left[ d\varphi
+h(r)dt\right] ^{2}.
\end{equation}

We define the throat for the line element by%
\begin{equation}
ds_{\Sigma }^{2}=-d\tau ^{2}+a^{2}d\psi ^{2}.
\end{equation}%
The\ throat must satisfy the Israel junction conditions so 
\begin{equation}
-f(a)^{2}\dot{t}^{2}+g\left( a\right) ^{2}\dot{a}^{2}=-1
\end{equation}%
and it is found that%
\begin{equation}
\dot{t}=\frac{1}{f}\sqrt{1+g^{2}\dot{a}^{2}}
\end{equation}%
and 
\begin{equation}
\ddot{t}=-\frac{-f^{\text{ }\prime }\dot{a}}{f^{2}}\sqrt{1+g^{2}\dot{a}^{2}}+%
\frac{\left( 2g^{2}\dot{a}\ddot{a}+2gg^{\prime }\dot{a}^{3}\right) }{2f\sqrt{%
1+g^{2}\dot{a}^{2}}}
\end{equation}%
in which a $dot$ stands for the derivative with respect to the proper time $%
\tau .$
For convenience we move to a comoving frame to eliminate cross term in the
induced metrics by introducing 
\begin{equation}
d\varphi +h\left( a\right) dt=d\psi .
\end{equation}
Then for interior and exterior of WH, it becomes
\begin{equation}
ds_{b}^{2}=-f(r)^{2}dt^{2}+g\left( r\right) ^{2}dr^{2}+r^{2}\left[ d\psi
+\omega (r)dt\right] ^{2},
\end{equation}
where 
\begin{equation}
\omega (r)=h\left( r\right) -h\left( a\right) .
\end{equation}
The parametric equation of the hypersurface $\Sigma $ is given by 
\begin{equation}
F\left( r,a\left( \tau \right) \right) =r-a\left( \tau \right) =0,
\end{equation}%
and the normal unit vectors to $\mathcal{M}_{\pm }$ defined by 
\begin{equation}
n_{\gamma }=\left( \pm \frac{1}{\sqrt{\Delta }}\frac{\partial F}{\partial
x^{\gamma }}\right) _{\Sigma },
\end{equation}%
where 
\begin{equation}
\Delta =\left\vert g^{\alpha \beta }\frac{\partial F}{\partial x^{\alpha }}%
\frac{\partial F}{\partial x^{\beta }}\right\vert .
\end{equation}
\begin{equation}
\Delta =g^{tt}\frac{\partial F}{\partial t}\frac{\partial F}{\partial t}%
+g^{rr}\frac{\partial F}{\partial r}\frac{\partial F}{\partial r}
\end{equation}%
\begin{equation}
\Delta =-\frac{1}{f^{2}}\left( -\frac{\dot{a}f}{\sqrt{\Theta }}\right) ^{2}+%
\frac{1}{g^{2}}
\end{equation}
in which $\Theta =1+g^{2}\dot{a}^{2},$%
\begin{equation}
\Delta =\frac{-\dot{a}^{2}g^{2}+\Theta }{\Theta g^{2}}=\frac{1}{g^{2}(1+g^{2}%
\dot{a}^{2})}
\end{equation}
\begin{equation}
\sqrt{\Delta }=\frac{1}{g\sqrt{\Theta }}
\end{equation}
so 
\begin{equation}
g=\frac{1}{\sqrt{\Delta }\sqrt{\Theta }}
\end{equation}
 The normal unit vector must satisfy $n^{\gamma }n_{\gamma }=1,$ so
that non-zero normal unit vectors are calculated \ as
\begin{equation}
n_{t}=\pm \frac{1}{\sqrt{\Delta }}\left( -\frac{da}{dt}\right) _{\Sigma }=-%
\frac{1}{\sqrt{\Delta }}\frac{\dot{a}}{\dot{t}}=-\frac{1}{\sqrt{\Delta }}%
\frac{\dot{a}f}{\sqrt{\Theta }}=-\dot{a}fg
\end{equation}
\begin{equation}
n_{r}=\frac{1}{\sqrt{\Delta }}=g\sqrt{\Theta }
\end{equation}
so corrresponding unit vectors are
\begin{equation}
n_{\gamma }=g\left( -\dot{a}f,\sqrt{\Theta },0\right) .
\end{equation}
Now we will calculate the extrinsic curvatures
\begin{equation}
K_{\tau \tau }=-n_{t}\left( \frac{\partial ^{2}t}{\partial \tau ^{2}}+\Gamma
_{\alpha \beta }^{t}\frac{\partial x^{\alpha }}{\partial \tau }\frac{%
\partial x^{\beta }}{\partial \tau }\right) _{\Sigma }-n_{r}\left( \frac{%
\partial ^{2}r}{\partial \tau ^{2}}+\Gamma _{\alpha \beta }^{r}\frac{%
\partial x^{\alpha }}{\partial \tau }\frac{\partial x^{\beta }}{\partial
\tau }\right) _{\Sigma }
\end{equation}%
\begin{equation}
K_{\tau \tau }=-n_{t}\left( \ddot{t}+2\Gamma _{tr}^{t}\dot{t}\dot{a}\right)
_{\Sigma }-n_{r}\left( \ddot{a}+\Gamma _{tt}^{r}\dot{t}^{2}+\Gamma _{rr}^{r}%
\dot{a}^{2}\right) _{\Sigma }
\end{equation}
where the Levi-Civita connections calculated $(\omega=0)$ ; \ 
\begin{equation}
\Gamma _{tr}^{t}|_{\Sigma }=\Gamma _{rt}^{t}|_{\Sigma }=\frac{f^{\prime }}{f}%
,
\end{equation}%
\begin{equation}
\Gamma _{tt}^{r}|_{\Sigma }=\frac{ff^{^{\text{ }\prime }}}{g^{2}},
\end{equation}%
\begin{equation}
\Gamma _{\psi \psi }^{r}|_{\Sigma }=-\frac{r}{g^{2}}
\end{equation}

\begin{equation}
\Gamma _{t\psi }^{r}|_{\Sigma }=-\frac{r^{2}\omega ^{\prime }}{2g^{2}},
\end{equation}%
\begin{equation}
\Gamma _{r\psi }^{t}|_{\Sigma }=\frac{-r^{2}\omega ^{\prime }}{2f^{2}}
\end{equation}%
\begin{equation}
\Gamma _{r\psi }^{\psi }|_{\Sigma }=\frac{1}{r}
\end{equation}%
and%
\begin{equation}
\Gamma _{rr}^{r}|_{\Sigma }=\frac{g^{\prime }}{g}.
\end{equation}
so 
\begin{equation}
K_{\tau \tau }=-n_{t}\left( \ddot{t}+2\frac{f^{\prime }}{f}\dot{t}\dot{a}%
\right) _{\Sigma }-n_{r}\left( \ddot{a}+\frac{ff^{^{\text{ }\prime }}}{g^{2}}%
\dot{t}^{2}+\frac{g^{\prime }}{g}\dot{a}^{2}\right) _{\Sigma }
\end{equation}
\begin{equation}
K_{\tau \tau }=\dot{a}fg\left( \frac{-f^{\text{ }\prime }\dot{a}}{f^{2}}%
\sqrt{\Theta }+\frac{\dot{a}\left( g^{2}\ddot{a}+gg^{\prime }\dot{a}%
^{2}\right) }{f\sqrt{\Theta }}+2\frac{f^{\prime }}{f^{2}}\sqrt{\Theta }%
\right)
\end{equation}%
\begin{equation*}
-g\sqrt{\Theta }\left( \ddot{a}+\frac{f^{^{\text{ }\prime }}\Theta }{g^{2}f}+%
\frac{g^{\prime }}{g}\dot{a}^{2}\right)
\end{equation*}
\begin{equation}
K_{\tau \tau }=\frac{\dot{a}^{2}g}{\sqrt{\Theta }}\left( g^{\prime }\dot{a}%
^{2}+g\ddot{a}\right) -g\sqrt{\Theta }\left( \ddot{a}+\frac{f^{\prime }}{%
fg^{2}}+\frac{g^{\prime }}{g}\dot{a}^{2}\right)
\end{equation}
\begin{equation}
K_{\tau \tau }=\frac{-g}{\sqrt{\Theta }}\left[ \ddot{a}+\frac{f^{\text{ }%
\prime }}{fg^{2}}-\left( \frac{f^{\prime }}{f}+\frac{g^{\prime }}{g}\right) 
\dot{a}^{2}\right]
\end{equation}
Also it is found that the phi,phi component of the extrinsic curvature is 
\begin{equation}
K_{\psi \psi }=-n_{t}\left( \frac{\partial ^{2}t}{\partial \psi ^{2}}+\Gamma
_{\alpha \beta }^{t}\frac{\partial x^{\alpha }}{\partial \psi }\frac{%
\partial x^{\beta }}{\partial \psi }\right) _{\Sigma }-n_{r}\left( \frac{%
\partial ^{2}r}{\partial \psi ^{2}}+\Gamma _{\alpha \beta }^{r}\frac{%
\partial x^{\alpha }}{\partial \psi }\frac{\partial x^{\beta }}{\partial
\psi }\right) _{\Sigma },
\end{equation}%
\begin{equation}
K_{\psi \psi }=-n_{r}\Gamma _{\psi \psi }^{r}=n_{r}\frac{a}{g^{2}}=\frac{a}{g%
}\sqrt{\Theta }
\end{equation}
and lastly the tau,phi component of the extrinsic curvature is also found as 
\begin{equation}
K_{\psi \tau }=K_{\tau \psi }=-n_{t}\left( \frac{\partial ^{2}t}{\partial
\tau \partial \psi }+\Gamma _{\alpha \beta }^{t}\frac{\partial x^{\alpha }}{%
\partial \tau }\frac{\partial x^{\beta }}{\partial \psi }\right) _{\Sigma
}-n_{r}\left( \frac{\partial ^{2}r}{\partial \tau \partial \psi }+\Gamma
_{\alpha \beta }^{r}\frac{\partial x^{\alpha }}{\partial \tau }\frac{%
\partial x^{\beta }}{\partial \psi }\right) _{\Sigma },
\end{equation}
\begin{equation}
=-n_{t}\left( \Gamma _{r\psi }^{t}\dot{a}\right) -n_{r}\left( \Gamma _{r\psi
}^{r}\dot{t}\right) ,
\end{equation}%
\begin{equation}
=-(\dot{a}^{2}g)\left( \frac{a^{2}}{2f}\omega ^{\prime }\right) +\Theta 
\frac{1}{f}\frac{a^{2}}{2g}\omega ^{\prime },
\end{equation}%
\begin{equation}
K_{\tau \psi }=\frac{a^{2}\omega ^{\prime }}{2fg}.
\end{equation}
Finally 
\begin{equation}
K_{\tau \tau }=\frac{-g}{\sqrt{\Theta }}\left[ \ddot{a}+\frac{f^{\text{ }%
\prime }}{fg^{2}}-\left( \frac{f^{\prime }}{f}+\frac{g^{\prime }}{g}\right) 
\dot{a}^{2}\right]
\end{equation}%
\begin{equation}
K_{\psi \psi }=\frac{a}{g}\sqrt{\Theta }
\end{equation}%
\begin{equation}
K_{\tau \psi }=\frac{a^{2}\omega ^{\prime }}{2fg}
\end{equation}%
in which $\omega (r)^{\prime }=\left[ h(r)-h(a)\right] ^{\prime
}=h(r)^{\prime }.$
They become
\begin{equation}
\text{ }K_{\tau }^{\tau }=\frac{g}{\sqrt{\Theta }}\left[ \ddot{a}+\frac{f^{%
\text{ }\prime }}{fg^{2}}-\left( \frac{f^{\prime }}{f}+\frac{g^{\prime }}{g}%
\right) \dot{a}^{2}\right] ,
\end{equation}%
\begin{equation}
\text{\ \ }K_{\psi }^{\psi }=\frac{1}{ag}\sqrt{\Theta },
\end{equation}%
\begin{equation}
\text{\ }K_{\tau }^{\psi }=\frac{\omega ^{\prime }}{2fg},
\end{equation}%
and 
\begin{equation}
K_{\psi }^{\tau }=-\frac{a^{2}}{2fg}\omega ^{\prime }.
\end{equation}
If we take $g=\frac{1}{f}$ with $f=\sqrt{U},$ one finds same as part 1.
\vspace{-0.75cm}
\section{Israel Junction Conditions For Thin-Shell}
\vspace{-0.75cm}
 Now we define the junction conditions by using the exterior
curvatures as follows%
\begin{equation}
\text{ }K_{\tau }^{\tau \pm }=\frac{g_{\pm }}{\sqrt{\Theta }}\left[ \ddot{a}+%
\frac{f_{\pm }^{\text{ }\prime }}{f_{\pm }g_{\pm }^{2}}-\left( \frac{f_{\pm
}^{\prime }}{f_{\pm }}+\frac{g_{\pm }^{\prime }}{g_{\pm }}\right) \dot{a}^{2}%
\right] ,
\end{equation}%
\begin{equation}
\text{\ \ }K_{\psi }^{\psi \pm }=\frac{1}{ag_{\pm }}\sqrt{\Theta },
\end{equation}%
\begin{equation}
\text{\ }K_{\tau }^{\psi \pm }=\frac{\omega ^{\prime }}{2f_{\pm }g_{\pm }},
\end{equation}%
and 
\begin{equation}
K_{\psi }^{\tau \pm }=-\frac{a^{2}}{2f_{\pm }g_{\pm }}\omega ^{\prime }.
\end{equation}
\begin{equation}
K^{\pm }=K_{i}^{i\pm }=\frac{g_{\pm }}{\sqrt{\Theta }}\left[ \ddot{a}+\frac{%
f_{\pm }^{\text{ }\prime }}{f_{\pm }g_{\pm }^{2}}-\left( \frac{f_{\pm
}^{\prime }}{f_{\pm }}+\frac{g_{\pm }^{\prime }}{g_{\pm }}\right) \dot{a}^{2}%
\right] +\frac{1}{ag_{\pm }}\sqrt{\Theta }
\end{equation}%
\begin{equation}
-8\pi GS_{i}^{j}=[K_{i}^{j}]-[K]\delta^{j}_{i}
\end{equation}
where $[K]$ is the trace of $[K_{i}^{j}]$ and $S_{i}^{j}$ is the surface
stress-energy tensor on $\sigma $, and $[A]=A^{+}-A^{-}$.
Also $S_{i}^{j}=\left( 
\begin{array}{cc}
S_{\tau }^{\tau } & S_{\psi }^{\tau } \\ 
S_{\tau }^{\psi } & S_{\psi }^{\psi }%
\end{array}%
\right) \ $
\begin{equation}
-8\pi GS_{\tau }^{\tau }=\text{ }K_{\tau }^{\tau }-K=\text{ }K_{\tau }^{\tau
}-\text{ }K_{\tau }^{\tau }-K_{\psi }^{\psi }=-K_{\psi }^{\psi }
\end{equation}
\begin{equation}
8\pi GS_{\tau }^{\tau }=K_{\psi }^{\psi }
\end{equation}%
\begin{equation}
8\pi GS_{\tau }^{\tau }=K_{\psi }^{\psi +}-K_{\psi }^{\psi -}
\end{equation}%
\begin{equation}
S_{\tau }^{\tau }=\frac{1}{8\pi Ga}\left( \frac{1}{g_{+}}\sqrt{1+g_{+}^{2}%
\dot{a}^{2}}-\frac{1}{g_{-}}\sqrt{1+g_{-}^{2}\dot{a}^{2}}\right) 
\end{equation}%
Then for other components%
\begin{equation}
-8\pi GS_{\psi }^{\psi }=K_{\psi }^{\psi }-K=K_{\psi }^{\psi }+\text{ }%
K_{\tau }^{\tau }-K_{\psi }^{\psi }=+\text{ }K_{\tau }^{\tau }
\end{equation}%
\begin{equation}
S_{\psi }^{\psi }=-\frac{1}{8\pi G}\left( K_{\tau }^{\tau +}-K_{\tau }^{\tau
-}\right) 
\end{equation}%
\begin{equation}
S_{\psi }^{\psi }=\frac{1}{8\pi G}[-\left( \ddot{a}+\frac{f_{+}^{\text{ }%
\prime }}{f_{+}g_{+}^{2}}-\left( \frac{f_{+}^{\prime }}{f_{+}}+\frac{%
g_{+}^{\prime }}{g_{+}}\right) \dot{a}^{2}\right) \frac{g_{+}}{\sqrt{%
1+g_{+}^{2}\dot{a}^{2}}}
\end{equation}%
\begin{equation*}
+\left( \ddot{a}+\frac{f_{-}^{\text{ }\prime }}{f_{-}g_{-}^{2}}-\left( \frac{%
f_{-}^{\prime }}{f_{-}}+\frac{g_{-}^{\prime }}{g_{-}}\right) \dot{a}%
^{2}\right) \frac{g_{-}}{\sqrt{1+g_{-}^{2}\dot{a}^{2}}}]
\end{equation*}
and the last component is 
\begin{equation}
S_{\tau }^{\psi }=S_{\psi }^{\tau }=-\frac{1}{8\pi G}\left( K_{\psi }^{\tau
+}-K_{\psi }^{\tau -}\right) =-\frac{a^{2}}{8\pi G}\left( -\frac{\omega
^{\prime }}{2f_{+}g_{+}}+\frac{\omega ^{\prime }}{2f_{-}g_{-}}\right) 
\end{equation}
\begin{equation}
=-\frac{a^{2}}{8\pi G}\left( \omega _{+}^{\prime }-\omega _{-}^{\prime
}\right) 
\end{equation}
The special condition of $\omega _{+}^{\prime }=\omega _{-}^{\prime },$ $%
\omega _{+}=\omega _{-}$ so $S_{\tau }^{\psi }=S_{\psi }^{\tau }=0.$%
Therefore it implies that the upper-shell and the lower-shell are
co-rotating. The surface stress-energy tensor is $S_{b}^{a}=\left( 
\begin{array}{cc}
-\sigma  & 0 \\ 
0 & p%
\end{array}%
\right) .$
One calculates the charge density and the surface pressure as follows%
\begin{equation}
\sigma =-\frac{1}{8\pi Ga}\left( \frac{1}{g_{+}}\sqrt{1+g_{+}^{2}\dot{a}^{2}}%
-\frac{1}{g_{-}}\sqrt{1+g_{-}^{2}\dot{a}^{2}}\right) 
\end{equation}%
and 
\begin{equation*}
p=\frac{1}{8\pi G}[-\left( \ddot{a}+\frac{f_{+}^{\text{ }\prime }}{%
f_{+}g_{+}^{2}}-\left( \frac{f_{+}^{\prime }}{f_{+}}+\frac{g_{+}^{\prime }}{%
g_{+}}\right) \dot{a}^{2}\right) \frac{g_{+}}{\sqrt{1+g_{+}^{2}\dot{a}^{2}}}
\end{equation*}%
\begin{equation}
+\left( \ddot{a}+\frac{f_{-}^{\text{ }\prime }}{f_{-}g_{-}^{2}}-\left( \frac{%
f_{-}^{\prime }}{f_{-}}+\frac{g_{-}^{\prime }}{g_{-}}\right) \dot{a}%
^{2}\right) \frac{g_{-}}{\sqrt{1+g_{-}^{2}\dot{a}^{2}}}]
\end{equation}
\chapter{THIN-SHELL WORMHOLES}
Thin-shell WHs is constructed with the exotic matter which is located on a
hypersurface so that it can be minimized. Constructing WHs with non-exotic
source is a difficult issue in GR. On this purpose, firstly , Visser use the
thin-shell method to construct WHs for minimizing the exotic matter on the
throat of the WHs. We need to introduce some conditions on the
energy-momentum tensor such as\cite{nor3,lake,counter}
\begin{figure}[h!]

\centering
\includegraphics[width=1.00\textwidth]{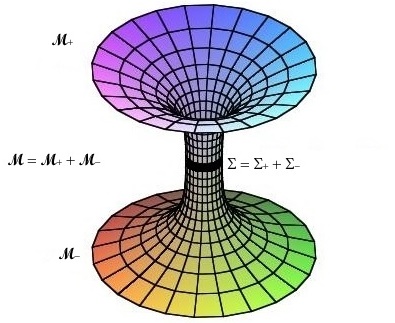} \label{fig:thin-shell
wormhole}
\caption{Thin-Shell WH}

\end{figure}
-Weak Energy Condition

This energy condition states that energy density of any matter distribution
must be non-negative, i.e., $3\sigma \geq 0$ and $3\sigma +p\geq 0$.

- Null Energy Condition

This condition implies that $3\sigma +p\geq 0$.

- Dominant Energy Condition

This condition holds if $3\sigma \geq 0$ and $3\sigma +2p\geq 0$.

- Strong Energy Condition

This condition demands $3\sigma +p\geq 0$ and $3\sigma +3p\geq 0.$
\vspace{-0.75cm}
\section{Mazharimousavi-Halilsoy Thin-Shell WH in 2+1 D}
\vspace{-0.75cm}
In this section \cite{ao2}, we introduce a SHBH (SHBH) investigated by
Mazharimousavi and Halilsoy recently \cite{charged}. The following action
describes the Einstein-Maxwell gravity that is minimally coupled to a scalar
field $\phi $ \ 
\begin{equation}
S=\int d^{3}r\sqrt{-g}\left( R-2\partial _{\mu }\phi \partial ^{\mu }\phi
-F^{2}-V\left( \phi \right) \right) ,  \label{eq1}
\end{equation}%
where $R$ denotes the Ricci scalar, $F=F_{\mu \nu }F^{\mu \nu }$ is the
Maxwell invariant, and $V\left( \phi \right) $ stands for the scalar ($\phi $%
) potential. From the action above, the SHBH solution is obtained as 
\begin{equation}
ds^{2}=-f(r)dt^{2}+\frac{4r^{2}dr^{2}}{f(r)}+r^{2}d\theta ^{2},  \label{eq2}
\end{equation}%
which 
\begin{equation}
f(r)=\frac{r^{2}}{l{^{2}}}-ur.  \label{eq3}
\end{equation}%
Here $u$ and $l$ are constants, and event horizon of the BH is located at $%
r_{h}=u\ell ^{2}$. It is clear that this BH possesses a non-asymptotically
flat geometry. Metric (\ref{eq2}) can be written in the form of 
\begin{equation}
ds^{2}=-\frac{r}{\ell ^{2}}\left( r-r_{h}\right) dt^{2}+\frac{4r\ell
^{2}dr^{2}}{\left( r-r_{h}\right) }+r^{2}d\theta ^{2}.
\end{equation}%
It is noted that the singularity located at $r=0$, which is also seen from
the Ricci and Kretschmann scalars: 
\begin{equation}
R=-\frac{2r+r_{h}}{4r^{3}\ell ^{2}},
\end{equation}%
\begin{equation}
K=\frac{4r^{2}-4r_{h}r+3r_{h}^{2}}{16r^{6}\ell ^{4}}.
\end{equation}%
Moreover, One obtains the scalar field and potential respectively as follows
\begin{equation}
\phi =\frac{\ln r}{\sqrt{2}},
\end{equation}
\begin{equation}
V\left( \phi \right) =\frac{\lambda _{1}+\lambda _{2}}{r^{2}},
\end{equation}%
in which $\lambda _{1,2}$ are constants. The corresponding Hawking
temperature is calculated as 
\begin{equation}
T_{H}=\frac{1}{4\pi }\left. \frac{\partial {f}}{\partial {r}}\right\vert
_{r=r_{h}}=\frac{1}{8\pi \ell ^{2}},
\end{equation}%
which is constant. Having a radiation with constant temperature is the
well-known isothermal process. It is worth noting that Hawking radiation of
the linear dilaton BHs exhibits similar isothermal behavior.
\vspace{-0.75cm}
\section{Stability of the Thin-Shell WH}
\vspace{-0.75cm}
In this section, we take two identical copies of the SHBHs with \cite{ao2} $%
(a\geq r)$: 
\begin{equation}
M^{\pm }=(x|r\geq 0),
\end{equation}%
and the manifolds are bounded by hypersurfaces $M^{+}$ and $M^{-}$, to get
the single manifold $M=M^{+}+M^{-}$, we glue them together at the surface of
the junction 
\begin{equation}
\Sigma ^{\pm }=(x|r=a).
\end{equation}%
where the boundaries $\Sigma $ are given. The spacetime on the shell is%
\begin{equation}
ds^{2}=-d\tau ^{2}+a(\tau )^{2}d\theta ^{2},
\end{equation}
where $\tau $ represents the proper time . Setting coordinates $\xi
^{i}=(\tau ,\theta )$, the extrinsic curvature formula connecting the two
sides of the shell is simply given by 
\begin{equation}
K_{ij}^{\pm }=-n_{\gamma }^{\pm }\left( \frac{\partial ^{2}x^{\gamma }}{%
\partial \xi ^{i}\partial \xi ^{j}}+\Gamma _{\alpha \beta }^{\gamma }\frac{%
\partial x^{\alpha }}{\partial \xi ^{i}}\frac{\partial x^{\beta }}{\partial
\xi ^{j}}\right) ,  \label{extcur}
\end{equation}%
where the unit normals ($n^{\gamma }n_{\gamma }=1)$ are 
\begin{equation}
n_{\gamma }^{\pm }=\pm \left\vert g^{\alpha \beta }\frac{\partial H}{%
\partial x^{\alpha }}\frac{\partial H}{\partial x^{\beta }}\right\vert
^{-1/2}\frac{\partial H}{\partial x^{\gamma }},  \label{normgen}
\end{equation}%
with $H(r)=r-a(\tau )$. The non zero components of $n_{\gamma }^{\pm }$ are
calculated as 
\begin{equation}
n_{t}=\mp 2a\dot{a},
\end{equation}%
\begin{equation}
n_{r}=\pm 2\sqrt{\frac{al^{2}(4\dot{a}^{2}l^{2}a-l^{2}u+a)}{(l^{2}u-a)}},
\end{equation}%
where the dot over a quantity denotes the derivative with respect to $\tau $%
. Then, the non-zero extrinsic curvature components yield 
\begin{equation}
K_{\tau \tau }^{\pm }=\mp \frac{\sqrt{-al^{2}}(8\dot{a}^{2}l^{2}a+8\ddot{a}%
l^{2}a^{2}-l^{2}u+2a)}{4a^{2}l^{2}\sqrt{-4\dot{a}^{2}l^{2}a-l^{2}u+a}},
\end{equation}%
\begin{equation}
K_{\theta \theta }^{\pm }=\pm \frac{1}{2a^{\frac{3}{2}}l}\sqrt{4\dot{a}%
^{2}l^{2}a-l^{2}u+a}.
\end{equation}
Since $K_{ij}$ is not continuous around the shell, we use the Lanczos
equation:
\begin{equation}
S_{ij}=-\frac{1}{8\pi }\left( [K_{ij}]-[K]g_{ij}\right) .  \label{ee}
\end{equation}
where $K$ is the trace of $K_{ij}$, $[K_{ij}]=K_{ij}^{+}-K_{ij}^{-}$ .
Firstly, $K^{+}=-K^{-}=[K_{ij}]$ while $[K_{ij}]=0.$ For the conservation of
the surface stress--energy $S_{j}^{ij}=0$ and $S_{ij}$ is stress
energy-momentum tensor at the junction which is given in general by 
\begin{equation}
S_{j}^{i}=diag(\sigma ,-p),  \label{enerji}
\end{equation}%
with the surface pressure $p$ and the surface energy density $\sigma $. Due
to the circular symmetry, we have 
\begin{equation}
K_{j}^{i}=[{K_{\tau }^{\tau }, 0, 0, K_{\theta }^{\theta } }].
\end{equation}%
Thus, from Eq.s (\ref{enerji}) and (\ref{ee}) one obtains the surface
pressure and surface energy density . Using the cut and paste technique, we
can demount the interior regions $r<a$ of the geometry, and links its
exterior parts. The energy density and pressure are 
\begin{equation}
\sigma =-\frac{1}{8\pi a^{\frac{3}{2}}l}\sqrt{4\dot{a}^{2}l^{2}a-l^{2}u+a},
\label{ed}
\end{equation}%
\begin{equation}
p=\frac{1}{16\pi a^{\frac{3}{2}}l}\frac{\left( 8\dot{a}^{2}l^{2}a+8\ddot{a}%
l^{2}a^{2}-l^{2}u+2a\right) }{\sqrt{4\dot{a}^{2}l^{2}a-l^{2}u+a}}.
\label{pre}
\end{equation}
Then for the static case ($a=a_{0}$), the energy and pressure quantities
reduce to 
\begin{equation}
\sigma _{0}=-\frac{1}{8\pi a_{0}^{\frac{3}{2}}l}\sqrt{-l^{2}u+a_{0}},
\end{equation}%
\begin{equation}
p_{0}=\frac{1}{16\pi a_{0}^{\frac{3}{2}}l}\frac{\left( -l^{2}u+2a_{0}\right) 
}{\sqrt{-l^{2}u+a_{0}}}.
\end{equation}
Once $\sigma \geq 0$ and $\sigma +p\geq 0$ hold, then WEC is satisfied.
Besides, $\sigma +p\geq 0$ is the condition of NEC. Furthermore, SEC is
conditional on $\sigma +p\geq 0$ and $\sigma +2p\geq 0$. It is obvious from
Eq. (24) that negative energy density violates the WEC, and consequently we
are in need of the exotic matter for constructing thin-shell WH. We note
that the total matter supporting the WH is given by
\begin{equation}
\Omega _{\sigma }=\int_{0}^{2\pi }\left. [\rho \sqrt{-g}]\right\vert
_{r=a_{0}}d\phi =2\pi a_{0}\sigma (a_{0})=-\frac{1}{4a_{0}^{\frac{1}{2}}|l|}%
\sqrt{-l^{2}u+a_{0}}.
\end{equation}
Stability of the WH is investigated using the linear perturbation so that
the EoS is 
\begin{equation}
p=\psi (\sigma ),
\end{equation}%
where $\psi (\sigma )$ is an arbitrary function of $\sigma $. Furthermore,
the energy conservation equation is introduced as follows 
\begin{equation}
S_{j;i}^{i}=-T_{\alpha \beta }\frac{\partial x^{\alpha }}{\partial \xi ^{j}}%
n^{\beta },
\end{equation}%
where $T_{\alpha \beta }$ is the bulk energy-momentum tensor. It can be
written in terms of the pressure and energy density: 
\begin{equation}
\frac{d}{d\tau }\left( \sigma a\right) +\psi \frac{da}{d\tau }=-\dot{a}%
\sigma .
\end{equation}%
From above equation, one reads 
\begin{equation}
\sigma ^{\prime }=-\frac{1}{a}(2\sigma +\psi ),
\end{equation}%
and its second derivative yields%
\begin{equation}
\sigma ^{\prime \prime }=\frac{2}{a^{2}}(\tilde{\psi}+3)(\sigma +\frac{\psi 
}{2}).
\end{equation}
where prime and tilde symbols denote derivative with respect to $a$ and $%
\sigma $, respectively. The equation of motion for the shell is in general
given by 
\begin{equation}
\dot{a}^{2}+V=0,  \label{eqmt}
\end{equation}%
where the effective potential $V$ is found from Eq. (\ref{ed} as 
\begin{equation}
V=\frac{1}{4l^{2}}-\frac{u}{4a}-16a^{2}\sigma ^{2}\pi ^{2}.
\end{equation}%
In fact, Eq. (\ref{eqmt}) is nothing but the equation of the oscillatory
motion in which the stability around the equilibrium point $a=a_{0}$ is
conditional on $V^{\prime \prime }(a_{0})\geq 0$. We finally obtain 
\begin{equation}
V^{\prime \prime }=\left. -\frac{1}{2a^{3}}\left[ 64\pi ^{2}a^{5}\left(
\left( \sigma \sigma ^{\prime }\right) ^{\prime }+4\sigma ^{\prime }\frac{%
\sigma }{a}+\frac{\sigma ^{2}}{a^{2}}\right) +u\right] \right\vert
_{a=a_{0}},
\end{equation}%
or equivalently, 
\begin{equation}
V^{\prime \prime }=\left. \frac{1}{2a^{3}}\{-64\pi ^{2}a^{3}\left[ (2\psi
^{\prime }+3)\sigma ^{2}+\psi (\psi ^{\prime }+3)\sigma +\psi ^{2}\right]
-u\}\right\vert _{a=a_{0}}.
\end{equation}
The equation of motion of the throat, for a small perturbation
becomes %
\begin{equation}
\dot{a}^{2}+\frac{V^{\prime \prime }(a_{0})}{2}(a-a_{0})^{2}=0.
\end{equation}
Note that for the condition of $V^{\prime \prime }(a_{0})\geq 0$%
, TSW is stable where the motion of the throat is oscillatory with
angular frequency $\omega =\sqrt{\frac{V^{\prime \prime }(a_{0})}{2}}$.
\vspace{-0.75cm}
\section{Some Models of EoS Supporting Thin-Shell WH}
\vspace{-0.75cm}
In this section, we use particular gas models (linear barotropic gas (LBG) ,
chaplygin gas (CG) , generalized chaplygin gas (GCG) and logarithmic gas
(LogG) ) to explore the stability of TSW.
\vspace{-0.75cm}
\subsection{Stability analysis of Thin-Shell WH via the LBG}
\vspace{-0.75cm}
The equation of state of LBG is given by 
\begin{equation}
\psi =\varepsilon _{0}\sigma ,
\end{equation}%
and hence 
\begin{equation}
\psi ^{\prime }(\sigma _{0})=\varepsilon _{0},
\end{equation}%
where $\varepsilon _{0}$ is a constant parameter. By changing the values of $%
l$ and $u$ in Eq. (35), we illustrate the stability regions for TSW, in
terms of $\varepsilon _{0}$ and $a_{0}$, as depicted in Fig.3.2.
\begin{figure}[h!]
\centering
\includegraphics[width=0.40\textwidth]{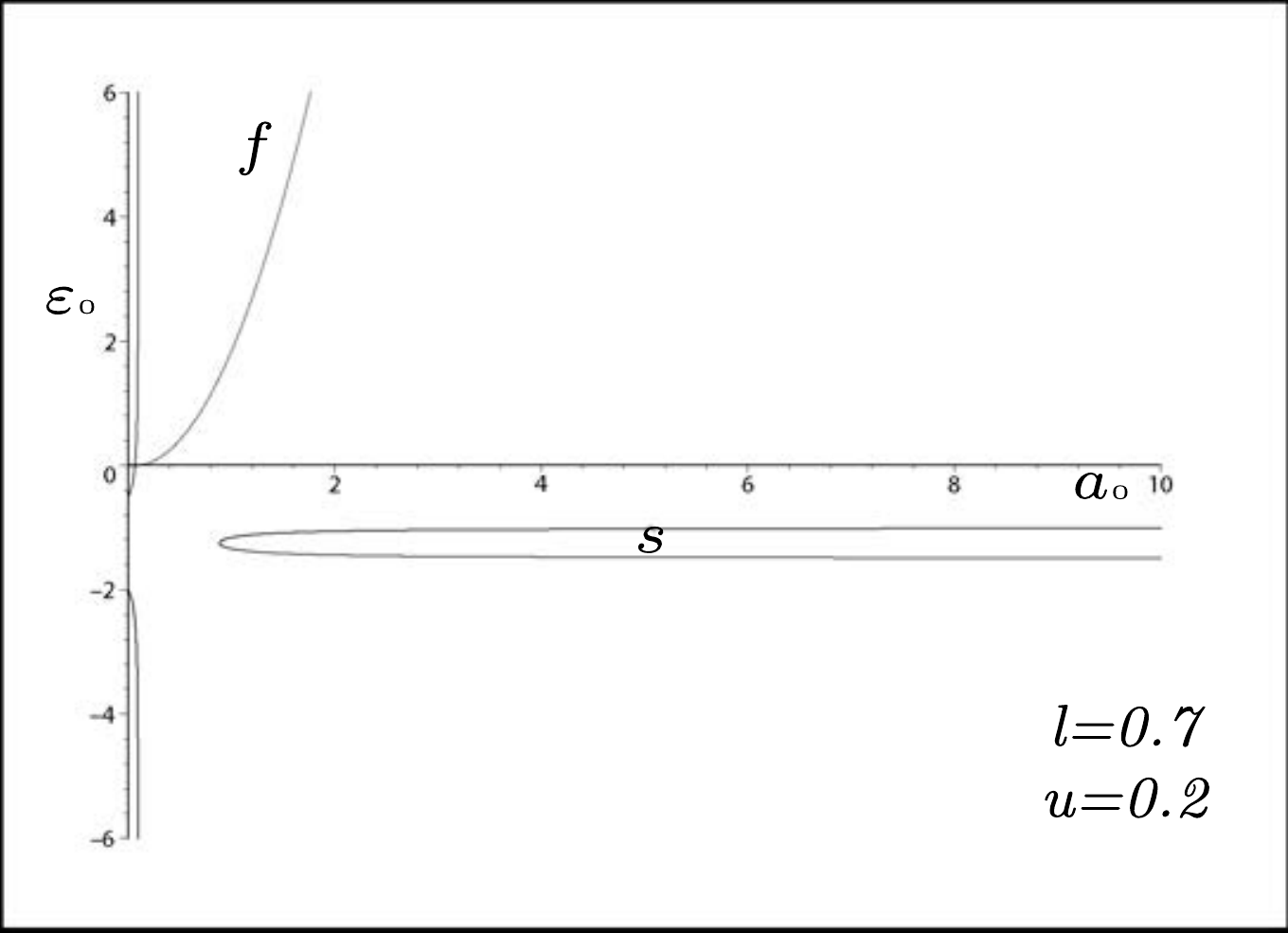} %
\includegraphics[width=0.40\textwidth]{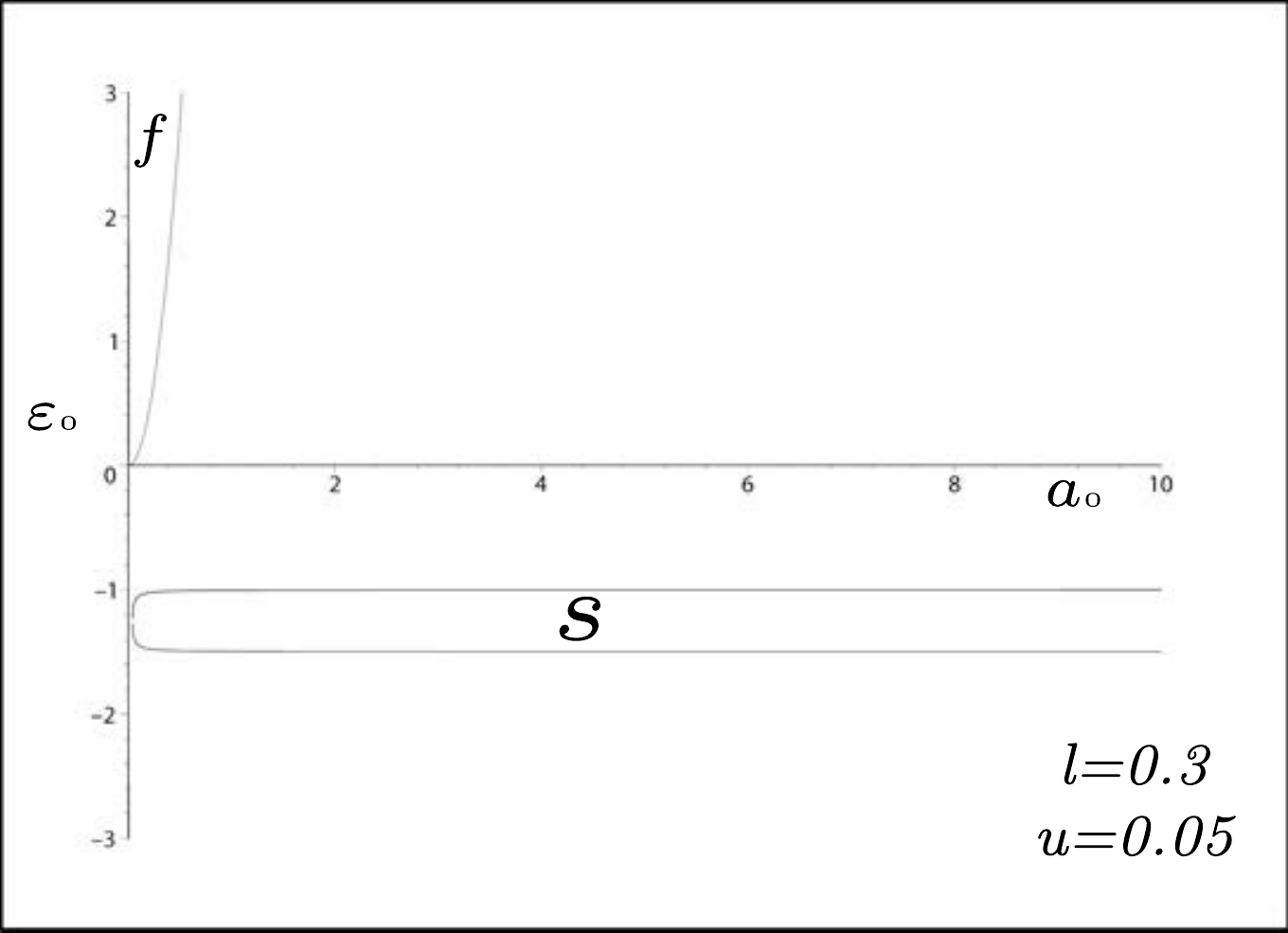} %
\includegraphics[width=0.40\textwidth]{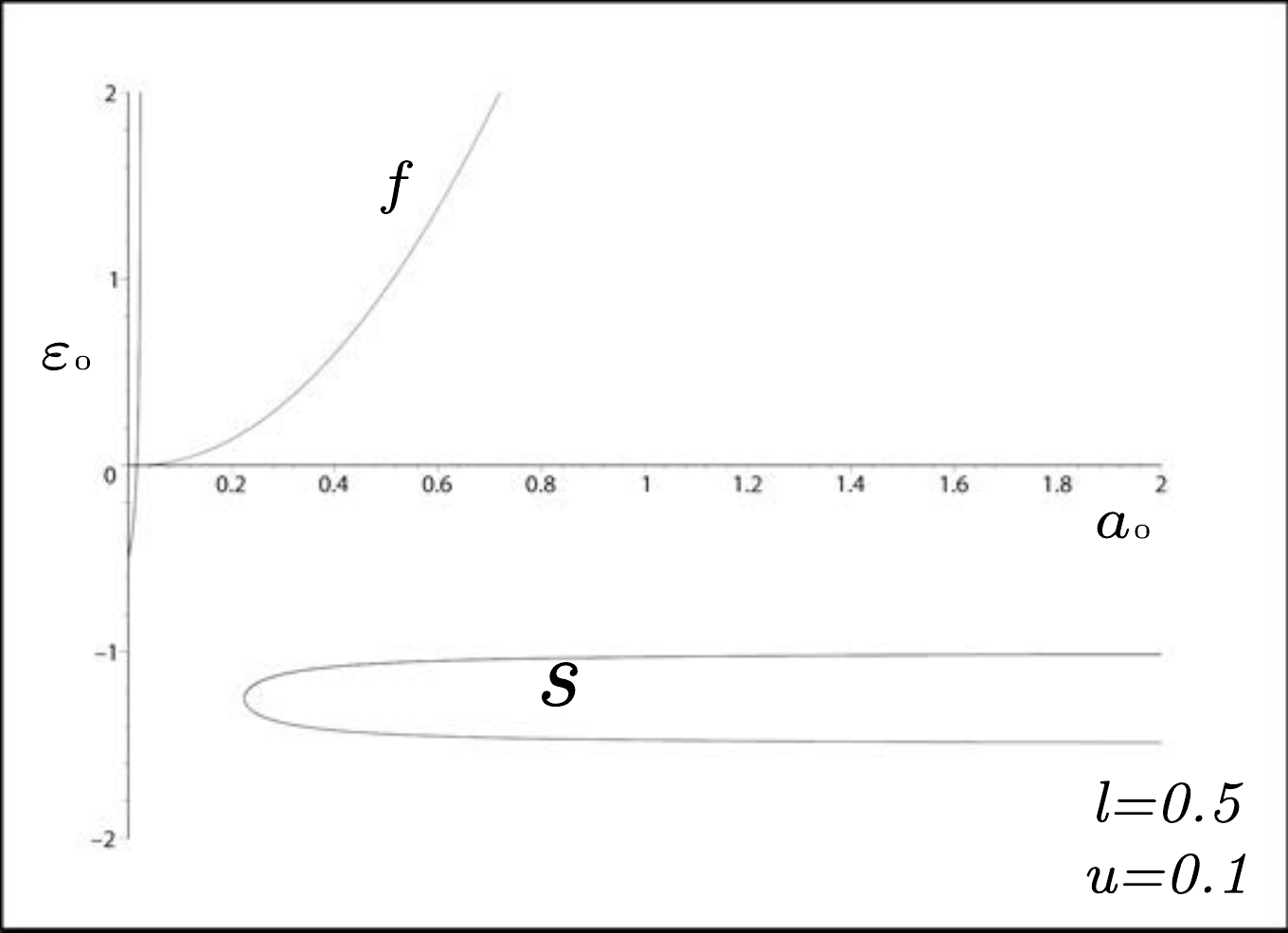} %
\includegraphics[width=0.40\textwidth]{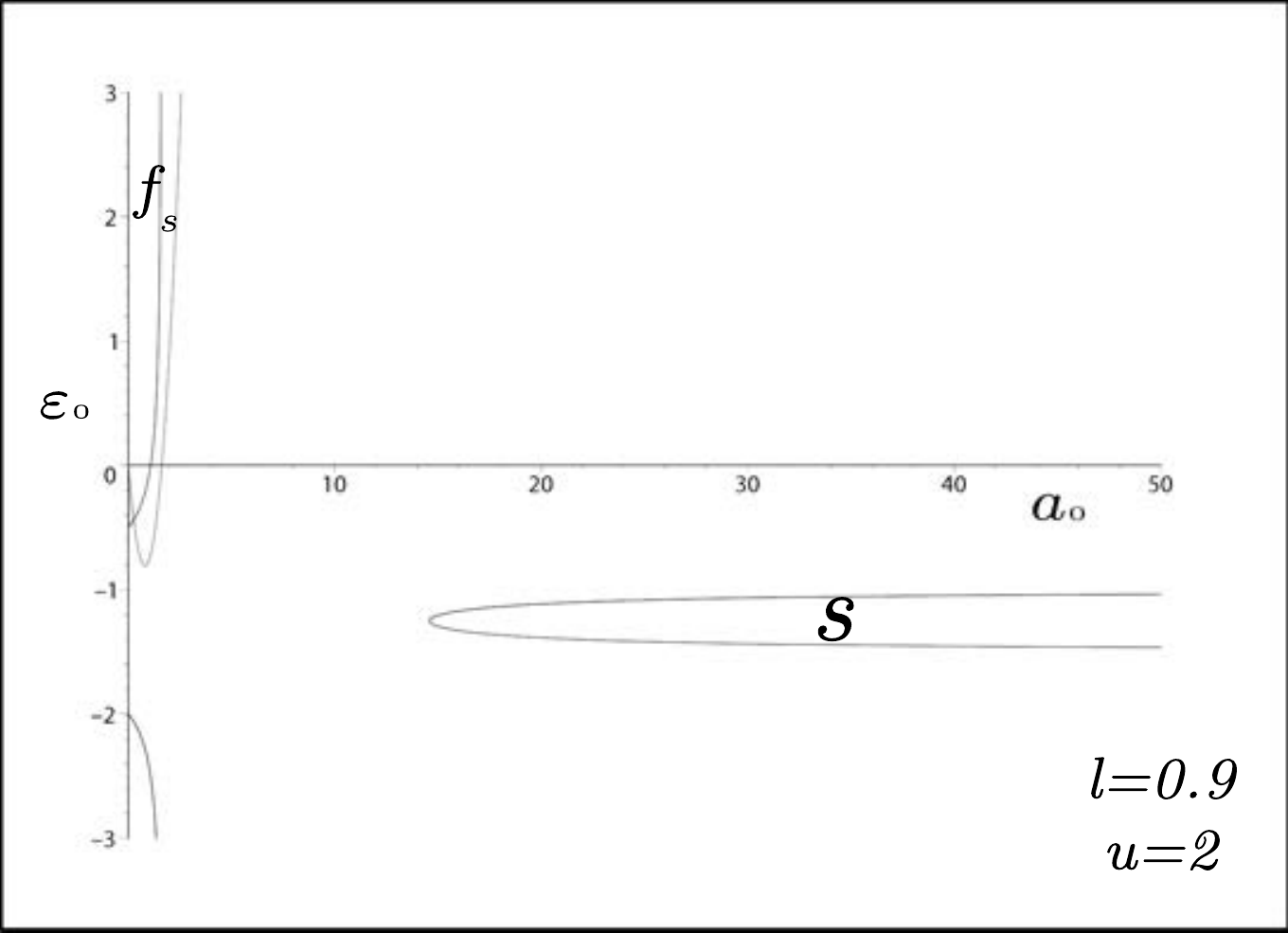} \label{fig:StabilityLBG} \caption{Stability Regions via the LBG}
\end{figure}
\vspace{-0.75cm}
\subsection{Stability analysis of Thin-Shell WH via CG}
\vspace{-0.75cm}
The equation of state of CG that we considered is given by 
\begin{equation}
\psi =\varepsilon _{0}(\frac{1}{\sigma }-\frac{1}{\sigma _{0}})+p_{0},
\end{equation}
and one naturally finds 
\begin{equation}
\psi ^{\prime }(\sigma _{0})=\frac{-\varepsilon _{0}}{\sigma _{0}^{2}}.
\end{equation}
After inserting Eq. (39) into Eq. (35), The stability regions for thin-shell
WH supported by CG is plotted in Fig.3.3.
\begin{figure}[h!]
\centering
\includegraphics[width=0.40\textwidth]{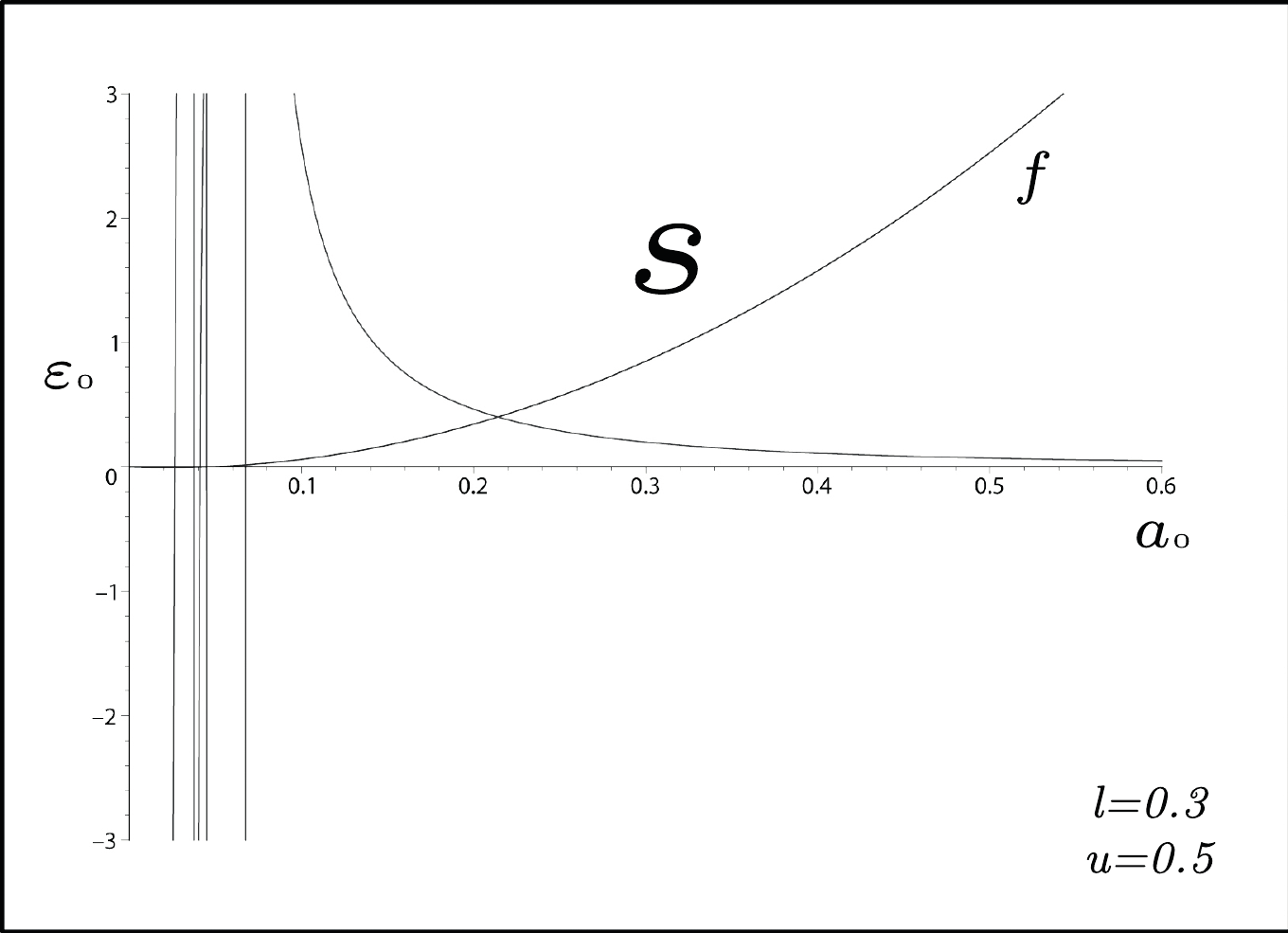} %
\includegraphics[width=0.40\textwidth]{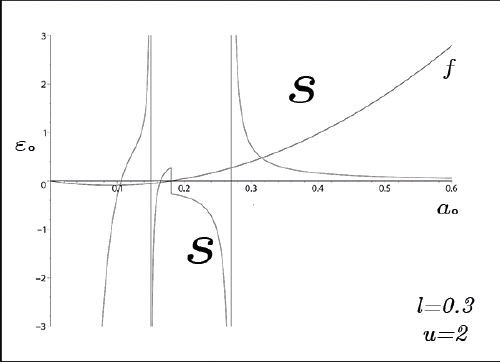} %
\includegraphics[width=0.40\textwidth]{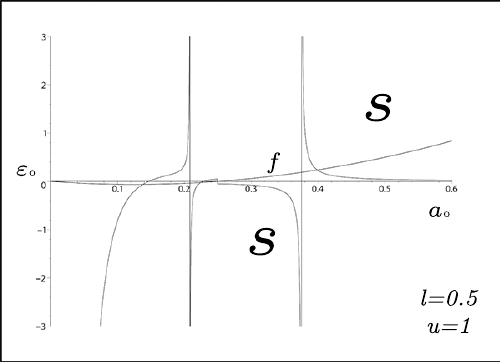} %
\includegraphics[width=0.40\textwidth]{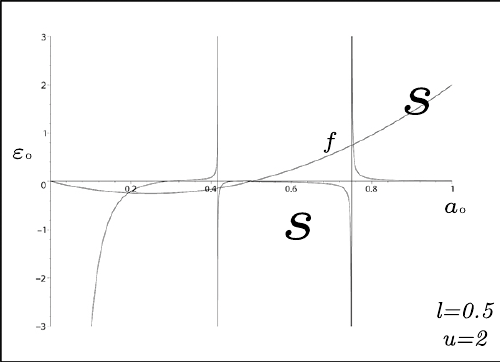} \label{fig:StabilityCG} \caption{Stability Regions via the CG}
\end{figure}
\vspace{-0.75cm}
\subsection{Stability analysis of Thin-Shell WH via GCG}
\vspace{-0.75cm}
By using the equation of state of GCG
\begin{equation}
\psi =p_{0}\left( \frac{\sigma _{0}}{\sigma }\right) ^{\varepsilon _{0}},
\end{equation}
and whence 
\begin{equation}
\psi ^{\prime }(\sigma _{0})=-\varepsilon _{0}\frac{p_{0}}{\sigma _{0}},
\end{equation}
Substituting Eq. (41) in Eq. (35), one can illustrate the stability regions
of thin-shell WH supported by GCG as seen in Fig.3.4. 
\begin{figure}[h!]
\centering
\includegraphics[width=0.40\textwidth]{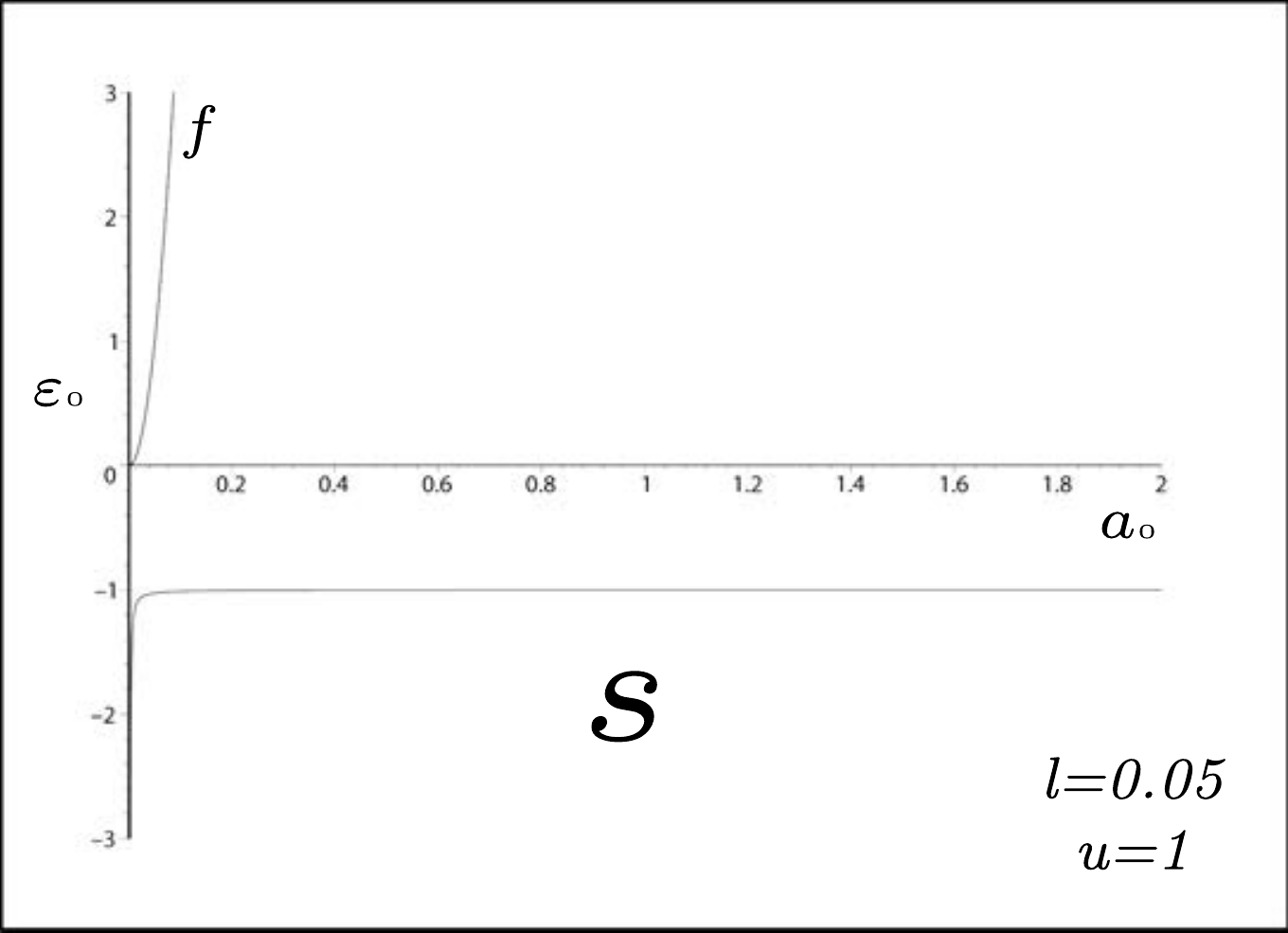} %
\includegraphics[width=0.40\textwidth]{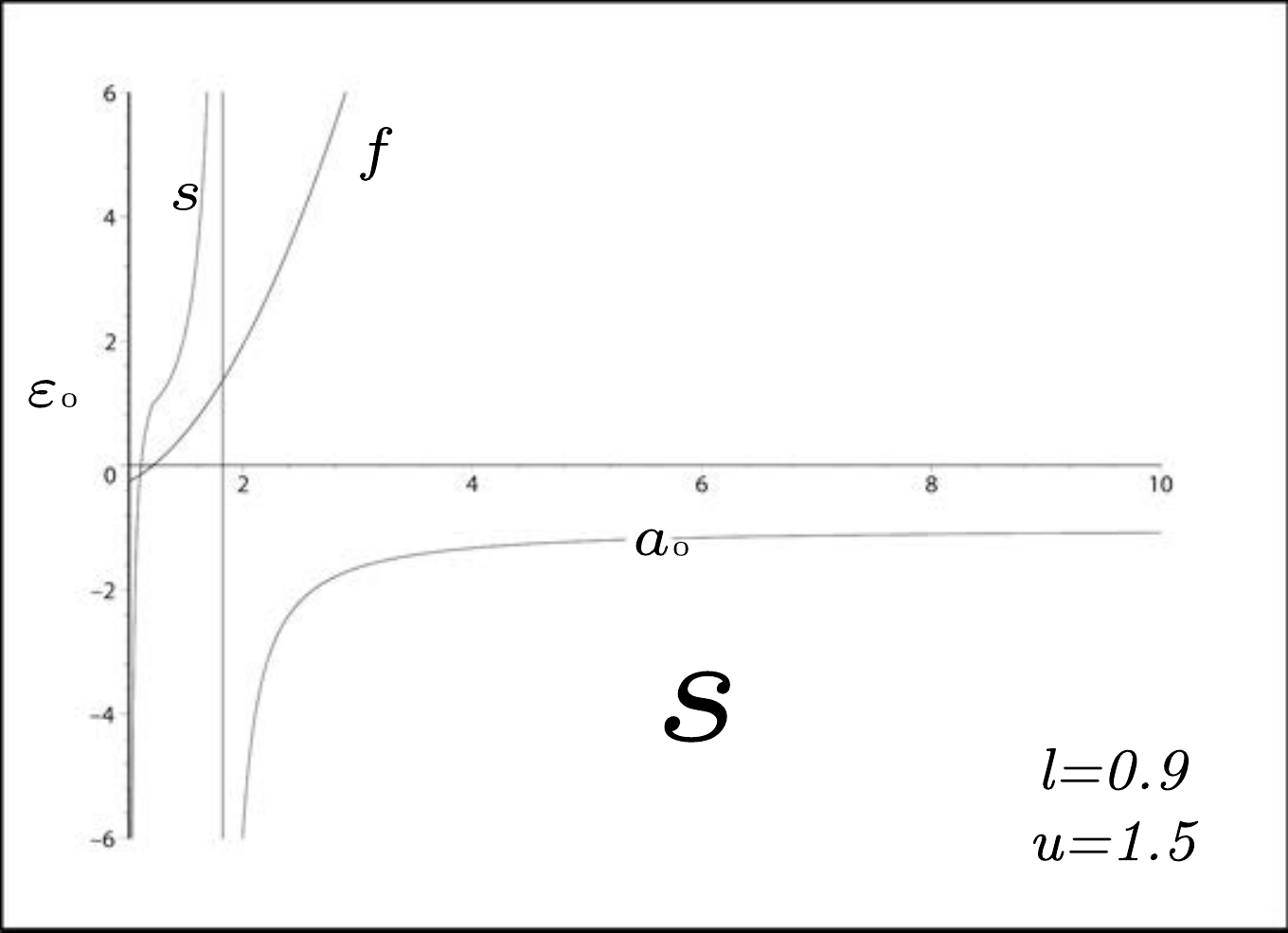} %
\includegraphics[width=0.40\textwidth]{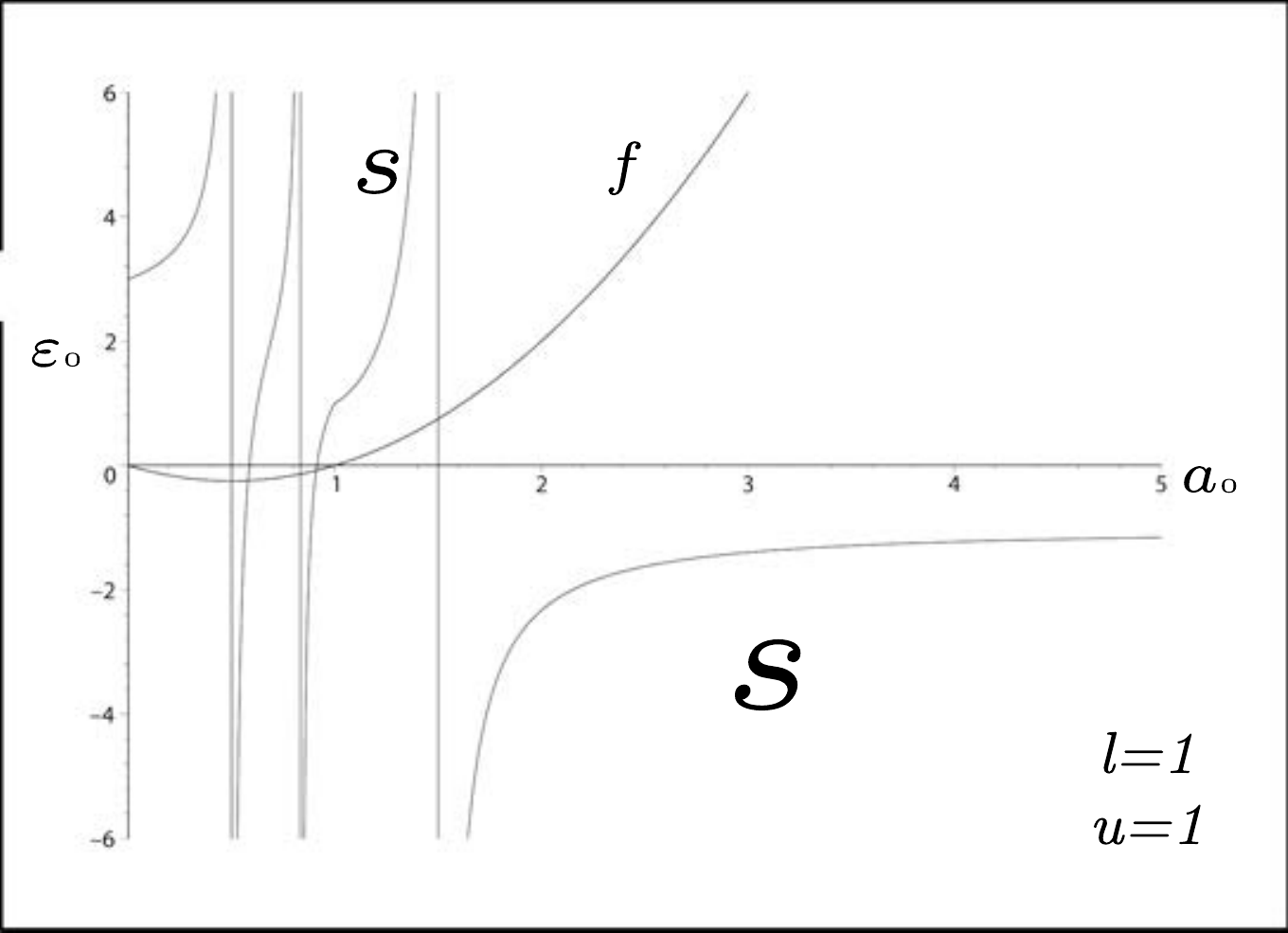} %
\includegraphics[width=0.40\textwidth]{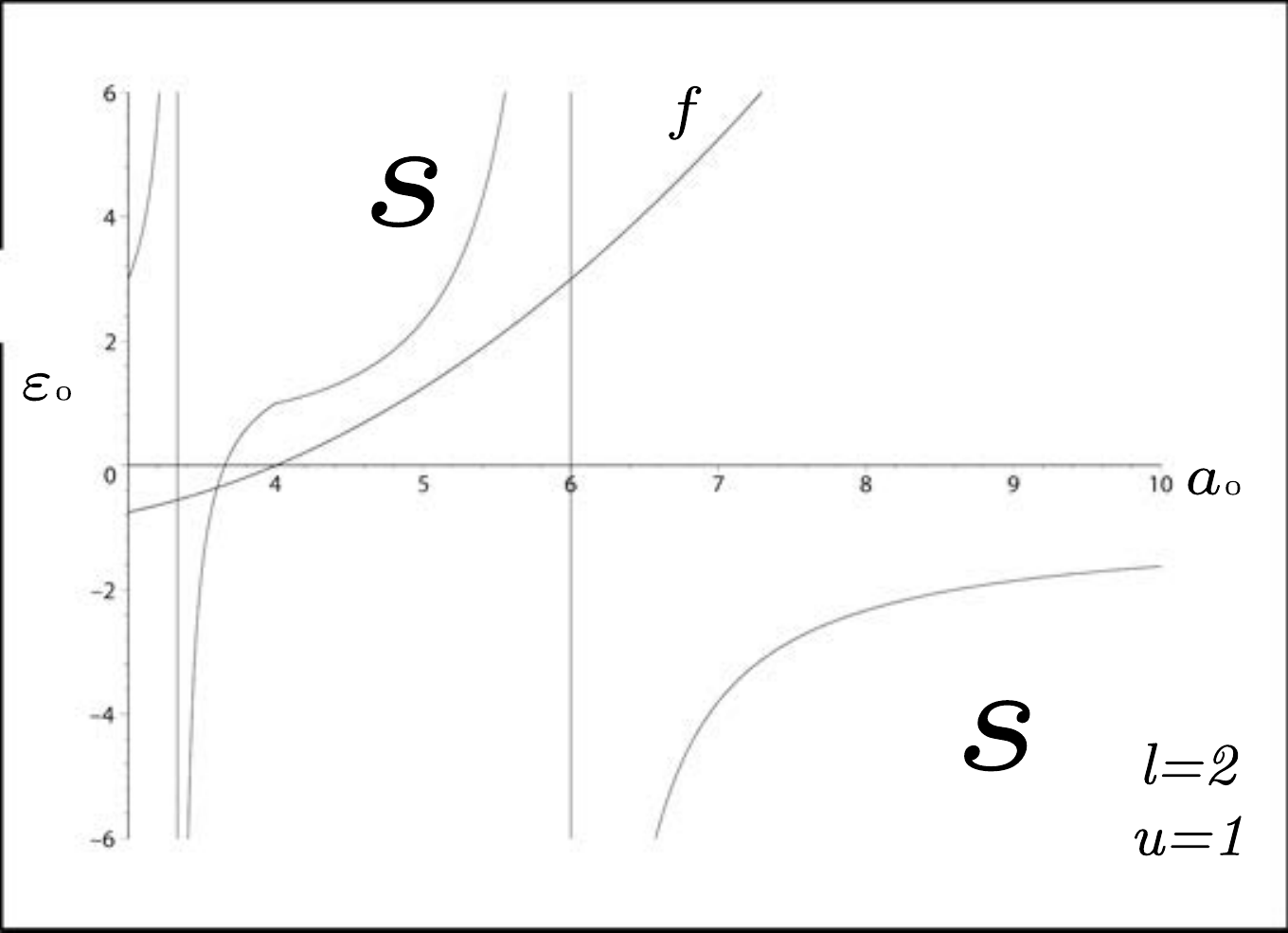} \label{fig:StabilityGCG} \caption{Stability Regions via the GCG}
\end{figure}
\vspace{-0.75cm}
\subsection{Stability analysis of Thin-Shell WH via LogG}
\vspace{-0.75cm}
In our final example, the equation of state for LogG is selected as follows (%
$\varepsilon _{0},$ $\sigma _{0},p_{0}$ are constants) 
\begin{equation}
\psi =\varepsilon _{0}\ln (\frac{\sigma }{\sigma _{0}})+p_{0},
\end{equation}%
which leads to 
\begin{equation}
\psi ^{\prime }(\sigma _{0})=\frac{\varepsilon _{0}}{\sigma _{0}}.
\end{equation}
After inserting the above expression into Eq. (35), we show the stability
regions of thin-shell WH supported by LogG in Fig.3.5.
\begin{figure}[h!]
\centering
\includegraphics[width=0.40\textwidth]{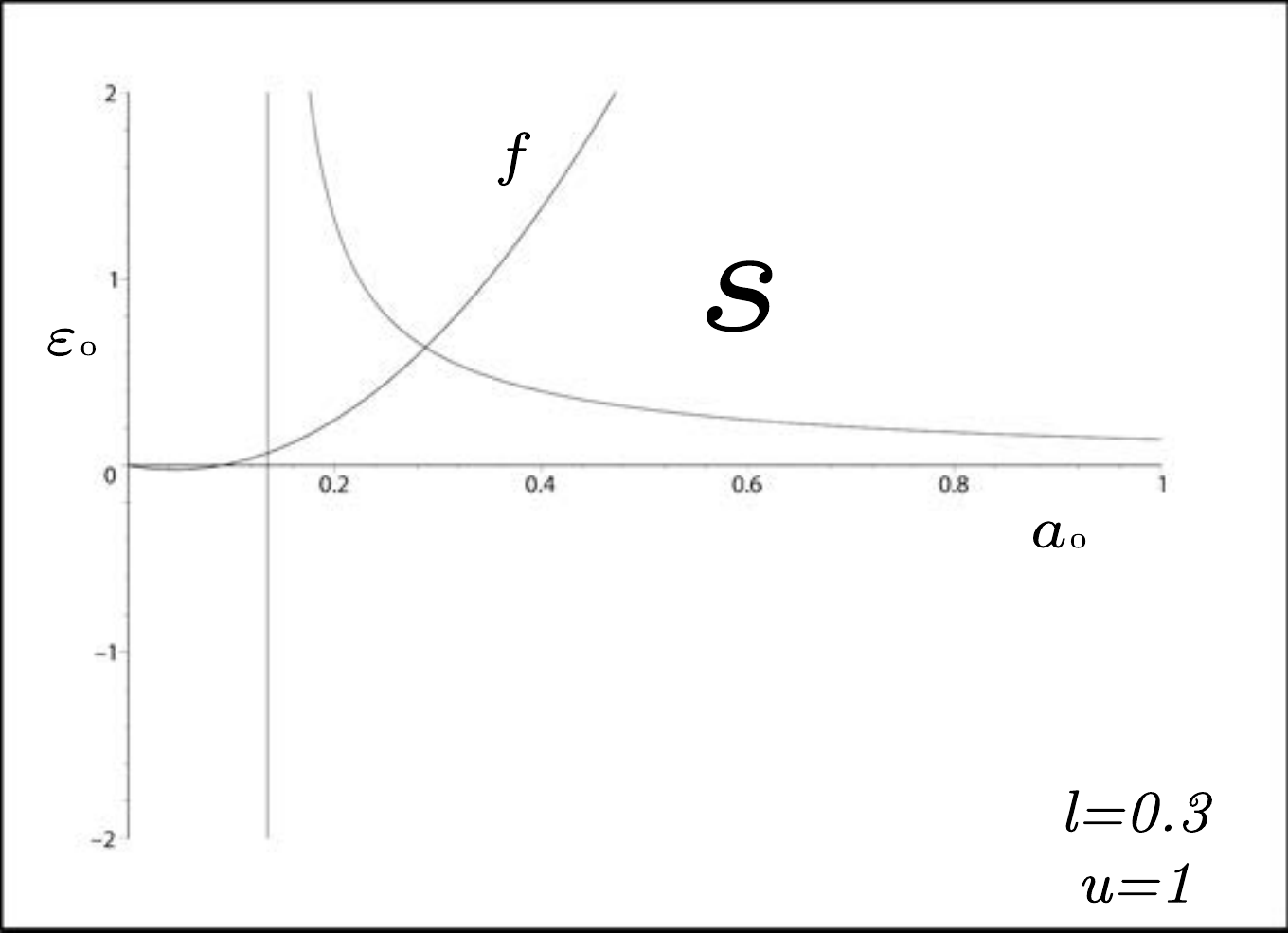} %
\includegraphics[width=0.40\textwidth]{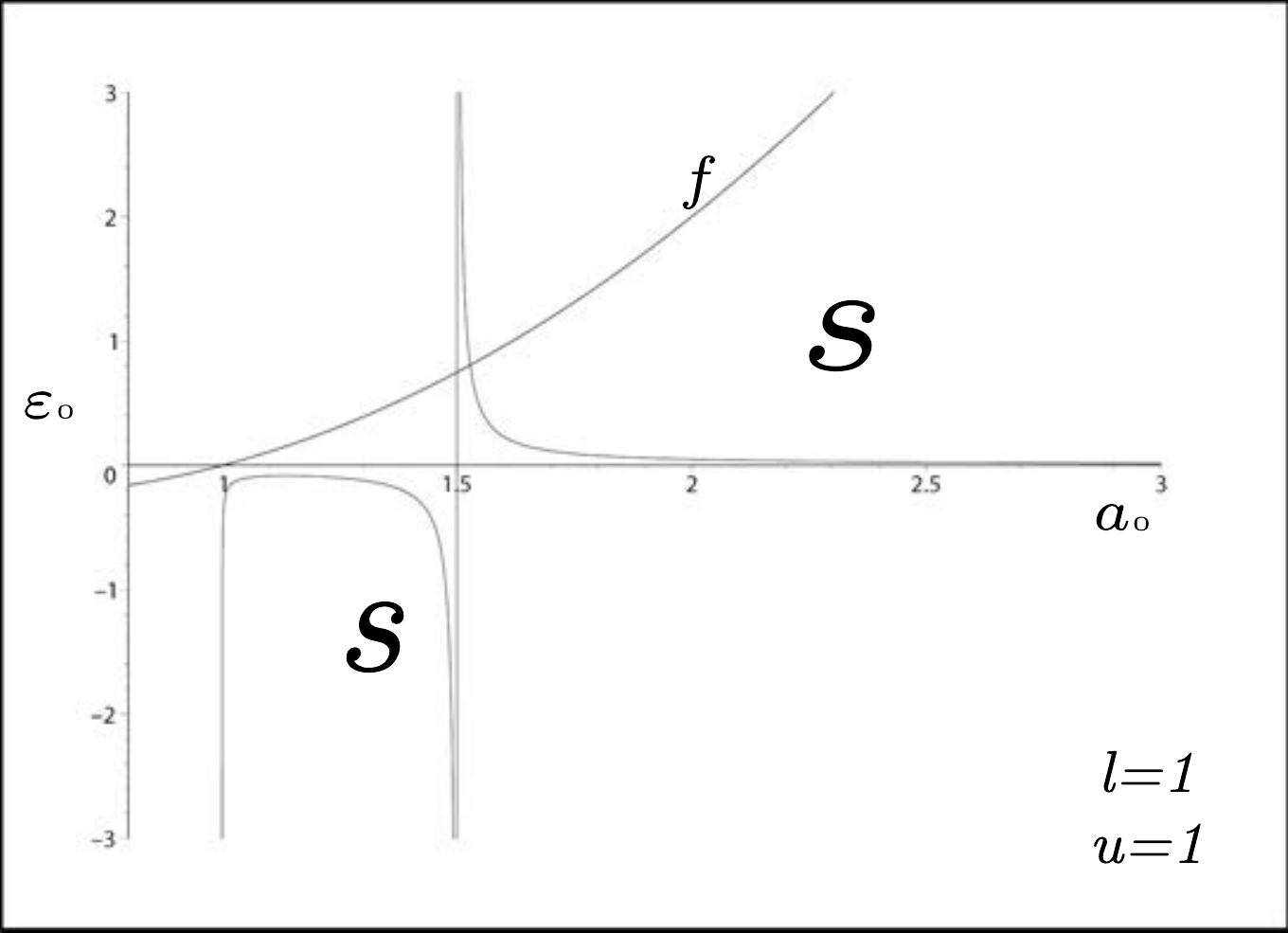} %
\includegraphics[width=0.40\textwidth]{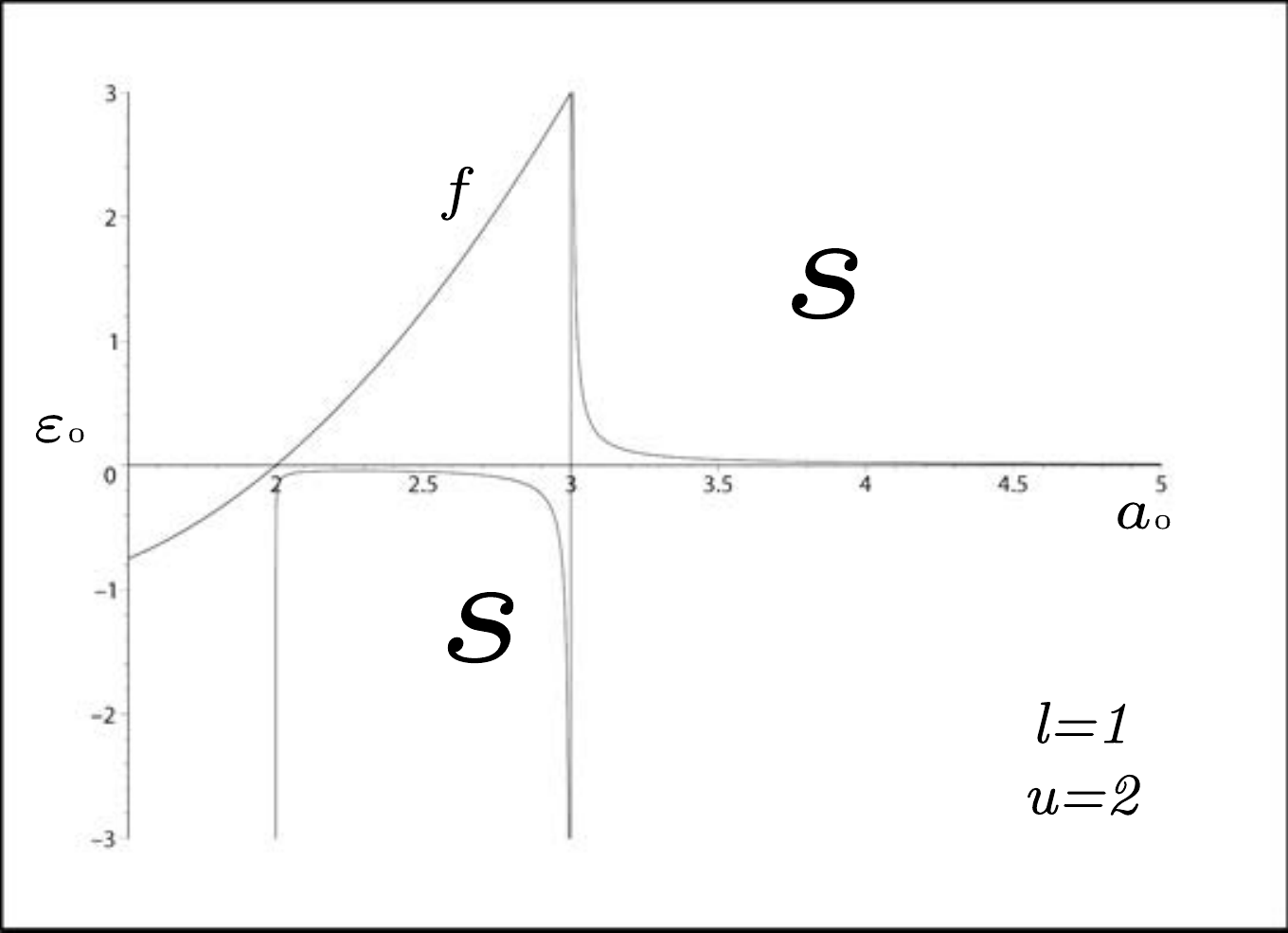} %
\includegraphics[width=0.40\textwidth]{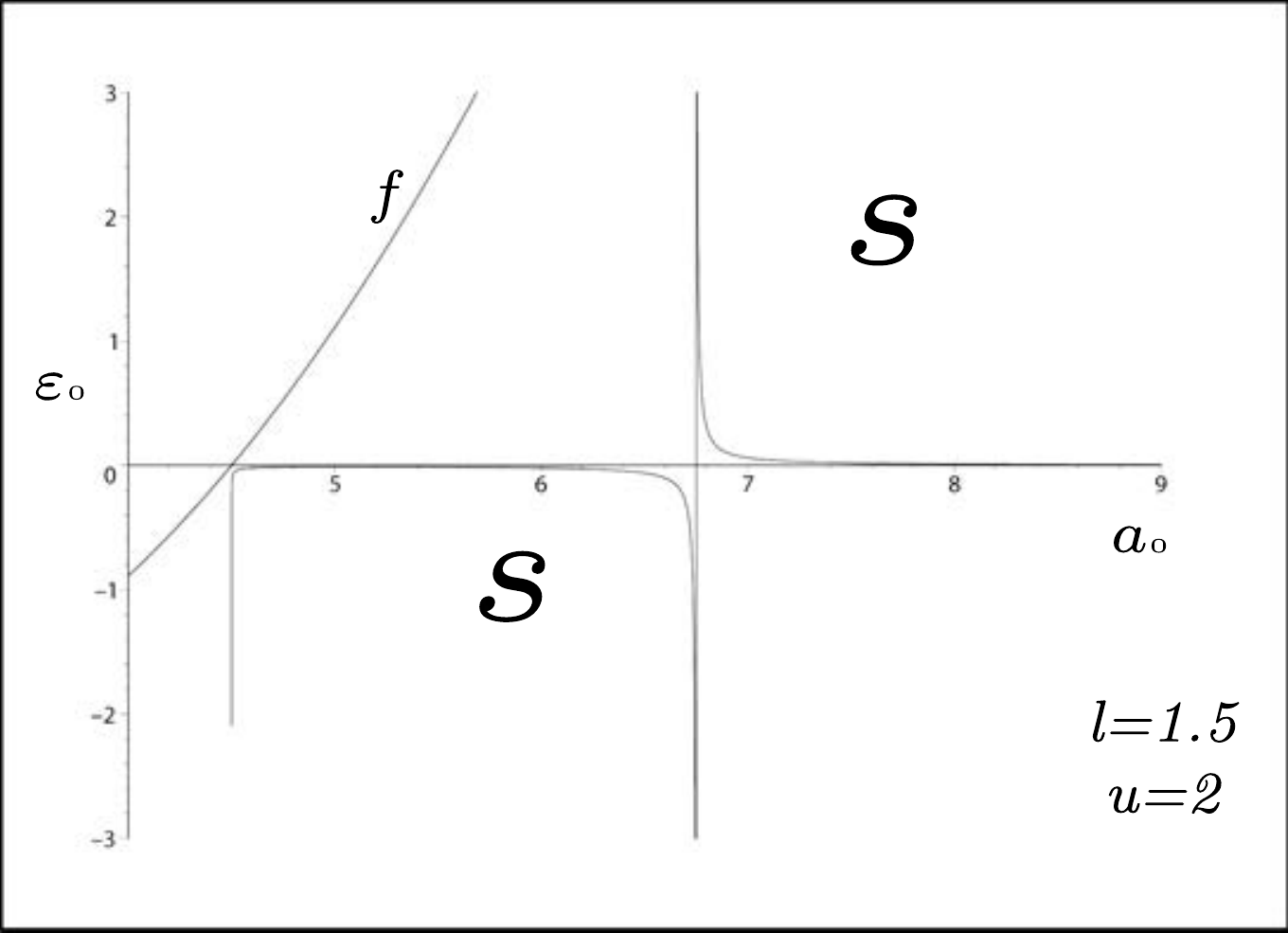} \label{fig:StabilityLogG} \caption{Stability Regions via the LogG}
\end{figure}
In summary, we have constructed thin-shell WH by gluing two copies of SHBH
via the cut and paste procedure. To this end, we have used the fact that the
radius of throat must be greater than the event horizon of the metric given:
($a_{0}>r_{h}$). We have used LBG, CG, GCG, and LogG EoS to the exotic
matter. Then, the stability analysis ($V^{\prime \prime }(a_{0})\geq 0$) is
plotted. We show the stability regions in terms $a_{0}$ and$\varepsilon _{0}$%
. The problem of the angular perturbation is out of scope for the present
paper. That's why we have only worked on the linear perturbation. However,
angular perturbation is in our agenda for the extension of this study. This
is going to be studied in the near future.
\vspace{-0.75cm}
\section{Hayward Thin-Shell WH in 3+1 D}
\vspace{-0.75cm}
The metric of the Hayward BH is given by \cite{ao1} 
\begin{equation}
ds^{2}=-\left( 1-\frac{2mr^{2}}{r^{3}+2ml^{2}}\right) dt^{2}+\left( 1-\frac{%
2mr^{2}}{r^{3}+2ml^{2}}\right) ^{-1}dr^{2}+r^{2}d\Omega ^{2}.
\label{haywardline}
\end{equation}%
with the metric function 
\begin{equation}
f\left( r\right) =\left( 1-\frac{2mr^{2}}{r^{3}+2ml^{2}}\right)
\end{equation}%
and 
\begin{equation}
d\Omega ^{2}=d\theta ^{2}+\sin ^{2}\theta d\phi ^{2}.
\end{equation}%
It is noted that $m$ and $l$ are free parameters. At large $r,$ the metric
function behaves%
\begin{equation}
\lim_{r\rightarrow \infty }f\left( r\right) \rightarrow 1-\frac{2m}{r}+%
\mathcal{O}\left( \frac{1}{r^{4}}\right) ,
\end{equation}%
whereas at small $r$ 
\begin{equation}
\lim_{r\rightarrow 0}f\left( r\right) \rightarrow 1-\frac{r^{2}}{l^{2}}+%
\mathcal{O}\left( r^{5}\right) .
\end{equation}%
One observes that for small $r$ \ the Hayward BH becomes a de Sitter BH and
for large $r$ it is a Schwarzschild spacetime. The event horizon of the
Hayward BH is calculated by using%
\begin{equation}
r^{3}-2mr^{2}+2ml^{2}=0,
\end{equation}%
and changing $r=m\rho $ and $l=m\lambda $ , it turns to%
\begin{equation}
\rho ^{3}-2\rho ^{2}+2\lambda ^{2}=0.
\end{equation}%
Note that for $\lambda ^{2}>\frac{16}{27}$ there is no horizon, for $\lambda
^{2}=\frac{16}{27}$ single horizon which is called a extremal BH and for $%
\lambda ^{2}<\frac{16}{27}$ double horizons. Hence the ratio $\frac{l}{m}$
is important parameter where the critical ratio is at $\left( \frac{l}{m}%
\right) _{crit.}=\frac{4}{3\sqrt{3}}$. Set $m=1$ where $f\left( r\right) =1-%
\frac{2r^{2}}{r^{3}+2l^{2}}.$
For the case of $l^{2}<\frac{16}{27}$ the event horizon is given by%
\begin{equation}
r_{h}=\frac{1}{3}\left( \sqrt[3]{\Delta }+\frac{4}{\sqrt[3]{\Delta }}%
+2\right)
\end{equation}%
with $\Delta =8-27l^{2}+3\sqrt{27l^{2}\left( 3l^{2}-2\right) }.$
The extremal BH case occurs at $l^{2}=\frac{16}{27}$ and the single horizon
occurs at $r_{h}=\frac{4}{3}.$ When $l^{2}\leq \frac{16}{27}$, the
temperature of Hawking is given by%
\begin{equation}
T_{H}=\frac{f^{\prime }\left( r_{h}\right) }{4\pi }=\frac{1}{4\pi }\left( 
\frac{3}{2}-\frac{2}{r_{h}}\right)
\end{equation}%
which clearly for $l^{2}=\frac{16}{27}$ vanishes and for $l^{2}<\frac{16}{27}
$ is positive so note that $r_{h}\geq \frac{4}{3}$. Entropy for the BH is
obtained by $S=\frac{\mathcal{A}}{4}$ with $\mathcal{A}=4\pi r_{h}^{2}$ to
find the heat capacity of the BH 
\begin{equation}
C_{l}=\left( T_{H}\frac{\partial S}{\partial T_{H}}\right) _{l}
\end{equation}%
and it is obtained as%
\begin{equation}
C_{l}=4\pi r_{h}^{3}\left( \frac{3}{2}-\frac{2}{r_{h}}\right) .
\end{equation}%
It is clearly positive.When the heat capacity of the BH is positive $C_{l}>0$%
, it shows the BH is stable according to thermodynamical laws.

To find the source of the Hayward BH, the action is considered as 
\begin{equation}
\mathcal{I}=\frac{1}{16\pi }\int d^{4}x\sqrt{-g}\left( R-\mathcal{L}\left(
F\right) \right)
\end{equation}%
where $R$ is the Ricci scalar and the nonlinear magnetic field Lagrangian
density is 
\begin{equation}
\mathcal{L}\left( F\right) =-\frac{24m^{2}l^{2}}{\left[ \left( \frac{2P^{2}}{%
F}\right) ^{3/4}+2ml^{2}\right] ^{2}}=-\frac{6}{l^{2}\left[ 1+\left( \frac{%
\beta }{F}\right) ^{3/4}\right] ^{2}}  \label{LHay}
\end{equation}%
with the Maxwell invariant$\ F=F_{\mu \nu }F^{\mu \nu }$with two constant
positive parameters $l$ and $\beta $. The analyses of the stability depends
on the fixing the $\beta $. Moreover, the magnetic field is%
\begin{equation}
\mathbf{F}=P\sin ^{2}\theta d\theta \wedge d\phi
\end{equation}%
where the charge of the magnetic monopole is $P$ . It implies 
\begin{equation}
F=\frac{2P^{2}}{r^{4}}.
\end{equation}%
with the line element given in Eq.(\ref{haywardline}). The
Einstein-Nonlinear Electrodynamics field equations are ($8\pi G=c=1$) 
\begin{equation}
G_{\mu }^{\nu }=T_{\mu }^{\nu }
\end{equation}%
in which 
\begin{equation}
T_{\mu }^{\nu }=-\frac{1}{2}\left( \mathcal{L}\delta _{\mu }^{\nu }-4F_{\mu
\lambda }F^{\lambda \nu }\mathcal{L}_{F}\right)  \label{1222hayward}
\end{equation}%
with $\mathcal{L}_{F}=\frac{\partial \mathcal{L}}{\partial F}.$
After using the nonlinear magnetic field Lagrangian $\mathcal{L}\left(
F\right) $ inside the Einstein equations, one finds $\beta =\frac{2P^{2}}{%
\left( 2ml^{2}\right) ^{4/3}}$ for the Hayward regular BH$.$ The limit of
the weak field of the $\mathcal{L}\left( F\right) $ is found by expanding it
around $F=0$,%
\begin{equation}
\mathcal{L}\left( F\right) =-\frac{6F^{3/2}}{l^{2}\beta ^{3/2}}+\frac{%
12F^{9/4}}{l^{2}\beta ^{9/4}}+\mathcal{O}\left( F^{3}\right) .
\end{equation}
Note that at the limit of the weak field, the lagrangian of the NED does not
reduce to the lagrangian of the linear Maxwell
\begin{equation}
\lim_{F\rightarrow 0}\mathcal{L}\left( F\right) \neq -F.
\end{equation}%
Hence, this spacetime has not any Reissner--Nordstr\"{o}m limit at weak
field.
\vspace{-0.75cm}
\section{Stability of Hayward Thin-Shell WH}
\vspace{-0.75cm}
We use the cut and past technique to constructe a thin-shell WH from the
Hayward BHs. We firstly take a thin-shell at $r=a$ where the throat is
outside of the horizon ($a>r_{h}$). Then we paste two copies of it at the
point of $r=a$. For this reason the thin-shell metric is taken as%
\begin{equation}
ds^{2}=-d\tau ^{2}+a\left( \tau \right) ^{2}\left( d\theta ^{2}+\sin
^{2}\theta d\phi ^{2}\right)  \label{elem}
\end{equation}%
where $\tau $ is the proper time on the shell. The Einstein equations on the
shell are 
\begin{equation}
\left[ K_{i}^{j}\right] -\left[ K\right] \delta _{i}^{j}=-S_{i}^{j}
\end{equation}%
where $\left[ X\right] =X_{2}-X_{1},$. It is noted that the extrinsic
curvature tensor is $K_{i}^{j}$. Moreover, $K$ stands for its trace. \ The
surface stresses, i.e., surface energy density $\sigma $ and surface
pressures $S_{\theta }^{\theta }=p=S_{\phi }^{\phi }$ , are determined by
the surface stress-energy tensor $S_{i}^{j}$.
 The energy and pressure densities are obtained as 
\begin{equation}
\sigma =-\frac{4}{a}\sqrt{f\left( a\right) +\dot{a}^{2}}
\end{equation}
\begin{equation}
p=2\left( \frac{\sqrt{f\left( a\right) +\dot{a}^{2}}}{a}+\frac{\ddot{a}%
+f^{\prime }\left( a\right) /2}{\sqrt{f\left( a\right) +\dot{a}^{2}}}\right)
.\end{equation}%
Then they reduce to simple form in a static configuration ($a=a_{0}$)%
\begin{equation}
\sigma _{0}=-\frac{4}{a_{0}}\sqrt{f\left( a_{0}\right) }
\end{equation}%
and 
\begin{equation}
p_{0}=2\left( \frac{\sqrt{f\left( a_{0}\right) }}{a_{0}}+\frac{f^{\prime
}\left( a_{0}\right) /2}{\sqrt{f\left( a_{0}\right) }}\right) .
\end{equation}%
Stability of such a WH is investigated by applying a linear perturbation
with the following EoS 
\begin{equation}
p=\psi \left( \sigma \right)
\end{equation}%
Moreover the energy conservation is%
\begin{equation}
S_{\;;j}^{ij}=0
\end{equation}%
which in closed form it equals to%
\begin{equation}
S_{\;,j}^{ij}+S^{kj}\Gamma _{kj}^{i\mu }+S^{ik}\Gamma _{kj}^{j}=0
\end{equation}%
after the line element in Eq.(\ref{elem}) is used, it opens to 
\begin{equation}
\frac{\partial }{\partial \tau }\left( \sigma a^{2}\right) +p\frac{\partial 
}{\partial \tau }\left( a^{2}\right) =0.
\end{equation}%
The 1-D equation of motion is 
\begin{equation}
\dot{a}^{2}+V\left( a\right) =0,
\end{equation}%
in which $V\left( a\right) $ is the potential, 
\begin{equation}
V\left( a\right) =f-\left( \frac{a\sigma }{4}\right) ^{4}.
\end{equation}%
The equilibrium point at $a=a_{0}$ means $V^{\prime }\left( a_{0}\right) =0$
and $V^{\prime \prime }\left( a_{0}\right) \geq 0.$ Then it is considered
that $f_{1}\left( a_{0}\right) =f_{2}\left( a_{0}\right) $, one finds $%
V_{0}=V_{0}^{\prime }=0.$ To obtain $V^{\prime \prime }\left( a_{0}\right)
\geq 0$ we use the given $p=\psi \left( \sigma \right) $ and it is found as
follows%
\begin{equation}
\sigma ^{\prime }\left( =\frac{d\sigma }{da}\right) =-\frac{2}{a}\left(
\sigma +\psi \right)
\end{equation}%
and 
\begin{equation}
\sigma ^{\prime \prime }=\frac{2}{a^{2}}\left( \sigma +\psi \right) \left(
3+2\psi ^{\prime }\right) ,
\end{equation}%
where $\psi ^{\prime }=\frac{d\psi }{d\sigma }.$ After we use $\psi
_{0}=p_{0},$ finally it is found that%
\begin{equation}
V^{\prime \prime }\left( a_{0}\right) =f_{0}^{\prime \prime }-\frac{1}{8}%
\left[ \left( \sigma _{0}+2p_{0}\right) ^{2}+2\sigma _{0}\left( \sigma
_{0}+p_{0}\right) \left( 1+2\psi ^{\prime }\left( \sigma _{0}\right) \right) %
\right]
\end{equation}
\vspace{-0.75cm}
\section{Some Models of EoS}
\vspace{-0.75cm}
In this section, we consider some specific models of matter such as Linear
gas (LG), Chaplygin gas (CG), generalized Chaplygin gas (GCG) , modified generalized Chaplygin gas (MGCG)
and logarithmic gas (LogG) to analyze the effect of the parameter of Hayward
in the stability of the constructed thin-shell WH.
\vspace{-0.75cm}
\subsection{Linear Gas}
\vspace{-0.75cm}
\begin{figure}[h!]

\includegraphics[width=150mm,scale=1]{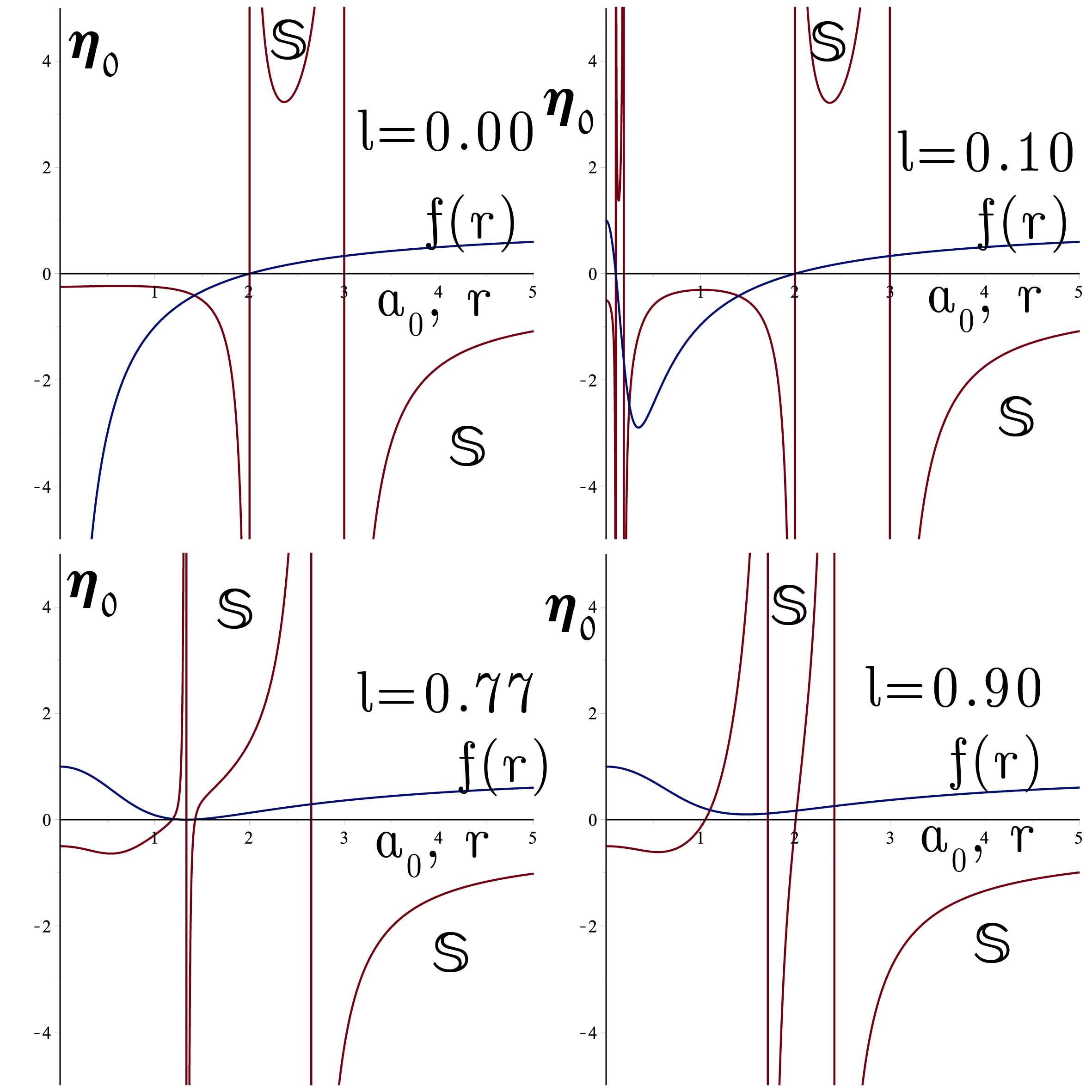}
\caption{Stability of Thin-Shell WH supported by LG.} \label{fig.LG}
\end{figure}
For a LG, EoS is choosen as%
\begin{equation}
\psi =\eta _{0}\left( \sigma -\sigma _{0}\right) +p_{0}
\end{equation}%
in which $\eta _{0}$ is a constant and $\psi ^{\prime }\left( \sigma
_{0}\right) =\eta _{0}.$ Fig.\ref{fig.LG} shows the stability regions in terms of $%
\eta _{0}$ and $a_{0}$ \ with different Hayward's parameter. It is noted that the $S$
shows the stable regions.
\vspace{-0.75cm}
\subsection{Chaplygin Gas}
\vspace{-0.75cm}
\begin{figure}[h!]

\includegraphics[width=150mm,scale=1]{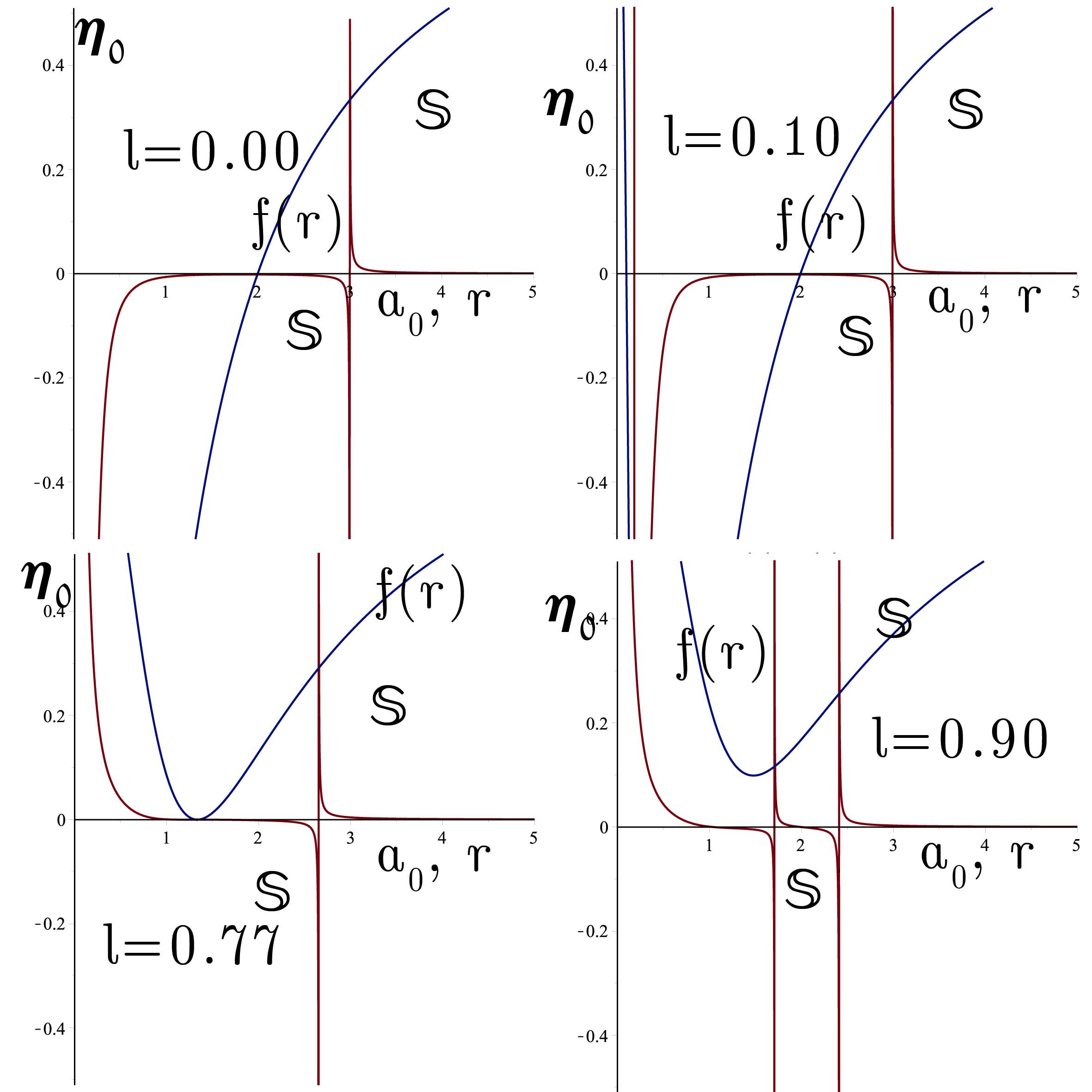}
\caption{Stability of Thin-Shell WH supported by CG.}
 \label{fig.CG}
\end{figure}
For CG, we choose the EoS as follows%
\begin{equation}
\psi =\eta _{0}\left( \frac{1}{\sigma }-\frac{1}{\sigma _{0}}\right) +p_{0}
\end{equation}%
where $\eta _{0}$ is a constant and $\psi ^{\prime }\left( \sigma
_{0}\right) =-\frac{\eta _{0}}{\sigma _{0}^{2}}.$ In Fig.\ref{fig.CG}, the stability
regions are shown in terms of $\eta _{0}$ and $a_{0}$ for different values
of $\ell .$ The effect of Hayward's constant is to increase the stability of the Thin-Shell WH.
\vspace{-0.75cm}
\subsection{Generalized Chaplygin Gas}
\vspace{-0.75cm}
\begin{figure}[h!]

\includegraphics[width=150mm,scale=1]{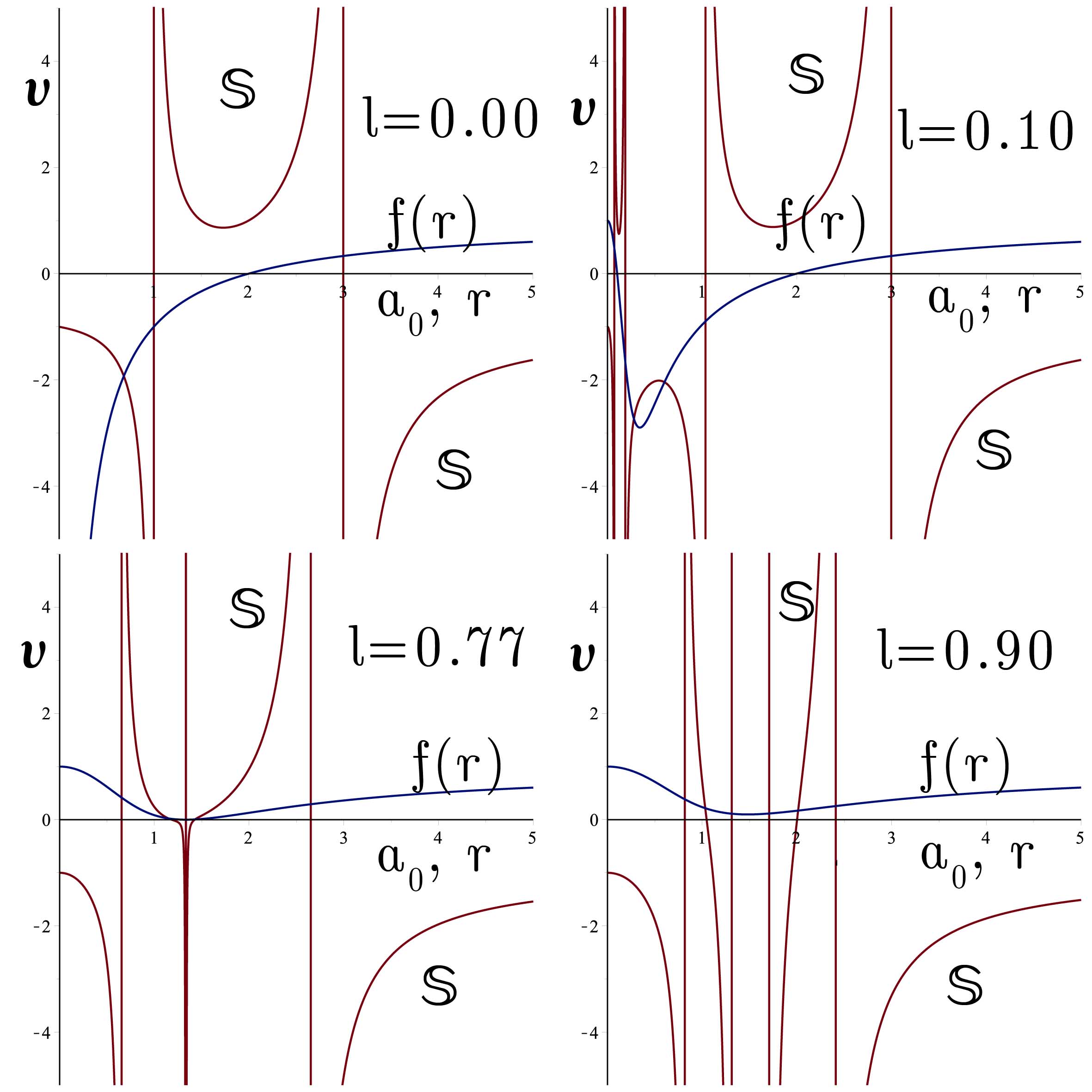}
\caption{Stability of Thin-Shell WH supported by GCG.}
\label{fig.GCG}
\end{figure}
The EoS of the GCG is taken as%
\begin{equation}
\psi \left( \sigma \right) =\eta _{0}\left( \frac{1}{\sigma ^{\nu }}-\frac{1%
}{\sigma _{0}^{\nu }}\right) +p_{0}
\end{equation}%
where $\nu $ and $\eta _{0}$ are constants. We check the effect of parameter 
$\nu $ in the stability and $\psi $ becomes 
\begin{equation}
\psi \left( \sigma \right) =p_{0}\left( \frac{\sigma _{0}}{\sigma }\right)
^{\nu }.
\end{equation}%
We find $\psi ^{\prime }\left( \sigma _{0}\right) =-\frac{p_{0}}{\sigma _{0}}%
\nu $. In Fig.\ref{fig.GCG}, the stability regions are shown in terms of $\nu $ and $%
a_{0}$ with various values of $\ell .$
\vspace{-0.75cm}
\subsection{Modified Generalized Chaplygin Gas}
\vspace{-0.75cm}
\begin{figure}[h!]

\includegraphics[width=150mm,scale=1]{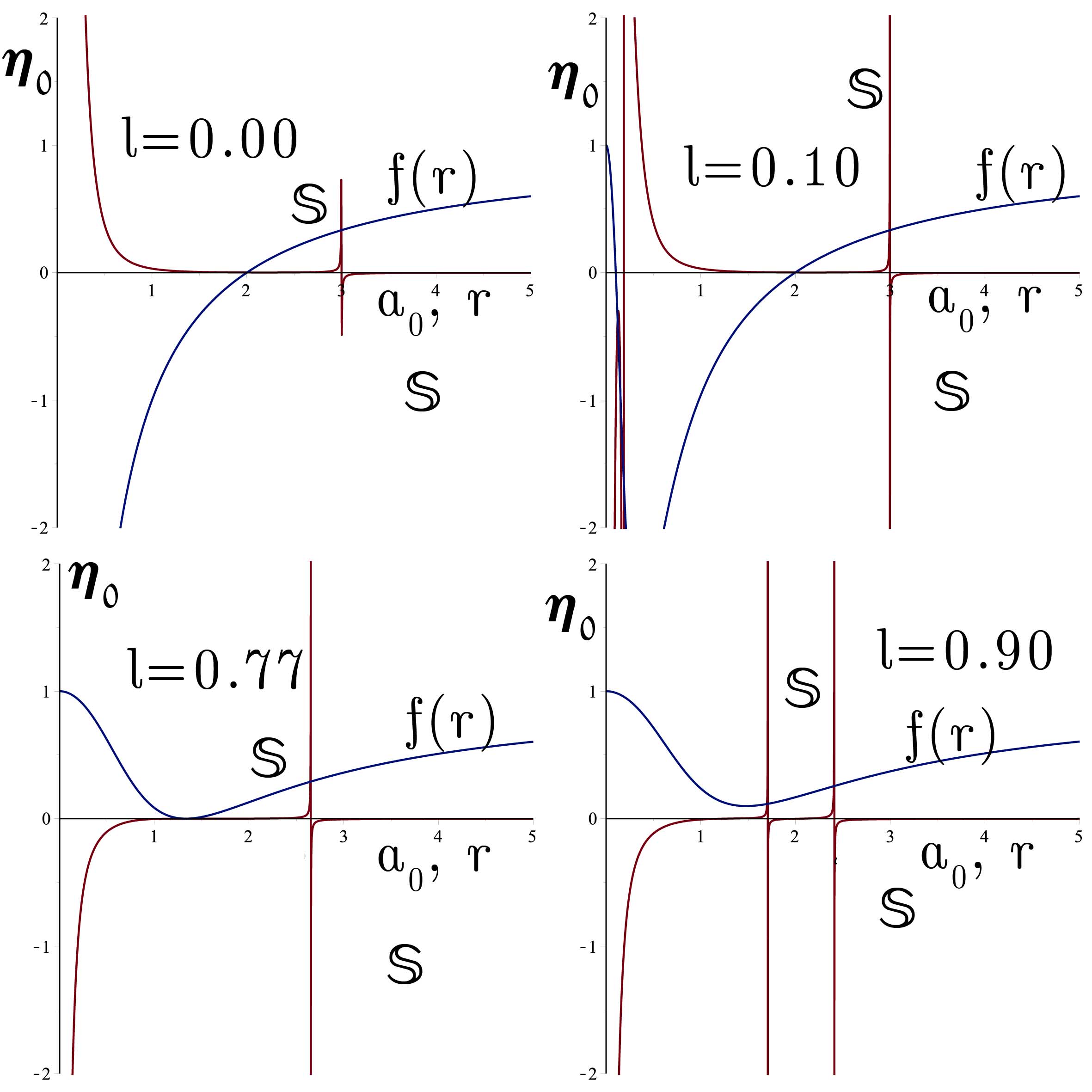}
\caption{Stability of Thin-Shell WH supported by MGCG.}
\label{fig.MGCG}
\end{figure}
In this case, the MGCG \ is 
\begin{equation}
\psi \left( \sigma \right) =\xi _{0}\left( \sigma -\sigma _{0}\right) -\eta
_{0}\left( \frac{1}{\sigma ^{\nu }}-\frac{1}{\sigma _{0}^{\nu }}\right)
+p_{0}
\end{equation}%
in which $\xi _{0}$, $\eta _{0}$ and $\nu $ are free parameters. Therefore,%
\begin{equation}
\psi ^{\prime }\left( \sigma _{0}\right) =\xi _{0}+\eta _{0}\frac{\eta
_{0}\nu }{\sigma _{0}^{\nu +1}}.
\end{equation}%
To go further we set $\xi _{0}=1$ and $\nu =1$. In Fig.\ref{fig.MGCG}, the stability
regions are plotted in terms of $\eta _{0}$ and $a_{0}$ with various values
of $\ell $. The effect of Hayward's constant is to increase the stability of the Thin-Shell WH.
\vspace{-0.75cm}
\subsection{Logarithmic Gas}
\vspace{-0.75cm}
\begin{figure}[h!]

\includegraphics[width=150mm,scale=1]{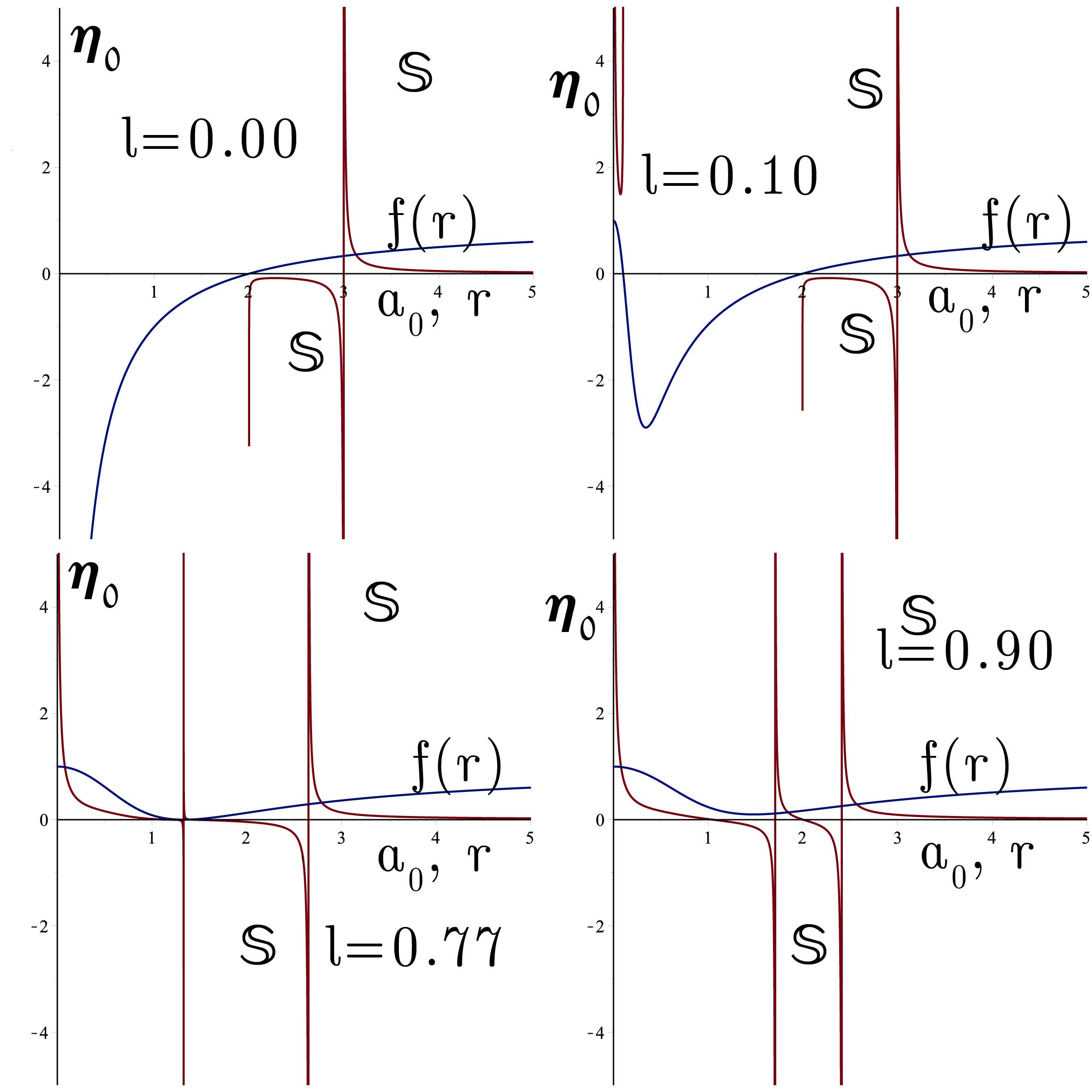}
\caption{Stability of Thin-Shell WH supported by LogG.}
 \label{fig.LogG}
\end{figure}
Lastly LogG is\ choosen by follows%
\begin{equation}
\psi \left( \sigma \right) =\eta _{0}\ln \left\vert \frac{\sigma }{\sigma
_{0}}\right\vert +p_{0}
\end{equation}%
in which $\eta _{0}$ is a constant. For LogG, we find that 
\begin{equation}
\psi ^{\prime }\left( \sigma _{0}\right) =\frac{\eta _{0}}{\sigma _{0}}.
\end{equation}%
In Fig.\ref{fig.LogG}, the stability regions are plotted to show the effect of
Hayward's parameter clearly. The effect of Hayward's constant is to increase the stability of the Thin-Shell WH.
\vspace{-0.75cm}
\section{Perturbation of Small Velocity}
\vspace{-0.75cm}
Now we check the stability of the Hayward WH using the small velocity
perturbations( $a=a_{0}$). The EoS is calculated by $f\left( a\right) $ and $%
a$%
\begin{equation}
p=-\frac{1}{2}\left( 1+\frac{af^{\prime }\left( a\right) }{2f\left( a\right) 
}\right) \sigma .
\end{equation}%
One finds the 1-D motion 
\begin{equation}
\ddot{a}-\frac{f^{\prime }}{2f}\dot{a}^{2}=0.
\end{equation}%
After integrating both sides,%
\begin{equation}
\dot{a}=\dot{a}_{0}\frac{\sqrt{f}}{\sqrt{f_{0}}}
\end{equation}%
then it is obtained by taking the second integral%
\begin{equation}
\int_{a0}^{a}\frac{da}{\sqrt{f\left( a\right) }}=\frac{\dot{a}_{0}}{\sqrt{%
f_{0}}}\left( \tau -\tau _{0}\right) .  \label{lastint}
\end{equation}%
It is noted that $\dot{a}_{0}=0$ is the equilibrium point, however we
consider that there is an initial small velocity after perturbation, which
is called $\dot{a}_{0}.$ We firstly consider the Schwarzschild BH as an
example. For the Schwarzschild spacetime, $f\left( a\right) =1-\frac{2m}{a}$
yields%
\begin{equation}
\frac{\dot{a}_{0}}{\sqrt{f_{0}}}\left( \tau -\tau _{0}\right) =a\sqrt{f}%
-a_{0}\sqrt{f_{0}}+m\ln \left( \frac{a-m+a\sqrt{f}}{a_{0}-m+a_{0}\sqrt{f_{0}}%
}\right) .
\end{equation}%
It is non-oscillatory so it means that the throat isn`t stable. Now we give
the another example of the Hayward thin-shell WH. For this case, using the
Hayward metric and the relation given in Eq.(\ref{lastint}),expanding to the
second order of $\ell$, admits 
\begin{multline}
\frac{\dot{a}_{0}}{\sqrt{f_{0}}}\left( \tau -\tau _{0}\right) \tilde{=}a%
\sqrt{f}-a_{0}\sqrt{f_{0}}+m\ln \left( \frac{a-m+a\sqrt{f}}{a_{0}-m+a_{0}%
\sqrt{f_{0}}}\right) + \\
2\ell ^{2}\left( \frac{2a^{2}-2am-m^{2}}{3ma^{2}\sqrt{f}}-\frac{%
2a_{0}^{2}-2a_{0}m-m^{2}}{3ma_{0}^{2}\sqrt{f_{0}}}\right) .
\end{multline}%
Here again we find that there is not stability of the throat against the
small velocity perturbation. Therefore this motion is not also oscillatory.
The throat's acceleration is $\ddot{a}=\frac{f^{\prime }}{2f}\dot{a}^{2}$.
Note that for the Schwarzschild and Hayward spacetimes we show that it is
positive. Therefore, the corresponding thin-shell WH is not stable.

In this section we constructe thin-shell WHs from the Hayward BH. Firstly it
is showed that a magnetic monopole field in the\ NED is the source of the
Hayward BH. On the thin-shell \ we use the different type of EoS with the
form $p=\psi \left( \sigma \right) $ and plot possible stable regions. We
show the stable and unstable regions on the plots. Stability simply depends
on the condition of $V^{\prime \prime }\left( a_{0}\right) >0$. We show that
the parameter $\ell $, which is known as Hayward parameter has a important
role. Moreover, for higher $\ell $ value \ the stable regions are increased.
It is checked the small velocity perturbations for the throat. It is found
that throat of the thin-shell WH is not stable against such kind of
perturbations. Hence, energy density of the WH is found negative so that we
need exotic matter.
\vspace{-0.75cm}
\section{Rotating BTZ Thin-Shell Wormholes}
\vspace{-0.75cm}
For constructing the thin-shell WH, we study 2 copies of a 2+1-dimensional
background spacetime given. These copies are used at a radius $\ r=a$, where
the throat of the WH will be located larger than the event horizon. All
steps for thin-shell by using the Isreal junction conditions are also valid
for the constructing of WH, only difference appears when \ calculating the
energt density and pressure of WH. Then we have the jump if the extrinsic
curvature components at the surface $r=a$, furthermore, we take the $%
U_{\left( o\right) },U_{\left( i\right) }$ and also $\omega _{\left(
i\right) },\omega _{\left( o\right) }$ for the WH case, and calculate
associated linear energy density and pressure for WH.
\begin{figure}[h!]

\centering
\includegraphics[width=0.80\textwidth]{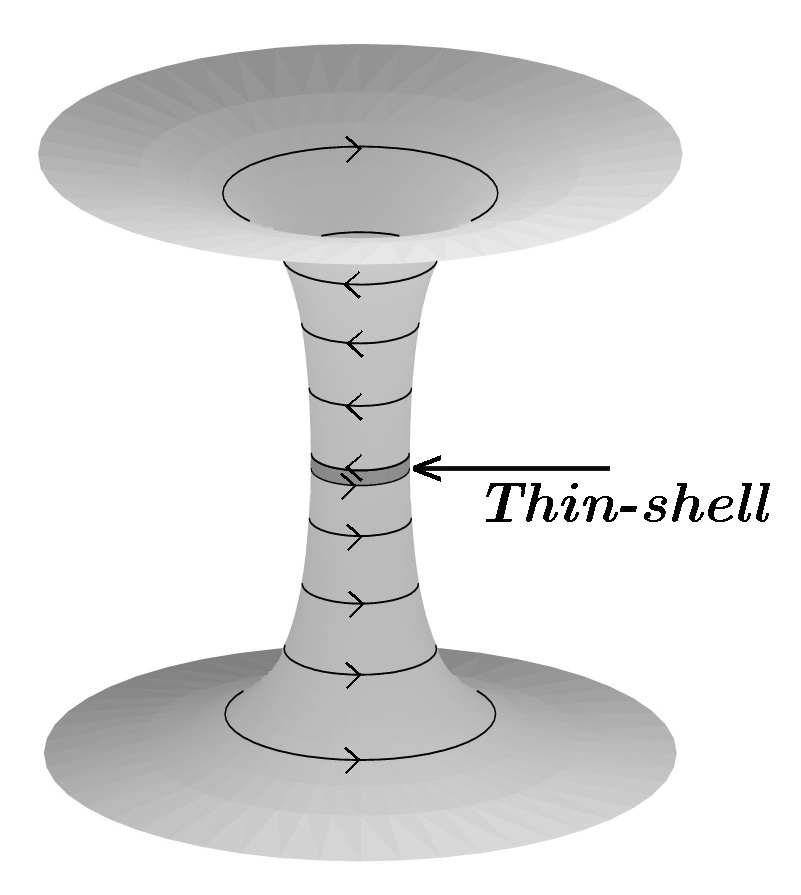} \label{fig:thin-shell WH} 
\caption{Rotating Thin-shell WH}
\end{figure}
Once more we add that all functions are evaluated at $r=a.$Therefore it gives%
\begin{equation}
\sigma =-S_{\tau }^{\tau }=-\frac{1}{8\pi G}\left( K_{\phi \left( o\right)
}^{\phi }+K_{\phi \left( i\right) }^{\phi }\right) =-\frac{1}{8\pi Ga}\left( 
\sqrt{U_{\left( o\right) }+\dot{a}^{2}}+\sqrt{U_{\left( i\right) }+\dot{a}%
^{2}}\right) ,  \label{s111}
\end{equation}%
\begin{equation}
p=S_{\phi }^{\phi }=\frac{1}{8\pi G}\left( K_{\tau \left( o\right) }^{\tau
}+K_{\tau \left( i\right) }^{\tau }\right) =\frac{1}{8\pi G}\left( \frac{%
U_{\left( o\right) }\left( \ddot{a}+U_{\left( o\right) }^{\prime }/2\right) 
}{\sqrt{U_{\left( o\right) }+\dot{a}^{2}}}+\frac{U_{\left( i\right) }\left( 
\ddot{a}+U_{\left( i\right) }^{\prime }/2\right) }{\sqrt{U_{\left( i\right)
}+\dot{a}^{2}}}\right) 
\end{equation}%
and%
\begin{equation}
q=S_{\tau }^{\phi }=-\frac{1}{8\pi G}\left( K_{\tau \left( o\right) }^{\phi
}+K_{\tau \left( i\right) }^{\phi }\right) =\frac{a^{2}}{8\pi G}\left( \frac{%
\omega _{\left( o\right) }^{\prime }}{U_{\left( o\right) }}\left( U_{\left(
o\right) }+2\dot{a}^{2}\right) +\frac{\omega _{\left( i\right) }^{\prime }}{%
U_{\left( i\right) }}\left( U_{\left( i\right) }+2\dot{a}^{2}\right) \right)
.
\end{equation}%
To finalize this section we add that at the equilibrium point where $a\left(
\tau \right) =a_{0}$ one finds%
\begin{equation}
\sigma _{0}=-\frac{1}{8\pi Ga_{0}}\left( \sqrt{U_{\left( o\right) }}+\sqrt{%
U_{\left( i\right) }}\right) ,  \label{sigma00}
\end{equation}%
\begin{equation}
p_{0}=\frac{1}{16\pi G}\left( \sqrt{U_{\left( o\right) }}U_{\left( o\right)
}^{\prime }+\sqrt{U_{\left( i\right) }}U_{\left( i\right) }^{\prime }\right) 
\label{p00}
\end{equation}%
and%
\begin{equation}
q_{0}=\frac{a^{2}}{8\pi G}\left( \omega _{\left( o\right) }^{\prime }+\omega
_{\left( i\right) }^{\prime }\right) ,  \label{q00}
\end{equation}%
in which all functions are evaluated at $a=a_{0}.$
\vspace{-0.75cm}
\section{Linearized Stability of Wormhole}
\vspace{-0.75cm}
Furthermore, the equation of motion for the thin-shell is 
\begin{equation}
\dot{a}^{2}+V_{eff}=0
\end{equation}%
with the effective potential ($8\pi G=1$)%
\begin{equation}
V_{eff}=\frac{1}{2}\left( U_{\left( o\right) }+U_{\left( i\right) }\right) +%
\frac{\left( U_{\left( o\right) }+U_{\left( i\right) }\right) ^{2}}{\left(
2a\sigma \right) ^{2}}-\left( \frac{a\sigma }{2}\right) ^{2}.  \label{veffwh}
\end{equation}%
It is noted that stability of WH solution depends upon the conditions of $%
V_{eff}^{\prime \prime }\left( a_{0}\right) >0$ and $V_{eff}^{\prime }\left(
a_{0}\right) =V_{eff}\left( a_{0}\right) =0$ 
\begin{equation}
V_{eff}\left( a\right) \sim \frac{1}{2}V_{eff}^{\prime \prime }\left(
a_{0}\right) \left( a-a_{0}\right) ^{2}.
\end{equation}%
Let's introduce $x=a-a_{0}$ and write the equation of motion again%
\begin{equation}
\dot{x}^{2}+\frac{1}{2}V_{eff}^{\prime \prime }\left( a_{0}\right) x^{2}=0
\end{equation}%
which after a derivative with respect to time it reduces to%
\begin{equation}
\ddot{x}+\frac{1}{2}V_{eff}^{\prime \prime }\left( a_{0}\right) x=0.
\end{equation}%
Hence our main aim is to discover the behaviour of $V_{eff}^{\prime \prime
}\left( a_{0}\right) $ and point out the conditions of its positive value
for the stability by also the help of the Eq.(\ref{energy}).

It is now interesting to reduce our general results to some specific
examples. such as rotating AdS-BTZ thinshell WH and rotating BTZ thinshell
WH.
\vspace{-0.75cm}
\section{Rotating BTZ Thin-shell Wormhole}
\vspace{-0.75cm}
Our last WH example is on very recent work where they introduce a counter
rotating thin shell WHs as follows \cite{counter}. In $r<a$ and $r>a$ the
spacetimes are given by rotating BTZ whose line elements are written as with
\begin{equation}
U=U_{+}=U_{-}=-M+\frac{r^{2}}{\ell ^{2}}+\frac{J^{2}}{4r^{2}}
\end{equation}%
\ \ which are the rotating BTZ. Here $M$ and $J$ are the integration
constants which respectively correspond to the mass and angular momentum,
and $Q$ is the charge carried by the BH. The exterior curvatures are derived
as follows
\begin{equation}
K_{\tau }^{\tau \left( \pm \right) }=\pm \frac{2\ddot{R}+f^{\prime }}{2\sqrt{%
\dot{R}^{2}+f}},
\end{equation}%
\begin{equation}
K_{\tau }^{\psi \left( \pm \right) }=\mp \frac{N_{\pm }^{\varphi }\left(
R\right) }{R}
\end{equation}%
and%
\begin{equation}
K_{\psi }^{\psi \left( \pm \right) }=\pm \frac{\sqrt{\dot{R}^{2}+f}}{R}.
\end{equation}%
Then it is found that%
\begin{equation}
k_{i}^{j}=\left( 
\begin{array}{cc}
\frac{2\ddot{R}+f^{\prime }}{\sqrt{\dot{R}^{2}+f}} & -\frac{\left[
N_{+}^{\varphi }\left( R\right) +N_{-}^{\varphi }\left( R\right) \right] }{R}
\\ 
-\frac{\left[ N_{+}^{\varphi }\left( R\right) +N_{-}^{\varphi }\left(
R\right) \right] }{R} & \frac{2\sqrt{\dot{R}^{2}+f}}{R}%
\end{array}%
\right) .
\end{equation}%
Note that $S_{i}^{j}$ is diagonal so that $N_{+}^{\varphi }\left( R\right)
+N_{-}^{\varphi }\left( R\right) =0$ which consequently admits $%
J_{+}+J_{-}=0.$ It means that between the upper and lower shells there is a
counterrotating.

\ After some calculations, the energy density and the pressures is
calculated as%
\begin{equation}
\sigma =-\frac{1}{8\pi G}-\frac{2\sqrt{\dot{a}^{2}+U}}{a}
\label{energy density}
\end{equation}%
and 
\begin{equation}
p=\frac{1}{8\pi G}\frac{2\ddot{a}+U^{\prime }}{\sqrt{\dot{a}^{2}+U}}.
\end{equation}
The static case where $\dot{a}=0$ and $\ddot{a}=0$) 
\begin{equation}
\sigma _{0}=-\frac{2\sqrt{U_{0}}}{8\pi Ga_{0}}
\end{equation}%
and 
\begin{equation}
P_{0}=\frac{U_{0}^{\prime }}{8\pi G\sqrt{U_{0}}}.
\end{equation}
We \ have already imposed a generic potential equation in Eq.(\ref{veffwh}%
). Now, Eq. (\ref{veffwh}) can be written for the rotating BTZ thinshell WH
as \ 
\begin{equation}
\dot{a}^{2}=-V_{eff}
\end{equation}
with $V_{eff}$ given by \ 
\begin{equation}
V_{eff}=U-\frac{\left( 8\pi G\sigma \right) ^{2}R^{2}}{4}
\end{equation}
Using the more general form of the pressure as a function of $a$ and $\sigma 
$ i.e., 
\begin{equation}
p=\psi \left( a,\sigma \right) ,
\end{equation}
Next, at equilibrium point $a=a_{0}$ it is obtained that 
\begin{equation}
V_{eff}^{\prime \prime }\left( a_{0}\right) =U_{0}^{\prime \prime }-\frac{%
U_{0}^{\prime 2}a_{0}^{2}+2\psi _{0}^{\prime }U_{0}\left(
2U_{0}-U_{0}^{\prime }a_{0}\right) }{2U_{0}a_{0}^{2}}
\end{equation}
One can show that $V_{eff}\left( a_{0}\right) =V_{eff}^{\prime }\left(
a_{0}\right) =0$.
\vspace{-0.75cm}
\subsection{Phantomlike\ EoS}
\vspace{-0.75cm}
The EoS of a LG is given by $\frac{d\xi }{d\sigma }=\beta $ in which $\beta $
is a constant parameter. One observes that for the cases 
\begin{equation}
\frac{\left( 2f_{0}f_{0}^{\prime \prime }-f_{0}^{\prime 2}\right) R_{0}^{2}}{%
2f_{0}\left( 2f_{0}-f_{0}^{\prime }R_{0}\right) }<\beta \text{ \ \ \ \ \ for
\ \ \ }\frac{2f_{0}}{f_{0}^{\prime }}>R_{0}
\end{equation}%
and%
\begin{equation}
\frac{2f_{0}R_{0}^{2}f_{0}^{\prime \prime }-f_{0}^{\prime 2}R_{0}^{2}}{%
2f_{0}\left( 2f_{0}-f_{0}^{\prime }R_{0}\right) }>\beta \text{ \ \ \ \ \ for
\ \ \ \ }\frac{2f_{0}}{f_{0}^{\prime }}<R_{0}\text{ \ \ }
\end{equation}%
$V_{eff}^{\prime \prime }\left( R_{0}\right) >0$ and the equilibrium is
stable. In Fig. 3.12 we plot the stability region with respect to $\beta $ and $%
R_{0}.$ In the same figure the result for different $J$ are compared. As one
can see, increasing the value of $J$ increases the region of stability
\begin{figure}[h!]

\centering
\includegraphics[width=0.80\textwidth]{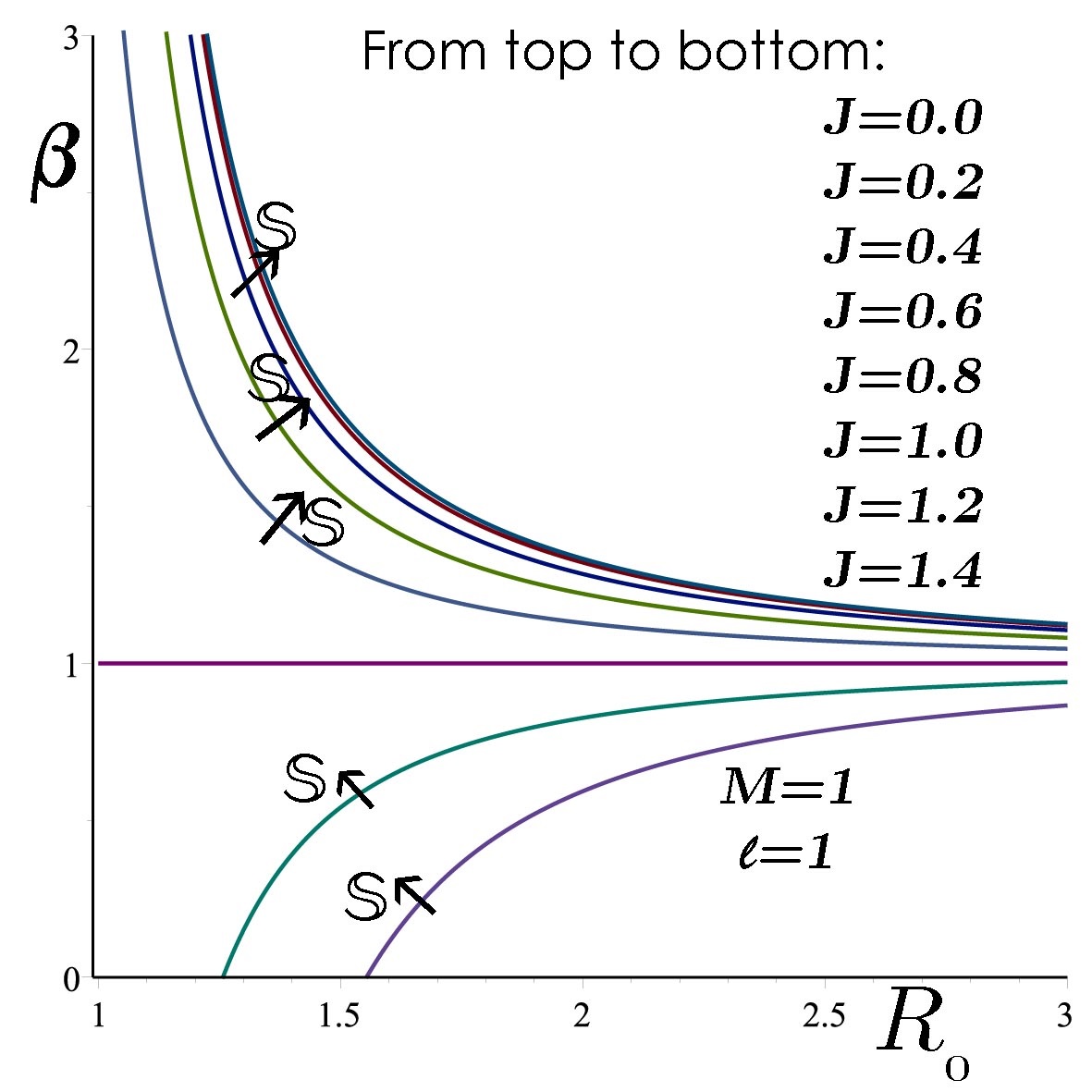} \label%
{fig:stabilityrotating} \caption{Stability of the rotating thin-shell WH}

\end{figure}
\chapter{CONCLUSION}
In this thesis, we have studied the stability of thin-shells around the BHs
and the stability of the thin-shell WHs. Firstly, we have introduced
generally the BHs, WHs and their properties in the theory of the GR.
Furthermore, we have calculated the Hawking Radiation from the traversable
WHs. In Chapter 2, we have studied the rotating thin-shells methods and gave
some examples. Especially, the Bardeen BH with a charged thin-shell is used
to produce thin-shell matching inside and outside. We have created the
thin-shell by using the Israel junction conditions which we derived earlier.
Then as a result of the thin-shell, the effect of the charge is disappeared
to outside. Moreover, we have checked the stability regions of the
thin-shell and we have plotted the areas of the stability. Experimentally,
there is a possibility that thin-shells can cancel the charge's effects for
the outside observers. Our example is in the 2+1 dimensions but it can be
found similar effects for the higher dimensions.

In Chapter 3, we have defined the methods of the constructing thin-shell WHs
and we have studied two example of thin-shell WHs in 2+1 dimensions and 3+1
dimensions by using same method which is gluing two copies of bulks via the
cut and paste procedure. To this end, we have used the fact that the radius
of throat must be greater than the event horizon of the metric given: ($%
a_{0}>r_{h}$). We have adopted LBG, CG, GCG, and LogG gas equation of states
to the exotic matter locating at the throat. Then, the stability analysis
has become the study of checking positivity of the second derivative of an
effective potential at the throat radius $a_{0}$: $V^{\prime \prime
}(a_{0})\geq 0$. In all cases, we have managed to find the stability regions
in terms of the throat radius $a_{0}$ and constant parameter $\varepsilon
_{0}$, which are associated with the EoS employed. Hence, in our studies the
exotic matter is needed to support construction of the throat of the
thin-shell.

In last section, we have studied the rotating thin-shell WHs in
2+1-dimensions. This time we use the counterrotating shells around the
throat which has no effect on the gluing procedure and Lanczos energy
conditions. It can be thought that the upper and lower pressures cancel each
others and leaving no cross terms of the pressure component. Stability of a
counterrotating TSW for a gas of linear equation of state turns out to make
the TSW more stable. For very fast rotation the stability region grows much
larger in the parameter space. For a velocity dependent perturbation,
however, it is shown that TSW is no more stable. That is, perturbation of
the throat radius ($R_{0}$) that depends on initial speed ($\dot{R}_{0}\neq
0 $), no matter how small, doesn't return to the equilibrium radius $R_{0}$
again. Although our work is confined to the simple $2+1-$dimensional
spacetime it is our belief that similar behaviours are exhibited also by the
higher dimensional TSWs. However, even for the $3+1-$dimensional spacetime cases it is not easy task to check the stability because there are more degrees of freedom in $3+1-$ dimensional spacetime.


\begin{thebibliography}{22}

\vspace{-0.75cm}

\bibitem{ligo} Berti, E. (2016). The first sounds of merging black holes. \textit{APS Physics}, 9, 17. doi:10.1103/Physics.9.17

\bibitem{ligo1} LIGO Scientific, \& Virgo Collaborations (Abbott, B.P.,(Caltech) et al.) (2016). 	
GW150914: The advanced LIGO detectors in the era of first discoveries. \textit{Phys. Rev. Lett.}, 116, no.13, 131103. doi: 10.1103/PhysRevLett.116.131103 

\bibitem{3} Misner, C. W., Thorne, K. S., \& Wheeler, J. A. (1973). \textit{Gravitation}. W. H. Freeman and Co., San Francisco.

\bibitem{blag} Straub, O.,Vincent, F. H., Abramowicz, M. A., Gourgoulhon, E., \& Paumard, T. (2012). Modelling the black hole silhouette in Sgr A* with ion tori. \textit{ Astron. Astroph.}, 543, A83. doi: 10.1051/0004-6361/201219209 

\bibitem{erosen}Einstein, A., \& Rosen, N. (1935). The particle problem in the general theory of relativity. \textit{Phys. Rev.}, 48, 73. doi: 10.1103/PhysRev.48.73

\bibitem{sch}Schwarzschild, K. (1916). On the gravitational field of a mass point according to Einstein's theory. \textit{Sitzungsber Preuss Akad. Wiss. Berlin (Math.Phys.)}, 189-196, arXiv:physics/9905030.

\bibitem{flamm} Flamm, L. (1916). Comments on Einstein's theory of gravity. \textit{Physikalische Zeitschrift}, 17, 448.

\bibitem{wheeler} Fuller, R. W., \& Wheeler, J. A. (1962). Causality and multiply connected space-time. \textit{Phys. Rev.}, 128, 919. doi: 10.1103/PhysRev.128.919

\bibitem{morris} Morris, M. S., \& Thorne, K. S. (1988). Wormholes in space-time and their use for interstellar travel: a tool for teaching general relativity. \textit{Am. J. Phys.}, 56, 395. doi: 10.1119/1.15620

\bibitem{visser1} Hochberg, D., \& Visser, M. (1998). The null energy condition in dynamic wormholes. \textit{Phys. Rev. Lett.}, 81, 746. doi: 10.1103/PhysRevLett.81.746 

\bibitem{visser2} Hochberg, D., Molina-Paris, C., \& Visser, M. (1999). Tolman wormholes violate the strong energy condition. \textit{Phys. Rev. D}, 59, 044011. doi: 10.1103/PhysRevD.59.044011 

\bibitem{yurtsever} Morris, M. S., Thorne, K. S., \& Yurtsever, U. (1988). 
Wormholes, time Machines, and the weak energy condition. \textit{Phys. Rev. Lett.}, 61, 1446. doi: 10.1103/PhysRevLett.61.1446

\bibitem{witt} Friedman, J. L., Schleich, K., \& Witt, D. M. (1995). Topological censorship. \textit{Phys. Rev. Lett.}, 71, 1486. doi: 10.1103/PhysRevLett.71.1486 

\bibitem{MV1} Visser, M. (1989). Traversable wormholes: some simple examples. \textit{Phys. Rev. D}, 39, 3182. doi: 10.1103/PhysRevD.39.3182 

\bibitem{MV2} Visser, M. (1989). 
Traversable wormholes from surgically modified Schwarzschild space-times. \textit{ Nucl. Phys. B}, 328, 203. doi: 10.1016/0550-3213(89)90100-4 

\bibitem{MV3} Visser, M. (1995). \textit{Lorentzian Wormholes from Einstein to
Hawking}. American Institute of Physics, New York.

\bibitem{MV4}Poisson, E., \& Visser, M. (1995). Thin shell wormholes: linearization stability. \textit{Phys. Rev. D}, 52, 7318. doi: 10.1103/PhysRevD.52.7318 

\bibitem{MV5} Visser, M. (2003). Traversable wormholes with arbitrarily small energy condition violations. \textit{Phys. Rev. Lett.}, 90, 201102. doi: 10.1103/PhysRevLett.90.201102 

\bibitem{MV6plot}  Garcia, N. M., Lobo, F. S. N., \& Visser, M. (2012). Generic spherically symmetric dynamic thin-shell traversable wormholes in standard general relativity. \textit{Phys. Rev. D}, 86, 044026. doi: 10.1103/PhysRevD.86.044026 

\bibitem{loboreview0} Lobo, F. S. N. (2008). Exotic solutions in General Relativity: Traversable wormholes and 'warp drive' spacetimes. \textit{Classical and Quantum Gravity Research}, 1-78. Nova Sci. Pub.

\bibitem{loboreview1}  Lobo, F. S. N. (2005). Energy conditions, traversable wormholes and dust shells. \textit{Gen. Relativ. Gravit.}, 37, 12, 2023. doi: 10.1007/s10714-005-0177-x

\bibitem{loboreview2}  Lobo, F. S. N. (2016). From the Flamm-Einstein-Rosen bridge to the modern renaissance of traversable wormholes. \textit{Int. J. Mod. Phys. D}, 25, 1630017. doi: 	10.1142/S0218271816300172

\bibitem{israeljunction} Kaeonikhom, C., Singleton, D., Sushkov, S. V., \& Yongram, N. (2012). Dynamics of Dirac-Born-Infeld dark energy interacting with dark matter. \textit{Phys. Rev. D}, 86, 124049. doi: 10.1103/PhysRevD.86.124049 

\bibitem{sotiriou} Sotiriou, T. P., \& Liberati, S. (2006). Field equations from a surface term. \textit{Phys. Rev. D}, 74, 4, 044016. doi: 10.1103/PhysRevD.74.044016

\bibitem{cassimir} Casimir, H. B. G. (1948). 
On the attraction between two perfectly conducting plates. \textit{Kon. Ned. Akad. Wetensch. Proc.}, 51, 793. (1987) \textit{Front.Phys.} 65 342-344.

\bibitem{Hawking}  Hawking, S. W. (1975). Particle creation by black holes. \textit{Commun. Math. Phys.}, 43, 199. doi: 10.1007/BF02345020

\bibitem{hawking2}  Hawking, S. W., Perry, M. J., \& Strominger, A. (2016). Soft hair on black holes. \textit{Phys. Rev. Lett.} 116, no.23, 231301. doi: 10.1103/PhysRevLett.116.231301   

\bibitem{Damour} Damour, T., \& Ruffini, R. (1976). Black hole evaporation in the Klein-Sauter-Heisenberg-Euler formalism. \textit{Phys. Rev. D}, 14, 332. doi: 10.1103/PhysRevD.14.332

\bibitem{ES2}  Eiroa, E. F., \& Simeone, C. (2013). Charged shells in a (2+1)-dimensional spacetime. \textit{Phys. Rev. D}, 87, 064041. doi: 10.1103/PhysRevD.87.064041 

\bibitem{5} Poisson, E. (2004). \textit{A relativist's toolkit}. Cambridge University Press, Cambridge.

\bibitem{Wilczek1}  Parikh, M. K., \& Wilczek, F. (2000). 
Hawking radiation as tunneling.  \textit{Phys. Rev. Lett.}, 85, 5042. doi: 10.1103/PhysRevLett.85.5042 

\bibitem{Mann} Kerner, R., \&  Mann, R. B. (2006). Tunnelling, temperature and Taub-NUT black holes. \textit{Phys. Rev. D}, 73, 104010. doi: 10.1103/PhysRevD.73.104010 

\bibitem{review}Vanzo, L., Acquaviva, G., \& Di Criscienzo, R. (2011). Tunnelling methods and Hawking's radiation: achievements and prospects. \textit{Class. Quantum Grav.}, 28, 183001. doi: 10.1088/0264-9381/28/18/183001 

\bibitem{Kruglov1} Kruglov, S. I. (2014). 
Black hole emission of vector particles in (1+1) dimensions. \textit{Int. J. Mod. Phys. A}, 29, 1450118. doi: 10.1142/S0217751X14501188 

\bibitem{K2} Kruglov, S. I.  (2014), Black hole radiation of spin-1 particles in (1+2) dimensions. \textit{Modern Physics Letters A}, 29, 39, 1450203. doi: 10.1142/S0217732314502034.

\bibitem{kim}  Martin-Moruno, P., \& Gonzalez-Diaz, P. F. (2009). Thermal radiation from Lorentzian traversable wormholes. \textit{Phys. Rev. D}, 80, 024007. doi: 10.1103/PhysRevD.80.024007 

\bibitem{haw}  Hayward, S. A. (1994). General laws of black hole dynamics. \textit{Phys. Rev. D}, 49, 6467. doi: 10.1103/PhysRevD.49.6467

\bibitem{kim2} Martin-Moruno, P., \& Gonzalez-Diaz, P. F. (2011). Thermal radiation from lorentzian traversable wormholes. \textit{Journal of Physics: Conference Series}, 314, 012037. doi: 10.1088/1742-6596/314/1/012037

\bibitem{Diaz} Gonzalez-Diaz, P. F. (2010). Thermal processes in ringholes. \textit{Phys. Rev. D}, 82, 044016. doi: 10.1103/PhysRevD.82.044016

\bibitem{simeo} Bejarano, C., Eiroa, E. F., \& Simeone, C. (2014). General formalism for the stability of thin-shell wormholes in 2+1 dimensions. \textit{Eur. Phys. J. C}, 74, 3015. doi: 10.1140/epjc/s10052-014-3015-z 

\bibitem{bhar} Bhar, P., \& Banerjee, A. (2015). Stability of thin-shell wormholes from noncommutative BTZ black hole. \textit{Int. J. Mod. Phys. D}, 24, 1550034. doi: 10.1142/S0218271815500340 

\bibitem{lopes}  Lemos, J. P. S., Lopes, F. J., Minamitsuji, M., \& Rocha, J. V. (2015). Thermodynamics of rotating thin shells in the BTZ spacetime. \textit{Phys. Rev. D}, 92, 064012. doi: 10.1103/PhysRevD.92.064012 

\bibitem{lopes2} Lemos, J. P. S., Lopes, F. J., Minamitsuji, M. (2015). Rotating thin shells in (2 + 1)-dimensional asymptotically AdS spacetimes: mechanical properties, machian effects, and energy conditions. \textit{Int. J. Mod. Phys. D}, 24, 1542022. doi: 10.1142/S0218271815420225

\bibitem{col} Delsate, T., Rocha, J. V., \& Santarelli, R. (2014). Collapsing thin shells with rotation. \textit{Phys. Rev. D }, 89, 121501. doi: 10.1103/PhysRevD.89.121501 

\bibitem{MV45}  Mazharimousavi S. H., \& Halilsoy, M. (2013). Charge screening by thin-shells in a 2+1-dimensional regular black hole. \textit{Eur. Phys. J. C}, 73, 2527. doi: 10.1140/epjc/s10052-013-2527-2 

\bibitem{nor3} Mazharimousavi S. H., \& Halilsoy, M. (2015). Casimir dark Energy, stabilization of the extra dimensions and Gauss-Bonnet term. \textit{Eur. Phys. J. C}, 75, 6, 271. doi: 10.1140/epjc/s10052-014-3237-0 

\bibitem{MV46} Mazharimousavi S. H., Halilsoy, M., \& Amirabi, Z. (2011). Black hole and thin-shell wormhole solutions in Einstein-Hoffman-Born-Infeld theory, \textit{Phys. Lett.
A}, 375, 3649. doi: 10.1016/j.physleta.2011.08.036 

\bibitem{BTZ} Banados, M., Teitelboim, C., \& Zanelli, J. (1992). Black hole in three-dimensional spacetime. \textit{Phys. Rev. Lett.}, 69, 184. doi: 10.1103/PhysRevLett.69.1849 

\bibitem{eid} Eid, A. (2015). Linearized stability of Reissner Nordstrom de-Sitter thin shell wormholes. \textit{New Astronomy}, 39, 72. doi: 10.1016/j.newast.2015.03.003

\bibitem{BTZ1} Martinez, C., Teitelboim, C., \& Zanelli, J. (2000). Charged rotating black hole in three spacetime dimensions. \textit{Phys. Rev. D}, 61, 104013. 	doi: 10.1103/PhysRevD.61.104013

\bibitem{ISRAEL1} Israel, W. (1966). 
Singular hypersurfaces and thin shells in general relativity. \textit{Nuovo Cimento}, 44B, 1. doi: 10.1007/BF02712210, 10.1007/BF02710419

\bibitem{lake} Ishak, M., \& Lake, K. (2002). Stability of transparent spherically symmetric thin shells and wormholes. Phys. Rev. D, 65, 044011. doi: 10.1103/PhysRevD.65.044011 

\bibitem{counter} Mazharimousavi S. H., \& Halilsoy, M. (2014). Counter-rotational effects on stability of 2+1 -dimensional thin-shell wormholes. \textit{Eur. Phys. J. C}, 74, 9, 3073. doi: 10.1140/epjc/s10052-014-3073-2 

\bibitem{charged} Mazharimousavi S. H., \& Halilsoy, M. (2015). Einstein-Maxwell gravity coupled to a scalar field in 2+1-dimensions. \textit{Eur. Phys. J. Plus}, 130, 8, 158. doi: 10.1140/epjp/i2015-15158-5

\bibitem{ao17}   Ovgun, A. (2016), Acceleration of universe by nonlinear magnetic monopole fields.  arXiv:1604.01837.

\bibitem{ao16} Sakalli, I., Ovgun, A., \& Jusufi, K. (2016), GUP assisted Hawking radiation of rotating acoustic black holes. arXiv:1602.04304.

\bibitem{ao15} Sakalli, I., \& Ovgun, A. (2016), Hawking radiation of mass generating particles from dyonic Reissner Nordstrom black hole. arXiv:1601.04040.

\bibitem{ao14}  Jusufi, K., \& Ovgun, A. (2016), Tunneling of massive vector particles from rotating charged black strings. \textit{Astrophys. Space Sci.}, 361, no.7, 207. doi: 10.1007/s10509-016-2802-4 

\bibitem{ao13} Ovgun, A., \& Halilsoy, M. (2016), Existence of traversable wormholes in the spherical stellar systems. \textit{Astrophys. Space Sci.} 361, no.7, 214. doi: 10.1007/s10509-016-2803-3 

\bibitem{ao12} Halilsoy, M., \& Ovgun, A. (2015), Particle collision near 1+1 dimensional Horava-Lifshitz black holes.  arXiv:1504.03840.

\bibitem{ao11} Halilsoy, M., \& Ovgun, A. (2015), Particle acceleration by static black holes in a model of f(R) gravity. arXiv:1507.00633.

\bibitem{ao10} Ovgun, A., \& Jusufi, K. (2016), Massive vector particles tunneling from noncommutative charged black holes and their GUP-corrected thermodynamics. \textit{Eur. Phys. J. Plus}, 131, no.5, 177. doi: 10.1140/epjp/i2016-16177-4 

\bibitem{ao9}  Ovgun, A. (2016). Entangled particles tunneling from a Schwarzschild black hole immersed in an electromagnetic universe with GUP. \textit{Int. J. Theor. Phys.} 55, 6, 2919. doi: 10.1007/s10773-016-2923-0 

\bibitem{ao8} Sakalli, I., \& Ovgun, A. (2016). Quantum tunneling of massive spin-1 particles from non-stationary metrics. \textit{Gen. Rel. Grav.}, 48, 1, 1. doi: 10.1007/s10714-015-1997-y 

\bibitem{ao7} Sakalli, I., \& Ovgun, A. (2015). Gravitinos tunneling from traversable Lorentzian wormholes. \textit{Astrophys. Space Sci}, 359, 32. doi: 10.1007/s10509-015-2482-5 

\bibitem{ao6} Sakalli, I., \& Ovgun, A. (2015). Tunnelling of vector particles from Lorentzian wormholes in 3+1 dimensions. \textit{Eur. Phys. J. Plus}, 130, 6, 110. doi: 10.1140/epjp/i2015-15110-9 

\bibitem{ao5} Sakalli, I., \& Ovgun, A. (2015). Hawking radiation of spin-1 particles from three dimensional rotating hairy black hole. \textit{J. Exp. Theor. Phys.}, 121, 3, 404. doi: 10.1134/S1063776115090113 

\bibitem{ao4} Sakalli, I., \& Ovgun, A. (2015). Uninformed Hawking radiation. \textit{Europhys. Lett.}, 110, 1, 10008. doi: 10.1209/0295-5075/110/10008 

\bibitem{ao3} Sakalli, I., Ovgun, A., \& Mirekhtiary, S. F. (2014). Gravitational lensing effect on the Hawking radiation of dyonic black holes. \textit{Int. J. Geom. Meth. Mod. Phys.}, 11, 8, 1450074. doi: 10.1142/S0219887814500741 

\bibitem{ao2} Ovgun, A., \& Sakalli, I. (2016), On a particular thin-shell wormhole. (to be appear in Theoretical and Mathematical Physics), arXiv:1507.03949.

\bibitem{ao1} Halilsoy, M., Ovgun, A., \& Mazharimousavi, S. B. (2014). Thin-shell wormholes from the regular Hayward black hole. \textit{Eur. Phys. J. C}, 74, 2796. doi: 10.1140/epjc/s10052-014-2796-4


\end{thebibliography}
\end{document}